\documentclass[preprint,prd,aps,showpacs,showkeys,nofootinbib]{revtex4}
\usepackage{graphicx}
\begin{document}


\title{The system of partial differential equations for the $C_{_0}$ function}

\author{Tai-Fu Feng\footnote{email:fengtf@hbu.edu.cn}$^{a}$,
Chao-Hsi Chang\footnote{email:zhangzx@itp.ac.cn}$^{b,c,d}$,
Jian-Bin Chen$^{e}$\footnote{email:chenjianbin@tyut.edu.cn},
Hai-Bin Zhang$^{a}$\footnote{email:hbzhang@hbu.edu.cn}}

\affiliation{$^a$Department of Physics, Hebei University, Baoding, 071002, China}
\affiliation{$^b$Key Laboratory of Theoretical Physics, Institute of Theoretical Physics,
Chinese Academy of Science, Beijing, 100190, China}
\affiliation{$^c$CCAST (World Laboratory), P.O.Box 8730, Beijing, 100190, China}
\affiliation{$^d$School of Physical Sciences, University of Chinese
Academy of Sciences, Beijing 100049, China}
\affiliation{$^e$Department of Physics, Taiyuan University of Technology, Taiyuan, 030024, China}

\begin{abstract}
We present an approach to analyze the scalar integrals of any Feynman diagrams in detail here.
This method not only completely recovers some well-known results in the literature, but also produces
some new results on the $C_{_0}$ function.
The approach can be employed to evaluate the coefficient of arbitrary power
of $\varepsilon$ in the expansion of a scalar integral, where $D=4-2\varepsilon$ denotes the time-space dimension.
\end{abstract}


\maketitle

\section{Introduction\label{sec1}}
\indent\indent
The discovery of the Higgs particle in the Large Hadron Collider (LHC) implies
the great success of the standard model (SM)\cite{CMS2012,ATLAS2012}.
With the increasing of luminosity of the collider, one of the targets
for particle physics now is to test the SM precisely and to search for
new physics (NP) beyond the SM\cite{CEPC-SPPC,ILC,HI-LHC}.

How to evaluate the scalar integrals exactly is an open problem to predict
the electroweak observables precisely in the SM.  The author of literature~\cite{V.A.Smirnov2012}
presents several methods to evaluate those scalar integrals.
Nevertheless each method mentioned in this literature
has its blemishes since it can only be applied to the Feynman diagrams with
special topology and kinematic invariants.

In dimensional regularization, any scalar integral can be expanded
around $\varepsilon=0$, here the time-space dimension $D=4-2\varepsilon$.
For example, the one-loop three-point function $C_{_0}$ is expanded as
\begin{eqnarray}
&&C_{_0}=C_{_{0}}^{(0)}+\sum\limits_{i=1}^\infty C_{_0}^{(i)}\varepsilon^i\;,
\label{C1m-1}
\end{eqnarray}
around $D=4$, and the well-known result of $C_{_{0}}^{(0)}$ is presented in the
literature\cite{tHooft1979}. However, evaluating $C_{_0}^{(i)}$
precisely is also necessary to obtain high order radiative corrections
to the amplitude self-consistently when the virtual interactions
originate from the counter terms\cite{Collins1984}.
In addition, the vector and tensor coefficients $C_{_i}$, $C_{_{00}}$,
$C_{_{ij}}$, $(i,\;j=1,\;2)$ adopted in the paper~\cite{Passarino1979}
can be expressed by the one-point function
$A_{_0}$, the two-point function $B_{_0}$, and the three-point function $C_{_0}$,
respectively.

There are many analytical results on the scalar integral $C_{_0}$ already
in the references. In Ref.~\cite{Davydychev1} the massless $C_{_0}$ function
is presented as the linear combination of the fourth kind Appell function
$F_{_4}$ whose arguments are the dimensionless ratios among the external momenta squared,
and is simplified further as the linear combination of the Gauss function $_2F_1$
through the quadratic transformation~\cite{Davydychev1993NPB} in the literature~\cite{Davydychev2000}.
With some special assumptions on the virtual masses,
the analytic expressions of the scalar integral $C_{_0}$ are given by
the multiple hypergeometric functions in Ref.~\cite{Davydychev3} through the corresponding
Mellin-Barnes representations. Taking the massless $C_{_0}$ function as
an example, the author of Ref~\cite{Davydychev1992JPA} presents an algorithm
to evaluate the scalar integrals of one-loop vertex-type Feynman diagrams.
Certainly, some analytic results of the $C_{_0}$ function can also be extracted from
the expressions for the scalar integrals of one-loop massive $N-$point
Feynman diagrams~\cite{Davydychev1991JMP,Davydychev1992JMP}. In addition,
the literature~\cite{Davydychev2006} also provides a geometrical interpretation
of the analytic expressions of the scalar integrals from one-loop $N-$point
Feynman diagrams. Using the recurrence relations respecting the time-space
dimension, the papers~\cite{Tarasov2000,Tarasov2003} formulate one-loop
two-point function $B_{_0}$ as the linear combination of the Gauss function
$_2F_1$, one-loop three-point function $C_{_0}$
with arbitrary external momenta and virtual masses as the linear combination
of the Appell function $F_{_1}$ with two arguments,
and one-loop four-point function $D_{_0}$ with arbitrary external momenta
and virtual masses as the linear combination of the Lauricella-Saran function $F_{_s}$
with three arguments, respectively. The expression for the scalar integral $C_{_0}$ is convenient
for analytic continuation and numerical evaluation because continuation of the Appell
functions has been analyzed thoroughly. Nevertheless, how to perform continuation
of the Lauricella-Saran function $F_{_s}$ outside its convergent domain is still a challenge.
Basing on the hypergeometric system of linear partial differential equations (PDEs),
we present an approach to evaluate the scalar integral of arbitrary multiloop Feynman
diagram systematically. Actually the system of PDEs satisfied by the
corresponding scalar integral is a holonomic integrable system which can be transformed
into the Pfaffian system of PDEs~\cite{M.E.Taylor12} by the algebra decomposition.
Then one can obtain the singularities of the holonomic hypergeometric system
and the number of independent solutions in certain parameter space.
Together with the analytic expressions in some convergent regions,
the holonomic hypergeometric system provides a new way to understand
the corresponding scalar integral. Taking the $C_{_0}$ function as an example, we elucidate
how to evaluate the coefficient of arbitrary power of $\varepsilon$ in the expansion of
a scalar integral around $D=4$. In fact, the corresponding analytic results
on the $C_{_0}$ function coincide with those well-known results mentioned above.
The evaluations of scalar integrals of any Feynman diagrams such as that presented in
Refs.~\cite{Davydychev1993,V.A.Smirnov1999,J.B.Tausk1999} are given elsewhere.

A holonomic hypergeometric system of linear PDEs
is given through the corresponding Mellin-Barnes representation of the concerned
scalar integral~\cite{M.Y.Kalmykov12}, where the system of linear PDEs is satisfied by the scalar integral
in the whole parameter space of the independent variables.
The scalar integral is written as the multiple hypergeometric
functions~\cite{L.J.Slater66} of the independent variables for some isolated singularities
of the integrand by residue theorem~\cite{J.Leray1959}.
In addition, the Horn's study of convergence\cite{Horn1889} predicts the
absolutely and uniformly convergent regions of those derived hypergeometric functions
exactly. Several convergent regions of the hypergeometric functions compose a set in which
each convergent region does not intersect with the others. Additionally
each convergent region of the set either does not intersect with,
or is a proper subset of other convergent regions which do not belong to the set.
Within each convergent region in the set, the scalar
integral can be written as the sum of those hypergeometric functions
whose convergent regions contain the concerned element entirely.

In view of mathematics, the most important point for a given multiple power series is how to
get its absolutely and uniformly convergent region, and how to continue
it to the whole parameter space.
Fortunately, the continuation can be achieved by the system of linear PDEs mentioned above.
As stated in our previous work\cite{Feng2018},
an effective Hamiltonian can be constructed if those linear PDEs are
the stationary conditions of the corresponding system. Then
we can numerically continue the scalar integral to the entire parameter space
with the finite element method.

Our presentation is organized as follows. We briefly mention some typical results
of the massive $C_{_0}$ function in section \ref{sec2} at first.  Then we present in detail
our analyses on the massless $C_{_0}$ function in section \ref{sec3},
the $C_{_0}$ function with one nonzero virtual mass in section \ref{sec4},
and the $C_{_0}$ function with three equally virtual masses in section \ref{sec5}, respectively.
In section \ref{sec6}, we recognize the system of linear PDEs as the
stationary conditions of a functional under some restrictions according to
Hamilton's principle, which is convenient to numerically continue
the scalar integral to the whole parameter space with the finite element method.
The conclusions are summarized in section \ref{sec7} and some tedious formulae are presented in the appendixes.

\section{Some results of the $C_{_0}$ function in general case\label{sec2}}
\indent\indent
The Mellin-Barnes representation of massive $C_{_0}$ function is generally written as
\begin{eqnarray}
&&C_{_0}(p_{_1}^2,\;p_{_2}^2,\;p_{_3}^2,\;m_{_1}^2,\;m_{_2}^2,\;m_{_3}^2)
\nonumber\\
&&\hspace{-0.5cm}=
\int{d^Dq\over(2\pi)^D}{1\over(q^2-m_{_3}^2)((q+p_{_1})^2-m_{_2}^2)
((q-p_{_2})^2-m_{_1}^2)}
\nonumber\\
&&\hspace{-0.5cm}=
-{i(-)^{D/2-3}(p_{_3}^2)^{D/2-3}\over(4\pi)^{D/2}(2\pi i)^5}
\int_{-i\infty}^{+i\infty}ds_{_1}\Big(-{m_{_1}^2\over p_{_3}^2}\Big)^{s_{_1}}\Gamma(-s_{_1})
\int_{-i\infty}^{+i\infty}ds_{_2}\Big(-{m_{_2}^2\over p_{_3}^2}\Big)^{s_{_2}}\Gamma(-s_{_2})
\nonumber\\
&&\hspace{0.0cm}\times
\int_{-i\infty}^{+i\infty}ds_{_3}\Big(-{m_{_3}^2\over p_{_3}^2}\Big)^{s_{_3}}\Gamma(-s_{_3})
\int_{-i\infty}^{+i\infty}dz_{_1}\int_{-i\infty}^{+i\infty}dz_{_2}
\Gamma(-z_{_1})\Gamma(-z_{_2})
\nonumber\\
&&\hspace{0.0cm}\times
\Gamma(3-{D\over2}+s_{_1}+s_{_2}+s_{_3}+z_{_1}+z_{_2})
\Big({p_{_1}^2\over p_{_3}^2}\Big)^{z_{_1}}\Big({p_{_2}^2\over p_{_3}^2}\Big)^{z_{_2}}
\nonumber\\
&&\hspace{0.0cm}\times
{\Gamma({D\over2}-2-s_{_1}-s_{_3}-z_{_2})\Gamma({D\over2}-2-s_{_2}-s_{_3}-z_{_1})
\Gamma(1+s_{_3}+z_{_1}+z_{_2})\over\Gamma(D-3-s_{_1}-s_{_2}-s_{_3})}
\nonumber\\
&&\hspace{-0.5cm}=
{i(p_{_3}^2)^{D/2-3}\over(4\pi)^{D/2}}F_{_{p_{_3}}}(\xi_{_{13}},\;\xi_{_{23}},\;\xi_{_{33}},\;
x_{_{13}},\;x_{_{23}})
\nonumber\\
&&\hspace{-0.5cm}=
{i(m_{_3}^2)^{D/2-3}\over(4\pi)^{D/2}}F_{_{m_{_3}}}(\eta_{_{13}},\;\eta_{_{23}},\;\eta_{_{33}},\;
y_{_{13}},\;y_{_{23}})
\label{1LoopC76}
\end{eqnarray}
with $p_{_3}^2=(p_{_1}+p_{_2})^2$, $\xi_{_{ij}}=-m_{_i}^2/p_{_j}^2$,
$\eta_{_{ij}}=-p_{_i}^2/m_{_j}^2=1/\xi_{_{ji}}$, $x_{_{ij}}=p_{_i}^2/p_{_j}^2$,
and $y_{_{ij}}=m_{_i}^2/m_{_j}^2\;(i,\;j=1,2,3)$, respectively.
Here the dimensionless functions $F_{_{p_{_3}}}$, $\xi_{_{33}}^{D/2-3}F_{_{m_{_3}}}$
comply with the holonomic hypergeometric system of linear PDEs
\begin{eqnarray}
&&\Big\{\Big[3-{D\over2}+\sum\limits_{i=1}^3\hat{\vartheta}_{_{\xi_{_{i3}}}}
+\sum\limits_{i=1}^2\hat{\vartheta}_{_{x_{_{i3}}}}\Big]\Big[1+\hat{\vartheta}_{_{\xi_{_{33}}}}
+\sum\limits_{i=1}^2\hat{\vartheta}_{_{x_{_{i3}}}}\Big]
\Big[4-D+\sum\limits_{i=1}^3\hat{\vartheta}_{_{\xi_{_{i3}}}}\Big]
\nonumber\\
&&\hspace{0.0cm}
-{1\over\xi_{_{33}}}\hat{\vartheta}_{_{\xi_{_{33}}}}
\prod\limits_{i=1}^3\Big[2-{D\over2}+\hat{\vartheta}_{_{\xi_{_{i3}}}}+\hat{\vartheta}_{_{\xi_{_{33}}}}
+\hat{\vartheta}_{_{x_{_{(3-i)3}}}}\Big]\Big\}F_{_{p_{_3}}}=0
\;,\nonumber\\
&&\Big\{\Big[3-{D\over2}+\sum\limits_{i=1}^3\hat{\vartheta}_{_{\xi_{_{i3}}}}
+\sum\limits_{i=1}^2\hat{\vartheta}_{_{x_{_{i3}}}}\Big]
\Big[4-D+\sum\limits_{i=1}^3\hat{\vartheta}_{_{\xi_{_{i3}}}}\Big]
\nonumber\\
&&\hspace{0.0cm}
+{1\over\xi_{_{j3}}}\hat{\vartheta}_{_{\xi_{_{j3}}}}
\Big[2-{D\over2}+\hat{\vartheta}_{_{\xi_{_{33}}}}+\hat{\vartheta}_{_{\xi_{_{j3}}}}
+\hat{\vartheta}_{_{x_{_{(3-j)3}}}}\Big]\Big\}F_{_{p_{_3}}}=0
\;,\nonumber\\
&&\Big\{\Big[3-{D\over2}+\sum\limits_{i=1}^3\hat{\vartheta}_{_{\xi_{_{i3}}}}
+\sum\limits_{i=1}^2\hat{\vartheta}_{_{x_{_{i3}}}}\Big]\Big[1+\hat{\vartheta}_{_{\xi_{_{33}}}}
+\sum\limits_{i=1}^2\hat{\vartheta}_{_{x_{_{i3}}}}\Big]
\nonumber\\
&&\hspace{0.0cm}
-{1\over x_{_{j3}}}\hat{\vartheta}_{_{x_{_{j3}}}}
\Big[2-{D\over2}+\hat{\vartheta}_{_{\xi_{_{33}}}}+\hat{\vartheta}_{_{\xi_{_{(3-j)3}}}}
+\hat{\vartheta}_{_{x_{_{j3}}}}\Big]\Big\}F_{_{p_{_3}}}=0,\;(j=1,\;2)\;,
\label{1LoopC77}
\end{eqnarray}
with the Euler operator $\hat{\vartheta}_{_x}=x\partial/\partial x$ etc.
Or equivalently, the functions $F_{_{m_{_3}}}$, $\eta_{_{33}}^{D/2-3}F_{_{p_{_3}}}$
satisfy the following holonomic system of linear PDEs
\begin{eqnarray}
&&\Big\{\Big[3-{D\over2}+\sum\limits_{i=1}^2\hat{\vartheta}_{_{y_{_{i3}}}}
+\sum\limits_{i=1}^3\hat{\vartheta}_{_{\eta_{_{i3}}}}\Big]\prod\limits_{i=1}^2
\Big[1+\hat{\vartheta}_{_{\eta_{_{i3}}}}
+\hat{\vartheta}_{_{\eta_{_{33}}}}+\hat{\vartheta}_{_{y_{_{(3-i)3}}}}\Big]
\nonumber\\
&&\hspace{0.0cm}
-{1\over\eta_{_{33}}}\hat{\vartheta}_{_{\eta_{_{33}}}}\Big[{D\over2}-1
+\sum\limits_{i=1}^3\hat{\vartheta}_{_{\eta_{_{i3}}}}\Big]
\Big[2-{D\over2}+\sum\limits_{i=1}^2\hat{\vartheta}_{_{y_{_{i3}}}}
+\hat{\vartheta}_{_{\eta_{_{33}}}}\Big]\Big\}F_{_{m_{_3}}}=0\;,
\nonumber\\
&&\Big\{\Big[3-{D\over2}+\sum\limits_{i=1}^2\hat{\vartheta}_{_{y_{_{i3}}}}
+\sum\limits_{i=1}^3\hat{\vartheta}_{_{\eta_{_{i3}}}}\Big]
\Big[1+\hat{\vartheta}_{_{\eta_{_{j3}}}}
+\hat{\vartheta}_{_{\eta_{_{33}}}}+\hat{\vartheta}_{_{y_{_{(3-j)3}}}}\Big]
\nonumber\\
&&\hspace{0.0cm}
+{1\over\eta_{_{j3}}}\hat{\vartheta}_{_{\eta_{_{j3}}}}\Big[{D\over2}-1
+\sum\limits_{i=1}^3\hat{\vartheta}_{_{\eta_{_{i3}}}}\Big]\Big\}F_{_{m_{_3}}}=0\;,
\nonumber\\
&&\Big\{\Big[3-{D\over2}+\sum\limits_{i=1}^2\hat{\vartheta}_{_{y_{_{i3}}}}
+\sum\limits_{i=1}^3\hat{\vartheta}_{_{\eta_{_{i3}}}}\Big]
\Big[1+\hat{\vartheta}_{_{\eta_{_{(3-j)3}}}}
+\hat{\vartheta}_{_{\eta_{_{33}}}}+\hat{\vartheta}_{_{y_{_{j3}}}}\Big]
\nonumber\\
&&\hspace{0.0cm}
-{1\over y_{_{j3}}}\hat{\vartheta}_{_{y_{_{j3}}}}\Big[2-{D\over2}
+\sum\limits_{i=1}^2\hat{\vartheta}_{_{y_{_{i3}}}}
+\hat{\vartheta}_{_{\eta_{_{33}}}}\Big]\Big\}F_{_{m_{_3}}}=0\;,\;\;(j=1,\;2)\;.
\label{1LoopC77-a}
\end{eqnarray}
Here we emphasize that the two systems in
Eq.(\ref{1LoopC77}) and Eq.(\ref{1LoopC77-a}) are equivalent.
Actually the Mellin-Barnes representation presented here is
equivalent to that of Eq.~(3.4) in Ref.~\cite{Davydychev1992JMP}.
In addition, the author of Ref.~\cite{Davydychev1992JMP} also
presents a single hypergeometric function originating from the Mellin-Barnes representation
in Eq.~(3.6). With our notation, the corresponding hypergeometric function is rewritten as
\begin{eqnarray}
&&C_{_{m_{_3}}}(p_{_1}^2,\;p_{_2}^2,\;p_{_3}^2,\;m_{_1}^2,\;m_{_2}^2,\;m_{_3}^2)
={i(m_{_3}^2)^{D/2-3}\over(4\pi)^{D/2}}
\sum\limits_{j_{_1}=0}^\infty\sum\limits_{j_{_2}=0}^\infty
\sum\limits_{n_{_1}=0}^\infty\sum\limits_{n_{_2}=0}^\infty\sum\limits_{n_{_3}=0}^\infty
\nonumber\\
&&\hspace{2.8cm}\times
{(-)^{n_{_1}+n_{_2}+n_{_3}}\over j_{_1}!j_{_2}!n_{_1}!n_{_2}!n_{_3}!}
\Big[1-y_{_{13}}\Big]^{j_{_1}}\Big[1-y_{_{23}}\Big]^{j_{_2}}\eta_{_{13}}^{n_{_1}}\eta_{_{23}}^{n_{_2}}
\eta_{_{33}}^{n_{_3}}
\nonumber\\
&&\hspace{2.8cm}\times
{\Gamma(3-{D\over2}+j_{_1}+j_{_2}+\sum\limits_{i=1}^3n_{_i})
\Gamma(1+j_{_1}+n_{_2}+n_{_3})\over\Gamma(3+j_{_1}+j_{_2}+2\sum\limits_{i=1}^3n_{_i})}
\nonumber\\
&&\hspace{2.8cm}\times
\Gamma(1+j_{_2}+n_{_1}+n_{_3})\Gamma(1+n_{_1}+n_{_2})\;,
\label{1LoopC77-b}
\end{eqnarray}
which obviously satisfies the holonomic hypergeometric systems in
Eq.(\ref{1LoopC77}) and Eq.(\ref{1LoopC77-a}).
Using the adjacent ratios of the coefficients, we define
\begin{eqnarray}
&&\Phi_{_{j_k}}(j_{_1},j_{_2},n_{_1},n_{_2},n_{_3})
={(j_{_1}+j_{_2}+n_{_1}+n_{_2}+n_{_3})(j_{_k}+n_{_{3-k}}+n_{_3})
\over j_{_k}(j_{_1}+j_{_2}+2n_{_1}+2n_{_2}+2n_{_3})}
\;,\nonumber\\
&&\Phi_{_{n_k}}(j_{_1},j_{_2},n_{_1},n_{_2},n_{_3})
={(j_{_1}+j_{_2}+n_{_1}+n_{_2}+n_{_3})(j_{_{3-k}}+n_{_{k}}+n_{_3})
(n_{_1}+n_{_2})\over n_{_k}(j_{_1}+j_{_2}+2n_{_1}+2n_{_2}+2n_{_3})^2}
\;,\nonumber\\
&&\Phi_{_{n_3}}(j_{_1},j_{_2},n_{_1},n_{_2},n_{_3})
={(j_{_1}+j_{_2}+n_{_1}+n_{_2}+n_{_3})\prod\limits_{i=1}^2
(j_{_{3-i}}+n_{_{i}}+n_{_3})\over n_{_3}(j_{_1}+j_{_2}+2n_{_1}+2n_{_2}+2n_{_3})^2}
,\;(k=1,\;2).
\label{1LoopC77-c}
\end{eqnarray}
The Cartesian hypersurface of the multiple hypergeometric
series in Eq.(\ref{1LoopC77-b}) is written as
\begin{eqnarray}
&&r_{_1}^2r_{_3}-r_{_1}r_{_2}r_{_3}+r_{_3}^2-r_{_1}r_{_3}^2-r_{_1}r_{_2}r_{_4}
+r_{_2}^2r_{_4}-2r_{_3}r_{_4}+r_{_1}r_{_3}r_{_4}+r_{_2}r_{_3}r_{_4}
\nonumber\\
&&+r_{_4}^2-r_{_2}r_{_4}^2+r_{_1}r_{_2}r_{_5}-2r_{_3}r_{_5}+r_{_1}r_{_3}r_{_5}-2r_{_4}r_{_5}
+r_{_2}r_{_4}r_{_5}+r_{_3}r_{_4}r_{_5}+r_{_5}^2=0\;.
\label{1LoopC77-d}
\end{eqnarray}
with $r_{_i}=|y_{_{i3}}-1|,\;r_{_{2+i}}=|\eta_{_{i3}}|,\;(i=1,\;2),\;
r_{_5}=|\eta_{_{33}}|$, respectively.
Correspondingly the absolutely and uniformly convergent domain of the series is
\begin{eqnarray}
&&\Omega_{_1}^{\prime,a}=
\left\{(\eta_{_{13}},\;\eta_{_{23}},\;\eta_{_{33}},\;y_{_{13}},\;y_{_{23}}):
\left|\begin{array}{l} |\eta_{_{13}}|<4,\;|\eta_{_{23}}|<4,\;|\eta_{_{33}}|<8,\\
|y_{_{13}}-1|<1,\;|y_{_{23}}-1|<1
\end{array}\right.\right\}\;.
\label{1LoopC77-e}
\end{eqnarray}

Through the recurrence relations with respect to the time-space dimension~\cite{Tarasov2000, Tarasov2003},
the scalar integral $C_{_0}$ can also be presented by the linear combination
of the first kind Appell hypergeometric function $F_{_1}$. Where the independent
variables of the hypergeometric functions are dimensionless ratios among
the parameters $r_{_{12}}$, $r_{_{23}}$, $r_{_{13}}$, $r_{_{123}}$, and $m_{_{i}},\;(i=1,\;2,\;3)$,
and the linear combination coefficients depend on the parameters
$\lambda_{_{12}}$, $\lambda_{_{23}}$, $\lambda_{_{13}}$, and $g_{_{123}}$.
With our notation above, those parameters are expressed as
\begin{eqnarray}
&&\lambda_{_{12}}=\kappa^2(m_{_{1}}^2,\;m_{_{2}}^2,\;p_{_{3}}^2),\;\;\;
\lambda_{_{13}}=\kappa^2(m_{_{1}}^2,\;p_{_{2}}^2,\;m_{_{3}}^2),
\nonumber\\
&&\lambda_{_{23}}=\kappa^2(p_{_{1}}^2,\;m_{_{2}}^2,\;m_{_{3}}^2),\;\;\;
g_{_{123}}=-2\kappa^2(p_{_{1}}^2,\;p_{_{2}}^2,\;p_{_{3}}^2),
\nonumber\\
&&\lambda_{_{123}}=8m_{_{1}}^2m_{_{2}}^2m_{_{3}}^2-2m_{_{1}}^2(m_{_{2}}^2+m_{_{3}}^2-p_{_{1}}^2)^2
\nonumber\\
&&\hspace{1.3cm}
-2m_{_{2}}^2(m_{_{1}}^2+m_{_{3}}^2-p_{_{2}}^2)^2-2m_{_{3}}^2(m_{_{1}}^2+m_{_{2}}^2-p_{_{3}}^2)^2
\nonumber\\
&&\hspace{1.3cm}
+2(m_{_{1}}^2+m_{_{2}}^2-p_{_{3}}^2)(m_{_{1}}^2+m_{_{3}}^2-p_{_{2}}^2)(m_{_{2}}^2+m_{_{3}}^2-p_{_{1}}^2)\;,
\nonumber\\
&&r_{_{12}}=-{\lambda_{_{12}}\over4p_{_3}^2},\;\;\;r_{_{13}}=-{\lambda_{_{13}}\over4p_{_2}^2},\;\;\;
r_{_{23}}=-{\lambda_{_{23}}\over4p_{_1}^2},\;\;\;r_{_{123}}={\lambda_{_{123}}\over g_{_{123}}}\;,
\label{1LoopC77-f}
\end{eqnarray}
with the K\"all\`en function squared
\begin{eqnarray}
&&\kappa^2(x,y,z)=2xy+2yz+2zx-x^2-y^2-z^2\;.
\label{1LoopC77-g}
\end{eqnarray}
Using the preparation above, one can check that the analytic expression in Eq.~(74)
of Ref.~\cite{Tarasov2003} complies with the holonomic hypergeometric systems in
Eq.(\ref{1LoopC77}).

In order to apply the residue theorem, we should choose five $\Gamma$ functions
in the numerator of the integrand to determine the poles because the Mellin-Barnes
representation in Eq.~(\ref{1LoopC76}) is a five-fold contour integral.
There are $C_9^5=126$ different choices in all.
When the poles are determined by
some specific choices of $\Gamma$ functions, for example, $\Gamma(-s_{_3})$,
$\Gamma(-z_{_1})$, $\Gamma(-z_{_2})$, $\Gamma(1+s_{_3}+z_{_1}+z_{_2})$,
and $\Gamma(3-{D\over2}+s_{_1}+s_{_2}+s_{_3}+z_{_1}+z_{_2})$,
the Jacobian of the transformation
\begin{eqnarray}
&&-s_{_3}^\prime=-s_{_3},\;\nonumber\\
&&-s_{_1}^\prime=3-{D\over2}+s_{_1}+s_{_2}+s_{_3}+z_{_1}+z_{_2},\;\nonumber\\
&&-s_{_2}^\prime=1+s_{_3}+z_{_1}+z_{_2},\;\nonumber\\
&&-z_{_1}^\prime=-z_{_1},\;\nonumber\\
&&-z_{_2}^\prime=-z_{_2}
\label{1LoopC77-1}
\end{eqnarray}
is zero. Correspondingly this special choice of poles in the
numerator of the integrand does not induce correction to the Feynman
integral at all~\cite{J.Leray1959}. For other choices of poles that
the Jacobian of the corresponding transformation is not zero, the Feynman integral
is written as a quintuple hypergeometric function.

In the hypergeometric function mentioned above, the coefficient of the power function of
independent variables is a fraction, whose numerator
and denominator are both products of several $\Gamma$ functions.
Where the independent variable of each $\Gamma$ function is a linear combination
of the time-space dimension and the summation indices, and the coefficients
of the summation indices are either positive or negative integers.
Because of this reason, the hypergeometric function is decomposed into the sum of several hypergeometric
series,  so that the independent variables of all $\Gamma$ functions
of each 'new' hypergeometric series have definite signs at $D=4$.
After the transformation of summation indices, we obtain
the absolutely and uniformly convergent region of each 'new' hypergeometric series
with Horn's study of convergence~\cite{Horn1889}.
The intersection of all convergent regions is the absolutely and uniformly
convergent region of the original hypergeometric function.

Some convergent regions of the quintuple hypergeometric functions compose a set,
and each convergent region does not intersect with the others in the set. Additionally each convergent region
in the set either does not intersect with,
or is a proper subset of the other convergent region which does not belong to the set.
In each convergent region of the set, the Feynman
integral can be written as the sum of those hypergeometric functions
whose convergent regions contain the concerned region of the set entirely.

In order to shorten the length of text, we apply the approach to
the Feynman integrals of the $C_{_0}$ function with some special assumptions
on the virtual masses.

\section{The scalar integral of massless $C_{_0}$ function\label{sec3}}
\indent\indent
Generally the Mellin-Barnes representation of massless $C_{_0}$ function is
simplified as
\begin{eqnarray}
&&C_{_0}(p_{_1}^2,\;p_{_2}^2,\;p_{_3}^2)=\int{d^Dq\over(2\pi)^D}{1\over q^2(q+p_{_1})^2(q-p_{_2})^2}
\nonumber\\
&&\hspace{2.6cm}=
{i(-)^{D/2}\over(4\pi)^{D/2}\Gamma(D-3)(2\pi i)^2}\int_{-i\infty}^{+i\infty}dz_{_1}
\int_{-i\infty}^{+i\infty}dz_{_2}(p_{_1}^2)^{z_{_1}}(p_{_2}^2)^{z_{_2}}
\nonumber\\
&&\hspace{3.1cm}\times
(p_{_3}^2)^{D/2-3-z_{_1}-z_{_2}}\Gamma(-z_{_1})\Gamma(-z_{_2})
\Gamma(3-{D\over2}+z_{_1}+z_{_2})
\nonumber\\
&&\hspace{3.1cm}\times
\Gamma(1+z_{_1}+z_{_2})\Gamma({D\over2}-2-z_{_1})
\Gamma({D\over2}-2-z_{_2})\;,
\label{Massless-1}
\end{eqnarray}
which coincides with Eq.~(24) of Ref.~\cite{Davydychev3} (i.e. Eq.~(2.5)
of Ref.~\cite{Davydychev1992JPA}). The analytic expression of the
massless $C_{_0}$ function is obtained by summing over residues of the integrand,
and the corresponding results are enumerated as following.
\begin{itemize}
\item[$\bullet$ 1(a):]After summing over the residues of $\Gamma(-z_{_1})$ and $\Gamma(-z_{_2})$,
we obtain the double hypergeometric function as
\begin{eqnarray}
&&C_{_{0,1a}}(p_{_1}^2,\;p_{_2}^2,\;p_{_3}^2)=
{i(-)^{D/2}(p_{_3}^2)^{D/2-3}\over(4\pi)^{D/2}\Gamma(D-3)}\sum\limits_{n_{_1}=0}^\infty
\sum\limits_{n_{_2}=0}^\infty{(-)^{n_{_1}+n_{_2}}\over n_{_1}!n_{_2}!}
\Big({p_{_1}^2\over p_{_3}^2}\Big)^{n_{_1}}\Big({p_{_2}^2\over p_{_3}^2}\Big)^{n_{_2}}
\nonumber\\
&&\hspace{3.1cm}\times
\Gamma(1+n_{_1}+n_{_2})\Gamma(3-{D\over2}+n_{_1}+n_{_2})\Gamma({D\over2}-2-n_{_1})
\nonumber\\
&&\hspace{3.1cm}\times
\Gamma({D\over2}-2-n_{_2})
\nonumber\\
&&\hspace{2.7cm}=
{i(-)^{D/2}(p_{_3}^2)^{D/2-3}\over(4\pi)^{D/2}}
{\Gamma^2(2-{D\over2})\Gamma^2({D\over2}-1)\over\Gamma(D-3)}
\sum\limits_{n_{_1}=0}^\infty\sum\limits_{n_{_2}=0}^\infty
\nonumber\\
&&\hspace{3.1cm}\times
A_{_{n_{_1},n_{_2}}}^{(1a)}x_{_{13}}^{n_{_1}}x_{_{23}}^{n_{_2}}\;.
\label{Massless-2}
\end{eqnarray}
The necessary condition $|p_{_3}^2|>\max(|p_{_1}^2|,\;|p_{_2}^2|)$
(i.e. $|x_{_{i3}}|<1$) should be satisfied to guarantee the uniform absolute-convergence
of the double series. Here the adjacent ratios of the coefficients are
\begin{eqnarray}
&&\Phi_{_1}^\prime(n_{_1},n_{_2})={A_{_{1+n_{_1},n_{_2}}}^{(1a)}\over A_{_{n_{_1},n_{_2}}}^{(1a)}}
={(1+n_{_1}+n_{_2})(3-D/2+n_{_1}+n_{_2})\over(1+n_{_1})(3-D/2+n_{_1})}
\;,\nonumber\\
&&\Phi_{_2}^\prime(n_{_1},n_{_2})={A_{_{n_{_1},1+n_{_2}}}^{(1a)}\over A_{_{n_{_1},n_{_2}}}^{(1a)}}
={(1+n_{_1}+n_{_2})(3-D/2+n_{_1}+n_{_2})\over(1+n_{_2})(3-D/2+n_{_2})}\;.
\label{Massless-3}
\end{eqnarray}
In order to investigate the absolutely and uniformly convergent
region of the double series in Eq.(\ref{Massless-2}), we define
\begin{eqnarray}
&&\Phi_{_1}(n_{_1},n_{_2})=\lim\limits_{\lambda\rightarrow\infty}
\Phi_{_1}^\prime(\lambda n_{_1},\lambda n_{_2})
={(n_{_1}+n_{_2})^2\over n_{_1}^2}
\;,\nonumber\\
&&\Phi_{_2}(n_{_1},n_{_2})=\lim\limits_{\lambda\rightarrow\infty}
\Phi_{_2}^\prime(\lambda n_{_1},\lambda n_{_2})
={(n_{_1}+n_{_2})^2\over n_{_2}^2}\;.
\label{Massless-4}
\end{eqnarray}
The Cartesian curve of the double power series in Eq.(\ref{Massless-2}) is
\begin{eqnarray}
&&\sqrt{r_{_1}}+\sqrt{r_{_2}}=1\;,
\label{Massless-5}
\end{eqnarray}
with $r_{_i}=|x_{_{i3}}|$. For convenience we denote the region
surrounded by the coordinate axes and the Cartesian curve in the
positive quadrant of the plane $Or_{_1}r_{_2}$ by $C$, and denote
the rectangle by $D$ in the positive quadrant of the plane $Or_{_1}r_{_2}$
bounded by the coordinate axes and the straight lines parallel
to the coordinate axes $r_{_1}=1/\Phi_{_1}(1,0)=1$, and $r_{_2}=1/\Phi_{_2}(0,1)=1$.
According to Horn's study of convergence of the hypergeometric series \cite{Horn1889},
one finds the well-known conclusion \cite{L.J.Slater66} that
the double power series in Eq.(\ref{Massless-2}) absolutely and uniformly converges
in the intersection of the regions $C$ and $D$ in the plane $Or_{_1}r_{_2}$.
In other words, the absolutely and uniformly convergent region of
the double power series in Eq.(\ref{Massless-2}) is written as
\begin{eqnarray}
&&\Omega_{_1}^\prime=
\left\{(x_{_{13}},\;x_{_{23}}):\left|\begin{array}{l}
|x_{_{13}}|<1,\;|x_{_{23}}|<1,\;\sqrt{|x_{_{13}}|}+\sqrt{|x_{_{23}}|}<1
\end{array}\right.\right\}\;.
\label{Convergent-region1}
\end{eqnarray}

\item[$\bullet$ 1(b):] For the isolated singularities defined by the poles of
$\Gamma(-z_{_1})$, $\Gamma({D\over2}-2-z_{_2})$, the corresponding
double hypergeometric function is written as
\begin{eqnarray}
&&C_{_{0,1b}}(p_{_1}^2,\;p_{_2}^2,\;p_{_3}^2)=
-{i(-)^{D/2}(p_{_3}^2)^{D/2-3}\over(4\pi)^{D/2}}
{\Gamma^2(2-{D\over2})\Gamma^2({D\over2}-1)\over\Gamma(D-3)}
x_{_{23}}^{D/2-2}
\nonumber\\
&&\hspace{3.1cm}\times
\sum\limits_{n_{_1}=0}^\infty\sum\limits_{n_{_2}=0}^\infty
{\Gamma(1+n_{_1}+n_{_2})
\Gamma({D\over2}-1+n_{_1}+n_{_2})\over n_{_1}!n_{_2}!\Gamma(3-{D\over2}+n_{_1})
\Gamma({D\over2}-1+n_{_2})}x_{_{13}}^{n_{_1}}x_{_{23}}^{n_{_{2}}}\;.
\label{Massless-1b}
\end{eqnarray}
Similarly the condition $|p_{_3}^2|>\max(|p_{_1}^2|,\;|p_{_2}^2|)$
should be satisfied to guarantee the double series converging.
Using Horn's theory of convergence, one obtains that
the double hypergeometric series above is absolutely and uniformly
convergent in the region Eq.(\ref{Convergent-region1}).

\item[$\bullet$ 1(c):] For the isolated singularities defined by the poles of
$\Gamma(-z_{_2})$, $\Gamma({D\over2}-2-z_{_1})$, the corresponding
double hypergeometric function is written as
\begin{eqnarray}
&&C_{_{0,1c}}(p_{_1}^2,\;p_{_2}^2,\;p_{_3}^2)=
-{i(-)^{D/2}(p_{_3}^2)^{D/2-3}\over(4\pi)^{D/2}}
{\Gamma^2(2-{D\over2})\Gamma^2({D\over2}-1)\over\Gamma(D-3)}
x_{_{13}}^{D/2-2}
\nonumber\\
&&\hspace{3.1cm}\times
\sum\limits_{n_{_1}=0}^\infty\sum\limits_{n_{_2}=0}^\infty
{\Gamma(1+n_{_1}+n_{_2})
\Gamma({D\over2}-1+n_{_1}+n_{_2})\over n_{_1}!n_{_2}!\Gamma({D\over2}-1+n_{_1})
\Gamma(3-{D\over2}+n_{_2})}x_{_{13}}^{n_{_1}}x_{_{23}}^{n_{_2}}\;.
\label{Massless-1c}
\end{eqnarray}
Certainly the essential condition $|p_{_3}^2|>\max(|p_{_1}^2|,\;|p_{_2}^2|)$
should be satisfied to guarantee the convergence of the double series.

\item[$\bullet$ 1(d):] For the isolated singularities defined by the poles of
$\Gamma({D\over2}-2-z_{_1})$, $\Gamma({D\over2}-2-z_{_2})$, the corresponding
double hypergeometric function is written as
\begin{eqnarray}
&&C_{_{0,1d}}(p_{_1}^2,\;p_{_2}^2,\;p_{_3}^2)=
{i(-)^{D/2}(p_{_3}^2)^{D/2-3}\over(4\pi)^{D/2}}
{\Gamma^2(2-{D\over2})\Gamma^2({D\over2}-1)\over\Gamma(D-3)}
(x_{_{13}}x_{_{23}})^{D/2-2}
\nonumber\\
&&\hspace{3.1cm}\times
\sum\limits_{n_{_1}=0}^\infty\sum\limits_{n_{_2}=0}^\infty
{\Gamma(D-3+n_{_1}+n_{_2})
\Gamma({D\over2}-1+n_{_1}+n_{_2})\over n_{_1}!n_{_2}!\Gamma({D\over2}-1+n_{_1})
\Gamma({D\over2}-1+n_{_2})}x_{_{13}}^{n_{_1}}x_{_{23}}^{n_{_2}}\;.
\label{Massless-1d}
\end{eqnarray}
Correspondingly the necessary condition $|p_{_3}^2|>\max(|p_{_1}^2|,\;|p_{_2}^2|)$
should be satisfied to guarantee the convergence of the double series.
The absolutely and uniformly convergent region of Eq.(\ref{Massless-1c})
and Eq.(\ref{Massless-1d}) is also provided in Eq.(\ref{Convergent-region1})
concretely.

\item[$\bullet$ 2(a):] For the isolated singularities defined by the poles of
$\Gamma(1+z_{_1}+z_{_2})$, $\Gamma(-z_{_2})$, the corresponding
double hypergeometric function is written as
\begin{eqnarray}
&&C_{_{0,2a}}(p_{_1}^2,\;p_{_2}^2,\;p_{_3}^2)=
C_{_{0,1c}}(p_{_3}\rightarrow p_{_1},\;x_{_{13}}\rightarrow x_{_{31}},\;
x_{_{23}}\rightarrow x_{_{21}})\;.
\label{Massless-2a}
\end{eqnarray}
The necessary condition $|p_{_1}^2|>\max(|p_{_2}^2|,\;|p_{_3}^2|)$
should be satisfied to guarantee the convergence of the double series.

\item[$\bullet$ 2(b):] For the isolated singularities defined by the poles of
$\Gamma(3-{D\over2}+z_{_1}+z_{_2})$, $\Gamma(-z_{_2})$, the corresponding
double hypergeometric function is written as
\begin{eqnarray}
&&C_{_{0,2b}}(p_{_1}^2,\;p_{_2}^2,\;p_{_3}^2)=C_{_{0,1a}}(
p_{_3}\rightarrow p_{_1},\;x_{_{13}}\rightarrow x_{_{31}},\;
x_{_{23}}\rightarrow x_{_{21}})\;.
\label{Massless-2b}
\end{eqnarray}

\item[$\bullet$ 2(c):] For the isolated singularities defined by the poles of
$\Gamma(1+z_{_1}+z_{_2})$, $\Gamma({D\over2}-2-z_{_2})$, the corresponding
double hypergeometric function is written as
\begin{eqnarray}
&&C_{_{0,2c}}(p_{_1}^2,\;p_{_2}^2,\;p_{_3}^2)=C_{_{0,1d}}(
p_{_3}\rightarrow p_{_1},\;x_{_{13}}\rightarrow x_{_{31}},\;
x_{_{23}}\rightarrow x_{_{21}})\;.
\label{Massless-2c}
\end{eqnarray}

\item[$\bullet$ 2(d):] For the isolated singularities defined by the poles of
$\Gamma(3-{D\over2}+z_{_1}+z_{_2})$, $\Gamma({D\over2}-2-z_{_2})$, the corresponding
double hypergeometric function is written as
\begin{eqnarray}
&&C_{_{0,2d}}(p_{_1}^2,\;p_{_2}^2,\;p_{_3}^2)=C_{_{0,1b}}(
p_{_3}\rightarrow p_{_1},\;x_{_{13}}\rightarrow x_{_{31}},\;
x_{_{23}}\rightarrow x_{_{21}})\;.
\label{Massless-2d}
\end{eqnarray}

\item[$\bullet$ 3(a):] For the isolated singularities defined by the poles of
$\Gamma(-z_{_1})$, and $\Gamma(1+z_{_1}+z_{_2})$, the corresponding
double hypergeometric function is written as
\begin{eqnarray}
&&C_{_{0,3a}}(p_{_1}^2,\;p_{_2}^2,\;p_{_3}^2)=C_{_{0,1b}}(
p_{_3}\rightarrow p_{_2},\;x_{_{13}}\rightarrow x_{_{12}},\;
x_{_{23}}\rightarrow x_{_{32}})\;.
\label{Massless-3a}
\end{eqnarray}
The condition $|p_{_2}^2|>\max(|p_{_1}^2|,\;|p_{_3}^2|)$
is necessary to guarantee the double series uniformly converging.

\item[$\bullet$ 3(b):] For the isolated singularities defined by the poles of
$\Gamma(-z_{_1})$, and $\Gamma(3-{D\over2}+z_{_1}+z_{_2})$, the corresponding
double hypergeometric function is written as
\begin{eqnarray}
&&C_{_{0,3b}}(p_{_1}^2,\;p_{_2}^2,\;p_{_3}^2)=C_{_{0,1a}}(
p_{_3}\rightarrow p_{_2},\;x_{_{13}}\rightarrow x_{_{12}},\;
x_{_{23}}\rightarrow x_{_{32}})\;.
\label{Massless-3b}
\end{eqnarray}

\item[$\bullet$ 3(c):] For the isolated singularities defined by the poles of
$\Gamma({D\over2}-2-z_{_1})$, and $\Gamma(1+z_{_1}+z_{_2})$, the corresponding
double hypergeometric function is written as
\begin{eqnarray}
&&C_{_{0,3c}}(p_{_1}^2,\;p_{_2}^2,\;p_{_3}^2)=C_{_{0,1d}}(
p_{_3}\rightarrow p_{_2},\;x_{_{13}}\rightarrow x_{_{12}},\;
x_{_{23}}\rightarrow x_{_{32}})\;.
\label{Massless-3c}
\end{eqnarray}

\item[$\bullet$ 3(d):] For the isolated singularities defined by the poles of
$\Gamma({D\over2}-2-z_{_1})$, and $\Gamma(3-{D\over2}+z_{_1}+z_{_2})$, the
corresponding double hypergeometric function is written as
\begin{eqnarray}
&&C_{_{0,3d}}(p_{_1}^2,\;p_{_2}^2,\;p_{_3}^2)=C_{_{0,1c}}(
p_{_3}\rightarrow p_{_2},\;x_{_{13}}\rightarrow x_{_{12}},\;
x_{_{23}}\rightarrow x_{_{32}})\;.
\label{Massless-3d}
\end{eqnarray}

\end{itemize}

Applying Horn's study of convergence to Eq.(\ref{Massless-2a}), Eq.(\ref{Massless-2b}),
Eq.(\ref{Massless-2c}), and Eq.(\ref{Massless-2d}), we obtain the absolutely and
uniformly convergent region of the double power series as
\begin{eqnarray}
&&\Omega_{_2}^\prime=\Omega_{_1}^\prime(x_{_{13}}\rightarrow x_{_{31}},\;
x_{_{23}}\rightarrow x_{_{21}})\;.
\label{Convergent-region2}
\end{eqnarray}
Similarly the absolutely and uniformly convergent region of the
double power series in Eq.(\ref{Massless-3a}), Eq.(\ref{Massless-3b}),
Eq.(\ref{Massless-3c}), and Eq.(\ref{Massless-3d}) is described by
\begin{eqnarray}
&&\Omega_{_3}^\prime=\Omega_{_1}^\prime(x_{_{13}}\rightarrow x_{_{12}},\;
x_{_{23}}\rightarrow x_{_{32}})\;.
\label{Convergent-region3}
\end{eqnarray}
Obviously the intersection $\Omega_{_i}^\prime\bigcap\Omega_{_j}^\prime$,
$i,\;j=1,\;2,\;3,\;i\neq j$ is empty set, i.e.
\begin{eqnarray}
&&\Omega_{_1}^\prime\bigcap\Omega_{_2}^\prime=\emptyset,\;
\Omega_{_1}^\prime\bigcap\Omega_{_3}^\prime=\emptyset,\;
\Omega_{_2}^\prime\bigcap\Omega_{_3}^\prime=\emptyset\;.
\label{intersection-1}
\end{eqnarray}

Based on the above-mentioned analyses, the scalar integral
$C_{_0}$ in the convergent region $\Omega_{_1}^\prime$ is presented as
\begin{eqnarray}
&&C_{_0}(p_{_1}^2,\;p_{_2}^2,\;p_{_3}^2)=C_{_{0,1a}}(p_{_1}^2,\;p_{_2}^2,\;p_{_3}^2)
+C_{_{0,1b}}(p_{_1}^2,\;p_{_2}^2,\;p_{_3}^2)
\nonumber\\
&&\hspace{3.1cm}
+C_{_{0,1c}}(p_{_1}^2,\;p_{_2}^2,\;p_{_3}^2)+C_{_{0,1d}}(p_{_1}^2,\;p_{_2}^2,\;p_{_3}^2)
\nonumber\\
&&\hspace{2.7cm}=
{i(-)^{D/2}(p_{_3}^2)^{D/2-3}\over(4\pi)^{D/2}}
{\Gamma^2(2-{D\over2})\Gamma^2({D\over2}-1)\over\Gamma(D-3)}f(x_{_{13}},\;x_{_{23}})\;,
\label{Massless-I1}
\end{eqnarray}
with
\begin{eqnarray}
&&f(x,\;y)={1\over\Gamma(3-{D\over2})}\;F_{_4}
\left(\left.\begin{array}{cc}1,&3-{D\over2}\\
3-{D\over2},&3-{D\over2}\end{array}\right|x,\;y\right)
\nonumber\\
&&\hspace{2.4cm}
-{y^{D/2-2}\over\Gamma(3-{D\over2})}\;F_{_4}
\left(\left.\begin{array}{cc}1,&{D\over2}-1\\
3-{D\over2},&{D\over2}-1\end{array}\right|x,\;y\right)
\nonumber\\
&&\hspace{2.4cm}
-{x^{D/2-2}\over\Gamma(3-{D\over2})}\;F_{_4}
\left(\left.\begin{array}{cc}1,&{D\over2}-1\\
{D\over2}-1,&3-{D\over2}\end{array}\right|x,\;y\right)
\nonumber\\
&&\hspace{2.4cm}
+{\Gamma(D-3)\over\Gamma({D\over2}-1)}
(xy)^{D/2-2}
\;F_{_4}\left(\left.\begin{array}{cc}D-3,&{D\over2}-1\\
{D\over2}-1,&{D\over2}-1\end{array}\right|x,\;y\right)\;,
\label{Massless-I2}
\end{eqnarray}
which is consistent with the Eq.~(7) of Ref.~\cite{Davydychev1} exactly,
where $F_{_4}$ denotes the Appell function.
Similarly in the convergent region $\Omega_{_2}^\prime$,
the massless $C_{_0}$ function is given as
\begin{eqnarray}
&&C_{_0}(p_{_1}^2,\;p_{_2}^2,\;p_{_3}^2)=C_{_{0,2a}}(p_{_1}^2,\;p_{_2}^2,\;p_{_3}^2)
+C_{_{0,2b}}(p_{_1}^2,\;p_{_2}^2,\;p_{_3}^2)
\nonumber\\
&&\hspace{3.1cm}
+C_{_{0,2c}}(p_{_1}^2,\;p_{_2}^2,\;p_{_3}^2)+C_{_{0,2d}}(p_{_1}^2,\;p_{_2}^2,\;p_{_3}^2)
\nonumber\\
&&\hspace{2.7cm}=
{\rm Eq}.~(\ref{Massless-I1})(p_{_3}\rightarrow p_{_1},\;x_{_{13}}\rightarrow x_{_{31}},\;
x_{_{23}}\rightarrow x_{_{21}})\;.
\label{Massless-II1}
\end{eqnarray}

In the parameter space $\Omega_{_3}^\prime$, the massless
$C_{_0}$ function is presented as
\begin{eqnarray}
&&C_{_0}(p_{_1}^2,\;p_{_2}^2,\;p_{_3}^2)=C_{_{0,3a}}(p_{_1}^2,\;p_{_2}^2,\;p_{_3}^2)
+C_{_{0,3b}}(p_{_1}^2,\;p_{_2}^2,\;p_{_3}^2)
\nonumber\\
&&\hspace{3.1cm}
+C_{_{0,3c}}(p_{_1}^2,\;p_{_2}^2,\;p_{_3}^2)+C_{_{0,3d}}(p_{_1}^2,\;p_{_2}^2,\;p_{_3}^2)
\nonumber\\
&&\hspace{2.7cm}=
{\rm Eq}.~(\ref{Massless-I1})(p_{_3}\rightarrow p_{_2},\;x_{_{13}}\rightarrow x_{_{12}},\;
x_{_{23}}\rightarrow x_{_{32}})\;.
\label{Massless-III1}
\end{eqnarray}

The emphasized point here is that the functions $f(x_{_{13}},\;x_{_{23}})$,
$x_{_{13}}^{D/2-3}f(x_{_{31}},\;x_{_{21}})$, and $x_{_{23}}^{D/2-3}f(x_{_{12}},\;x_{_{32}})$
all satisfy the system of PDEs\cite{M.Y.Kalmykov10,M.Y.Kalmykov12,M.Y.Kalmykov13,M.Y.Kalmykov17}
\begin{eqnarray}
&&\Big\{(\sum\limits_{i=1}^2\hat{\vartheta}_{_{x_{_{i3}}}}+1)
(\sum\limits_{i=1}^2\hat{\vartheta}_{_{x_{_{i3}}}}+3-{D\over2})
-{1\over x_{_{j3}}}\hat{\vartheta}_{_{x_{_{j3}}}}
(\hat{\vartheta}_{_{x_{_{j3}}}}+2-{D\over2})\Big\}f=0,\;(j=1,\;2).
\label{Massless-PEDs}
\end{eqnarray}
How to continue the solution to the whole parameter space
has already been presented in our previous work \cite{Feng2018}.
Actually the function $f(x,y)$ can be written in term of the Gauss
function $_2F_1$~\cite{Davydychev2000} by the well-known reduction of
the Appell function of the fourth kind~\cite{Davydychev1993NPB},
then the massless scalar integral is analytic continued through the transformations
of the Gauss function. Continuation through the holonomic hypergeometric
system of PDEs here can be applied to evaluate the scalar integral of any multiloop Feynman diagram.

\section{The $C_{_0}$ function with one nonzero virtual mass\label{sec4}}
\indent\indent
In this case, the Mellin-Barnes representation of the scalar integral is
\begin{eqnarray}
&&C_{_0}^a(p_{_1}^2,\;p_{_2}^2,\;p_{_3}^2,\;m^2)=
\int{d^Dq\over(2\pi)^D}{1\over(q^2-m^2)(q+p_{_1})^2(q-p_{_2})^2}
\nonumber\\
&&\hspace{3.5cm}=
-{i(-)^{D/2-3}\over(4\pi)^{D/2}(2\pi i)^3}
\int_{-i\infty}^{+i\infty}ds(-m^2)^{s}\Gamma(-s)
\nonumber\\
&&\hspace{4.0cm}\times
\int_{-i\infty}^{+i\infty}dz_{_1}(p_{_1}^2)^{z_{_1}}\Gamma(-z_{_1})
\int_{-i\infty}^{+i\infty}dz_{_2}(p_{_2}^2)^{z_{_2}}\Gamma(-z_{_2})
\nonumber\\
&&\hspace{4.0cm}\times
(p_{_3}^2)^{D/2-3-s-z_{_1}-z_{_2}}\Gamma(3-{D\over2}+s+z_{_1}+z_{_2})
\nonumber\\
&&\hspace{4.0cm}\times
\Gamma({D\over2}-2-s-z_{_2})\Gamma({D\over2}-2-s-z_{_1})
\nonumber\\
&&\hspace{4.0cm}\times
{\Gamma(1+s+z_{_1}+z_{_2})\over\Gamma(D-3-s)}\;.
\label{1M}
\end{eqnarray}
which coincides with Eq.~(26) of Ref.~\cite{Davydychev3} (i.e. Eq.~(4.3)
of Ref.~\cite{Davydychev1992JMP}).
Applying the residue theorem to the  singularities
of $\Gamma$ functions in the numerator of integrand in the Mellin-Barnes
representation, we formulate the scalar integral
as various hypergeometric functions whose absolutely and uniformly
convergent regions are analyzed through the Horn's study of convergence~\cite{Horn1889}.
If the scalar integral is expressed as
\begin{eqnarray}
&&C_{_{0}}^a(p_{_1}^2,\;p_{_2}^2,\;p_{_3}^2,\;m^2)=
{i(-)^{D/2}(p_{_3}^2)^{D/2-3}\over(4\pi)^{D/2}}
F_{_{a,p_{_3}}}(\xi_{_{33}},\;x_{_{13}},\;x_{_{23}})
\label{1Ma}
\end{eqnarray}
with $\xi_{_{33}}=-m^2/p_{_3}^2$, then the dimensionless function $F_{_{a,p_{_3}}}$ satisfies the
holonomic hypergeometric system of linear PDEs
\begin{eqnarray}
&&\Big\{(3-{D\over2}+\hat{\vartheta}_{\xi_{_{33}}}+\sum\limits_{i=1}^2\hat{\vartheta}_{x_{_{i3}}})
(1+\hat{\vartheta}_{\xi_{_{33}}}+\sum\limits_{i=1}^2\hat{\vartheta}_{x_{_{i3}}})
(4-D+\hat{\vartheta}_{\xi_{_{33}}})
\nonumber\\
&&\hspace{0.0cm}
-{1\over\xi_{_{33}}}\hat{\vartheta}_{\xi_{_{33}}}\prod\limits_{i=1}^2
(2-{D\over2}+\hat{\vartheta}_{\xi_{_{33}}}+\hat{\vartheta}_{x_{_{i3}}})\Big\}F_{_{a,p_{_3}}}=0
\;,\nonumber\\
&&\Big\{(3-{D\over2}+\hat{\vartheta}_{\xi_{_{33}}}+\sum\limits_{i=1}^2\hat{\vartheta}_{x_{_{i3}}})
(1+\hat{\vartheta}_{\xi_{_{33}}}+\sum\limits_{i=1}^2\hat{\vartheta}_{x_{_{i3}}})
\nonumber\\
&&\hspace{0.0cm}
-{1\over x_{_{j3}}}\hat{\vartheta}_{x_{_{j3}}}(2-{D\over2}+\hat{\vartheta}_{\xi_{_{33}}}
+\hat{\vartheta}_{x_{_{j3}}})\Big\}F_{_{a,p_{_3}}}=0\;,\;\;(j=1,\;2)\;.
\label{1M-PDEs}
\end{eqnarray}
Setting $m_{_1}=m_{_2}=0$ in the corresponding system of linear PDEs of Eq.~(\ref{1LoopC77}),
we find the first, fourth and fifth PDEs composing the system presented above.
Now we present our results in detail below.

\subsection{The hypergeometric functions and their convergent regions\label{sec3-1}}
\begin{itemize}
\item[$\bullet$ 1:] For the isolated singularities defined by the poles of
$\Gamma(-s)$, $\Gamma(-z_{_1})$, and $\Gamma(-z_{_2})$ in the numerator of integrand
of the Mellin-Barnes representation, the derived hypergeometric function is written as
\begin{eqnarray}
&&C_{_{0,1}}^a(p_{_1}^2,\;p_{_2}^2,\;p_{_3}^2,\;m^2)=
{i(-)^{D/2}(p_{_3}^2)^{D/2-3}\over(4\pi)^{D/2}}
{\Gamma^2(2-{D\over2})\Gamma^2({D\over2}-1)\over\Gamma(D-3)\Gamma(4-D)}
\nonumber\\
&&\hspace{4.2cm}\times
\sum\limits_{j=0}^\infty
\sum\limits_{n_{_1}=0}^\infty\sum\limits_{n_{_2}=0}^\infty
\xi_{_{33}}^{j}x_{_{13}}^{n_{_1}}x_{_{23}}^{n_{_2}}
{\Gamma(4-D+j)\over j!n_{_1}!n_{_2}!}
\nonumber\\
&&\hspace{4.2cm}\times
{\Gamma(3-{D\over2}+j+n_{_1}+n_{_2})\Gamma(1+j+n_{_1}+n_{_2})
\over\Gamma(3-{D\over2}+j+n_{_1})\Gamma(3-{D\over2}+j+n_{_2})}
\nonumber\\
&&\hspace{3.7cm}=
{i(-)^{D/2}(p_{_3}^2)^{D/2-3}\over(4\pi)^{D/2}}
{\Gamma^2(2-{D\over2})\Gamma^2({D\over2}-1)\over\Gamma(D-3)}
\nonumber\\
&&\hspace{4.2cm}\times
\sum\limits_{n_{_1}=0}^\infty\sum\limits_{n_{_2}=0}^\infty
x_{_{13}}^{n_{_1}}x_{_{23}}^{n_{_2}}
\nonumber\\
&&\hspace{4.2cm}\times
{\Gamma(3-{D\over2}+n_{_1}+n_{_2})\Gamma(1+n_{_1}+n_{_2})
\over n_{_1}!n_{_2}!\Gamma(3-{D\over2}+n_{_1})\Gamma(3-{D\over2}+n_{_2})}
\nonumber\\
&&\hspace{4.2cm}
+{i(-)^{D/2}(p_{_3}^2)^{D/2-3}\over(4\pi)^{D/2}}
{\Gamma^2(2-{D\over2})\Gamma^2({D\over2}-1)\over\Gamma(D-3)\Gamma(4-D)}
\nonumber\\
&&\hspace{4.2cm}\times
\sum\limits_{j=1}^\infty
\sum\limits_{n_{_1}=0}^\infty\sum\limits_{n_{_2}=0}^\infty
\xi_{_{33}}^{j}x_{_{13}}^{n_{_1}}x_{_{23}}^{n_{_2}}
{\Gamma(4-D+j)\over j!n_{_1}!n_{_2}!}
\nonumber\\
&&\hspace{4.2cm}\times
{\Gamma(3-{D\over2}+j+n_{_1}+n_{_2})\Gamma(1+j+n_{_1}+n_{_2})
\over\Gamma(3-{D\over2}+j+n_{_1})\Gamma(3-{D\over2}+j+n_{_2})}\;.
\label{1M1-1}
\end{eqnarray}
The condition $|p_{_3}^2|>\max(m^2,\;|p_{_1}^2|,\;|p_{_2}^2|)$ is necessary
to guarantee the convergence of the hypergeometric series. In addition,
$D=4$ is the second order pole of the first double hypergeometric
function, and the first order pole of the second triple hypergeometric
function, respectively. Certainly the expression
of Eq.~(\ref{1M1-1}) complies with the system of the PDEs in Eq.~(\ref{1M-PDEs}) explicitly.

The absolutely and uniformly convergent region of the hypergeometric functions
is given by
\begin{eqnarray}
&&\Omega_{_1}^a=\left\{(\xi_{_{33}},\;x_{_{13}},\;x_{_{23}}):\left|\begin{array}{l}
\sqrt{|x_{_{13}}|}+\sqrt{|x_{_{23}}|}<1,\;|\xi_{_{33}}|+|x_{_{13}}|<1,\\
|\xi_{_{33}}|+|x_{_{23}}|<1
\end{array}\right.\right\}\;.
\label{1M1-4}
\end{eqnarray}

\item[$\bullet$ 2:] For the isolated singularities defined by the poles of
$\Gamma(-s)$, $\Gamma(-z_{_1})$, and $\Gamma({D\over2}-2-s-z_{_2})$ in the numerator of integrand
of the Mellin-Barnes representation, the corresponding hypergeometric series is given as
\begin{eqnarray}
&&C_{_{0,2}}^a(p_{_1}^2,\;p_{_2}^2,\;p_{_3}^2,\;m^2)=
-{i(-)^{D/2}(p_{_3}^2)^{D/2-3}\over(4\pi)^{D/2}}
{\Gamma^2(2-{D\over2})\Gamma^2({D\over2}-1)\over\Gamma(D-3)\Gamma(4-D)}
\nonumber\\
&&\hspace{4.2cm}\times
\sum\limits_{j=0}^\infty\sum\limits_{n_{_1}=0}^\infty\sum\limits_{n_{_2}=0}^\infty
\xi_{_{33}}^jx_{_{13}}^{n_{_1}}x_{_{23}}^{D/2-2-j+n_{_2}}{\Gamma(4-D+j)\over j!n_{_1}!n_{_2}!}
\nonumber\\
&&\hspace{4.2cm}\times
{\Gamma(1+n_{_1}+n_{_2})\Gamma({D\over2}-1+n_{_1}+n_{_2})\over
\Gamma(3-{D\over2}+j+n_{_1})\Gamma({D\over2}-1-j+n_{_2})}
\nonumber\\
&&\hspace{3.7cm}=
-{i(-)^{D/2}(p_{_3}^2)^{D/2-3}\over(4\pi)^{D/2}}
{\Gamma^2(2-{D\over2})\Gamma^2({D\over2}-1)\over\Gamma(D-3)}
\nonumber\\
&&\hspace{4.2cm}\times
\sum\limits_{n_{_1}=0}^\infty\sum\limits_{n_{_2}=1}^\infty
x_{_{13}}^{n_{_1}}x_{_{23}}^{D/2-2+n_{_2}}
\nonumber\\
&&\hspace{4.2cm}\times
{\Gamma(1+n_{_1}+n_{_2})\Gamma({D\over2}-1+n_{_1}+n_{_2})\over
 n_{_1}!n_{_2}!\Gamma(3-{D\over2}+n_{_1})\Gamma({D\over2}-1+n_{_2})}
\nonumber\\
&&\hspace{4.2cm}
-{i(-)^{D/2}(p_{_3}^2)^{D/2-3}\over(4\pi)^{D/2}}
{\Gamma^2(2-{D\over2})\Gamma^2({D\over2}-1)\over\Gamma(D-3)\Gamma(4-D)}
\nonumber\\
&&\hspace{4.2cm}\times
\sum\limits_{n_{_1}=0}^\infty\sum\limits_{j=1}^\infty\sum\limits_{n_{_2}=1}^\infty
\xi_{_{33}}^jx_{_{13}}^{n_{_1}}x_{_{23}}^{D/2-2+n_{_2}}{\Gamma(4-D+j)\over j!n_{_1}!(j+n_{_2})!}
\nonumber\\
&&\hspace{4.2cm}\times
{\Gamma(1+j+n_{_1}+n_{_2})\Gamma({D\over2}-1+j+n_{_1}+n_{_2})\over
\Gamma(3-{D\over2}+j+n_{_1})\Gamma({D\over2}-1+n_{_2})}
\nonumber\\
&&\hspace{4.2cm}
-{i(-)^{D/2}(p_{_3}^2)^{D/2-3}\over(4\pi)^{D/2}}
{\Gamma(2-{D\over2})\Gamma({D\over2}-1)\over\Gamma(D-3)\Gamma(4-D)}
\nonumber\\
&&\hspace{4.2cm}\times
\sum\limits_{n_{_1}=0}^\infty\sum\limits_{n_{_2}=0}^\infty\sum\limits_{j=1}^\infty
\xi_{_{33}}^{j+n_{_2}}x_{_{13}}^{n_{_1}}x_{_{23}}^{D/2-2-j}
\nonumber\\
&&\hspace{4.2cm}\times
{(-)^{j}\Gamma(4-D+j+n_{_2})
\Gamma(2-{D\over2}+j)\over(j+n_{_2})!n_{_1}!n_{_2}!}
\nonumber\\
&&\hspace{4.2cm}\times
{\Gamma(1+n_{_1}+n_{_2})\Gamma({D\over2}-1+n_{_1}+n_{_2})\over
\Gamma(3-{D\over2}+j+n_{_1}+n_{_2})}
\nonumber\\
&&\hspace{4.2cm}
-{i(-)^{D/2}(p_{_3}^2)^{D/2-3}\over(4\pi)^{D/2}}
{\Gamma^2(2-{D\over2})\Gamma({D\over2}-1)\over\Gamma(D-3)\Gamma(4-D)}
\nonumber\\
&&\hspace{4.2cm}\times
\sum\limits_{n_{_1}=0}^\infty\sum\limits_{j=1}^\infty
\xi_{_{33}}^jx_{_{13}}^{n_{_1}}x_{_{23}}^{D/2-2}{\Gamma(4-D+j)\over(j!)^2n_{_1}!}
\nonumber\\
&&\hspace{4.2cm}\times
{\Gamma(1+j+n_{_1})\Gamma({D\over2}-1+j+n_{_1})\over
\Gamma(3-{D\over2}+j+n_{_1})}
\nonumber\\
&&\hspace{4.2cm}
-{i(-)^{D/2}(p_{_3}^2)^{D/2-3}\over(4\pi)^{D/2}}
{\Gamma^2(2-{D\over2})\Gamma({D\over2}-1)\over\Gamma(D-3)}
\nonumber\\
&&\hspace{4.2cm}\times
\sum\limits_{n_{_1}=0}^\infty
x_{_{13}}^{n_{_1}}x_{_{23}}^{D/2-2}{\Gamma({D\over2}-1+n_{_1})\over
\Gamma(3-{D\over2}+n_{_1})}\;.
\label{1M2-1}
\end{eqnarray}
The essential condition $|p_{_3}^2|>\max(m^2,\;|p_{_1}^2|,\;|p_{_2}^2|)$, $|p_{_2}^2|>m^2$
should be satisfied to guarantee the hypergeometric series converging.
In the above equation, one derives the first expression after applying the
residue theorem to the Mellin-Barnes representation.
The second expression decomposes the summation indices of the hypergeometric function
in the first expression, so that $D=4$ is the pole of the same fixed order
for each term of every hypergeometric series in the second expression.
Furthermore, we can check the second expression satisfying the system of linear PDEs
in Eq.~(\ref{1M-PDEs}) explicitly, this fact implies that the decomposition of
summation indices above is correct and reasonable.

Horn's theory of convergence predicts the absolutely and uniformly convergent
region of the hypergeometric functions as
\begin{eqnarray}
&&\Omega_{_{2}}^a=\left\{(\xi_{_{33}},\;x_{_{13}},\;x_{_{23}}):\left|\begin{array}{l}
|\xi_{_{33}}|<|x_{_{23}}|,\;\sqrt{|x_{_{13}}|}+\sqrt{|x_{_{23}}|}<1,\\
|\xi_{_{33}}|+|x_{_{13}}|<1,\;|\xi_{_{33}}|+|x_{_{23}}|<1
\end{array}\right.\right\}\;.
\label{1M2-2}
\end{eqnarray}

\item[$\bullet$ 3:] For the isolated singularities defined by the poles of
$\Gamma(-s)$, $\Gamma(-z_{_1})$, and $\Gamma(1+s+z_{_1}+z_{_2})$ in the numerator of integrand
of the Mellin-Barnes representation, the corresponding power series is
\begin{eqnarray}
&&C_{_{0,3}}^a(p_{_1}^2,\;p_{_2}^2,\;p_{_3}^2,\;m^2)=
-{i(-)^{D/2}(p_{_3}^2)^{D/2-3}\over(4\pi)^{D/2}}
{\Gamma^2(2-{D\over2})\Gamma^2({D\over2}-1)\over\Gamma(D-3)\Gamma(4-D)}
\nonumber\\
&&\hspace{4.2cm}\times
\sum\limits_{j=0}^\infty\sum\limits_{n_{_1}=0}^\infty\sum\limits_{n_{_2}=0}^\infty
\xi_{_{33}}^jx_{_{13}}^{n_{_1}}x_{_{23}}^{-1-j-n_{_1}-n_{_2}}
{(-)^{j}\Gamma(4-D+j)\over j!n_{_1}!n_{_2}!}
\nonumber\\
&&\hspace{4.2cm}\times
{\Gamma(1+j+n_{_1}+n_{_2})\Gamma({D\over2}-1+n_{_1}+n_{_2})
\over\Gamma(3-{D\over2}+j+n_{_1})\Gamma({D\over2}-1+n_{_2})}
\nonumber\\
&&\hspace{3.7cm}=
-{i(-)^{D/2}(p_{_2}^2)^{D/2-3}\over(4\pi)^{D/2}}
{\Gamma^2(2-{D\over2})\Gamma^2({D\over2}-1)\over\Gamma(D-3)}
\nonumber\\
&&\hspace{4.2cm}\times
\sum\limits_{n_{_1}=0}^\infty\sum\limits_{n_{_2}=0}^\infty
x_{_{12}}^{n_{_1}}x_{_{32}}^{D/2-2+n_{_2}}
\nonumber\\
&&\hspace{4.2cm}\times
{\Gamma(1+n_{_1}+n_{_2})\Gamma({D\over2}-1+n_{_1}+n_{_2})
\over n_{_1}!n_{_2}!\Gamma(3-{D\over2}+n_{_1})\Gamma({D\over2}-1+n_{_2})}
\nonumber\\
&&\hspace{4.2cm}
-{i(-)^{D/2}(p_{_2}^2)^{D/2-3}\over(4\pi)^{D/2}}
{\Gamma^2(2-{D\over2})\Gamma^2({D\over2}-1)\over\Gamma(D-3)\Gamma(4-D)}
\nonumber\\
&&\hspace{4.2cm}\times
\sum\limits_{j=1}^\infty\sum\limits_{n_{_1}=0}^\infty\sum\limits_{n_{_2}=0}^\infty
\xi_{_{32}}^jx_{_{12}}^{n_{_1}}x_{_{32}}^{D/2-2+n_{_2}}
{(-)^{j}\Gamma(4-D+j)\over j!n_{_1}!n_{_2}!}
\nonumber\\
&&\hspace{4.2cm}\times
{\Gamma(1+j+n_{_1}+n_{_2})\Gamma({D\over2}-1+n_{_1}+n_{_2})
\over\Gamma(3-{D\over2}+j+n_{_1})\Gamma({D\over2}-1+n_{_2})}\;.
\label{1M3-1}
\end{eqnarray}
The necessary condition $|p_{_2}^2|>\max(|p_{_1}^2|,\;|p_{_3}^2|,\;m^2)$
should be satisfied to guarantee the power series converging.
In addition, $D=4$ is the second order pole of the first double hypergeometric
function, and the first order pole of the second triple hypergeometric
function, respectively. Certainly the expression
of Eq.~(\ref{1M3-1}) complies with the system of linear PDEs in Eq.~(\ref{1M-PDEs}) explicitly.

Through Horn's study of convergence, the absolutely and uniformly convergent
region of the series is written as
\begin{eqnarray}
&&\Omega_{_{3}}^a=\left\{(\xi_{_{32}},\;x_{_{12}},\;x_{_{32}}):\left|\begin{array}{l}
|\xi_{_{32}}|+|x_{_{32}}|<1,\;\sqrt{|x_{_{12}}|}+\sqrt{|x_{_{32}}|}<1
\end{array}\right.\right\}\;.
\label{1M3-2}
\end{eqnarray}

\item[$\bullet$ 4:] For the isolated singularities defined by the poles of
$\Gamma(-s)$, $\Gamma(-z_{_1})$, and $\Gamma(3-{D\over2}+s+z_{_1}+z_{_2})$
in the numerator of integrand of the Mellin-Barnes representation,
the corresponding hypergeometric function is written as
\begin{eqnarray}
&&C_{_{0,4}}^a(p_{_1}^2,\;p_{_2}^2,\;p_{_3}^2,\;m^2)=
{i(-)^{D/2}(p_{_3}^2)^{D/2-3}\over(4\pi)^{D/2}}
{\Gamma^2(2-{D\over2})\Gamma^2({D\over2}-1)\over\Gamma(D-3)\Gamma(4-D)}
\nonumber\\
&&\hspace{4.2cm}\times
\sum\limits_{j=0}^\infty\sum\limits_{n_{_1}=0}^\infty\sum\limits_{n_{_2}=0}^\infty
\xi_{_{33}}^{j}x_{_{13}}^{n_{_1}}x_{_{23}}^{D/2-3-j-n_{_1}-n_{_2}}
{(-)^{j}\Gamma(4-D+j)\over j!n_{_1}!n_{_2}!}
\nonumber\\
&&\hspace{4.2cm}\times
{\Gamma(3-{D\over2}+j+n_{_1}+n_{_2})\Gamma(1+n_{_1}+n_{_2})
\over\Gamma(3-{D\over2}+j+n_{_1})\Gamma(3-{D\over2}+n_{_2})}
\nonumber\\
&&\hspace{3.7cm}=
{i(-)^{D/2}(p_{_2}^2)^{D/2-3}\over(4\pi)^{D/2}}
{\Gamma^2(2-{D\over2})\Gamma^2({D\over2}-1)\over\Gamma(D-3)}
\sum\limits_{n_{_1}=0}^\infty\sum\limits_{n_{_2}=0}^\infty
\nonumber\\
&&\hspace{4.2cm}\times
x_{_{12}}^{n_{_1}}x_{_{32}}^{n_{_2}}
{\Gamma(3-{D\over2}+n_{_1}+n_{_2})\Gamma(1+n_{_1}+n_{_2})
\over n_{_1}!n_{_2}!\Gamma(3-{D\over2}+n_{_1})\Gamma(3-{D\over2}+n_{_2})}
\nonumber\\
&&\hspace{4.2cm}
+{i(-)^{D/2}(p_{_2}^2)^{D/2-3}\over(4\pi)^{D/2}}
{\Gamma^2(2-{D\over2})\Gamma^2({D\over2}-1)\over\Gamma(D-3)\Gamma(4-D)}
\sum\limits_{j=1}^\infty\sum\limits_{n_{_1}=0}^\infty\sum\limits_{n_{_2}=0}^\infty
\nonumber\\
&&\hspace{4.2cm}\times
\xi_{_{32}}^{j}x_{_{12}}^{n_{_1}}x_{_{32}}^{n_{_2}}
{(-)^{j}\Gamma(4-D+j)\over j!n_{_1}!n_{_2}!}
\nonumber\\
&&\hspace{4.2cm}\times
{\Gamma(3-{D\over2}+j+n_{_1}+n_{_2})\Gamma(1+n_{_1}+n_{_2})
\over\Gamma(3-{D\over2}+j+n_{_1})\Gamma(3-{D\over2}+n_{_2})}\;.
\label{1M4-1}
\end{eqnarray}
The necessary condition $|p_{_2}^2|>\max(|p_{_3}^2|,\;|p_{_1}^2|,\;m^2)$
should be satisfied to guarantee the hypergeometric series converging.
The convergent region of the term is consistent with $\Omega_{_3}^a$
in Eq.(\ref{1M3-2}) exactly.
Furthermore, $D=4$ is the second order pole of the first double hypergeometric
function, and the first order pole of the second triple hypergeometric
function, respectively. Certainly the expression
of Eq.~(\ref{1M4-1}) complies with the system of linear PDEs in Eq.~(\ref{1M-PDEs}) explicitly.

\item[$\bullet$ 6:] For the isolated singularities originating from the poles of
$\Gamma(-s)$, $\Gamma({D\over2}-2-s-z_{_1})$, and $\Gamma(-z_{_2})$
in the numerator of integrand of the Mellin-Barnes representation,
the corresponding hypergeometric series is
\begin{eqnarray}
&&C_{_{0,6}}^a(p_{_1}^2,\;p_{_2}^2,\;p_{_3}^2,\;m^2)=
C_{_{0,2}}^a(x_{_{13}}\leftrightarrow x_{_{23}})\;.
\label{1M6-1}
\end{eqnarray}
The condition $|p_{_3}^2|>\max(m^2,\;|p_{_1}^2|,\;|p_{_2}^2|)$,
$|p_{_1}^2|>m^2$ should be met to guarantee the hypergeometric functions converging.
Applying Horn's study of convergence, one gets
the absolutely and uniformly convergent region of the hypergeometric series as
\begin{eqnarray}
&&\Omega_{_{6}}^a
=\left\{(\xi_{_{33}},\;x_{_{13}},\;x_{_{23}}):\left|\begin{array}{l}
|\xi_{_{33}}|<|x_{_{13}}|,\;\sqrt{|x_{_{13}}|}+\sqrt{|x_{_{23}}|}<1,\\
|\xi_{_{33}}|+|x_{_{13}}|<1,\;|\xi_{_{33}}|+|x_{_{23}}|<1
\end{array}\right.\right\}\;.
\label{1M6-2}
\end{eqnarray}

\item[$\bullet$ 7:]  For the isolated singularities defined by the poles of
$\Gamma(-s)$, $\Gamma({D\over2}-2-s-z_{_1})$, and $\Gamma({D\over2}-2-s-z_{_2})$
in the numerator of integrand of the Mellin-Barnes representation,
the corresponding hypergeometric series $C_{_{0,7}}^a$ is presented in Eq.~(\ref{1M7-1}),
where the condition $|p_{_3}^2|>\max(|p_{_1}^2|,\;|p_{_2}^2|),\; \min(|p_{_1}^2|,\;|p_{_2}^2|)>m^2$,
$|p_{_1}^2p_{_2}^2|>m^2|p_{_3}^2|$ should be met
to guarantee the power series converging.
After implementing the residue theorem, one derives the first expression in Eq.~(\ref{1M7-1}).
The second expression decomposes the summation indices of the hypergeometric series
in the first expression, so that $D=4$ is the pole of the same fixed order
for each term of every hypergeometric series in the second expression.
In addition, the second expression also satisfies the system of linear PDEs
in Eq.~(\ref{1M-PDEs}).

The absolutely and uniformly convergent region of the term is written as
\begin{eqnarray}
&&\Omega_{_{7}}^a=\left\{(\xi_{_{33}},\;x_{_{13}},\;x_{_{23}}):\left|\begin{array}{l}
\sqrt{|x_{_{13}}|}+\sqrt{|x_{_{23}}|}<1,\;|\xi_{_{33}}|+|x_{_{13}}|<1,\\
|\xi_{_{33}}|+|x_{_{23}}|<1,\;|\xi_{_{33}}|(1+|x_{_{23}}|)<|x_{_{13}}x_{_{23}}|,\\
|\xi_{_{33}}|(1+|x_{_{13}}|)<|x_{_{13}}x_{_{23}}|,\;\\
|\xi_{_{33}}|<(|x_{_{13}}-|\xi_{_{33}})(|x_{_{23}}|-|\xi_{_{33}}|)
\end{array}\right.\right\}\;.
\label{1M7-2}
\end{eqnarray}

\item[$\bullet$ 8:]  For the isolated singularities defined by the poles of
$\Gamma(-s)$, $\Gamma({D\over2}-2-s-z_{_1})$, and $\Gamma(1+s+z_{_1}+z_{_2})$
in the numerator of integrand of the Mellin-Barnes representation,
correspondingly the tedious hypergeometric series $C_{_{0,8}}^a$ is given in
Eq.~(\ref{1M8-1}), where the necessary condition $|p_{_2}^2|>\max(|p_{_1}^2|,\;|p_{_3}^2|)$, $|p_{_1}^2|>m^2$
should be satisfied to guarantee the triple series converging.
Applying the residue theorem, one derives the first expression of Eq.~(\ref{1M8-1}).
The second expression decomposes the summation indices of the hypergeometric series
in the first expression, so that $D=4$ is the pole of the same fixed order
for each term of every hypergeometric series in the second expression.
In addition, the second expression also satisfies the system of linear PDEs
in Eq.~(\ref{1M-PDEs}).

The absolutely and uniformly convergent region of the series is given by
\begin{eqnarray}
&&\Omega_{_{8}}^a=\left\{(\xi_{_{32}},\;x_{_{12}},\;x_{_{32}}):\left|\begin{array}{l}
\sqrt{|x_{_{12}}|}+\sqrt{|x_{_{32}}|}<1,\;|\xi_{_{32}}|+|x_{_{32}}|<1,\\
|\xi_{_{32}}|+|x_{_{12}}|<1,\;|\xi_{_{32}}|(1+|x_{_{32}}|)<|x_{_{12}}|,\\
|\xi_{_{32}}|(|x_{_{12}}|+|x_{_{32}}|)<|x_{_{12}}|,\\
|\xi_{_{32}}x_{_{32}}|<(|x_{_{12}}|-|\xi_{_{32}}|)(1-|\xi_{_{32}}|)
\end{array}\right.\right\}\;.
\label{1M8-2}
\end{eqnarray}

\item[$\bullet$ 9:] For the isolated singularities defined by the poles of
$\Gamma(-s)$, $\Gamma({D\over2}-2-s-z_{_1})$, and $\Gamma(3-{D\over2}+s+z_{_1}+z_{_2})$,
the corresponding power series $C_{_{0,9}}^a$ is given by Eq.~(\ref{1M9-1}), where
the condition $|p_{_2}^2|>\max(|p_{_1}^2|,\;|p_{_3}^2|)$, $|p_{_1}^2|>m^2$
should be satisfied to guarantee the triple series converging.
Applying the residue theorem, one derives the first expression of Eq.~(\ref{1M9-1}).
The second expression decomposes the summation indices of the hypergeometric series
in the first expression, so that $D=4$ is the pole of the same fixed order
for each term of every hypergeometric series in the second expression.
In addition, the second expression also satisfies the system of linear PDEs
in Eq.~(\ref{1M-PDEs}).
The convergent region of the series coincides with the region
$\Omega_{_8}^a$ in Eq.(\ref{1M8-2}) exactly.

\item[$\bullet$ 10:] For the isolated singularities defined by the poles of
$\Gamma(-s)$, $\Gamma(1+s+z_{_1}+z_{_2})$, and $\Gamma(-z_{_2})$,
the corresponding hypergeometric series is
\begin{eqnarray}
&&C_{_{0,10}}^a(p_{_1}^2,\;p_{_2}^2,\;p_{_3}^2,\;m^2)=C_{_{0,3}}^a(
p_{_2}\rightarrow p_{_1},\;\xi_{_{32}}\rightarrow\xi_{_{31}},\;
x_{_{12}}\rightarrow x_{_{21}},\;x_{_{32}}\rightarrow x_{_{31}})\;.
\label{1M10-1}
\end{eqnarray}
The condition $|p_{_1}^2|>\max(|p_{_2}^2|,\;|p_{_3}^2|,\;m^2)$ should
be satisfied to guarantee the hypergeometric series converging.
Horn's theory of convergence gives the absolutely and uniformly convergent region
of the series as
\begin{eqnarray}
&&\Omega_{_{10}}^a
=\left\{(\xi_{_{31}},\;x_{_{21}},\;x_{_{31}}):\left|\begin{array}{l}
|\xi_{_{31}}|+|x_{_{31}}|<1,\;\sqrt{|x_{_{21}}|}+\sqrt{|x_{_{31}}|}<1
\end{array}\right.\right\}\;.
\label{1M10-2}
\end{eqnarray}
Additionally $D=4$ is the second order pole of the first double hypergeometric
function, and the first order pole of the second triple hypergeometric
function, respectively. Certainly the expression
of Eq.~(\ref{1M4-1}) also complies with the system of linear PDEs in Eq.~(\ref{1M-PDEs}).

\item[$\bullet$ 11:] For the isolated singularities defined by the poles of
$\Gamma(-s)$, $\Gamma(1+s+z_{_1}+z_{_2})$, and $\Gamma({D\over2}-2-s-z_{_2})$,
the derived hypergeometric series $C_{_{0,11}}^a$ is
\begin{eqnarray}
&&C_{_{0,11}}^a(p_{_1}^2,\;p_{_2}^2,\;p_{_3}^2,\;m^2)=C_{_{0,8}}^a(
p_{_2}\rightarrow p_{_1},\;\xi_{_{32}}\rightarrow\xi_{_{31}},\;
x_{_{12}}\rightarrow x_{_{21}},\;x_{_{32}}\rightarrow x_{_{31}})\;,
\label{1M11-1}
\end{eqnarray}
where the condition $|p_{_1}^2|>\max(|p_{_2}^2|,\;|p_{_3}^2|)$, $|p_{_2}^2|>m^2$
should be met to guarantee the hypergeometric series converging.
The absolutely and uniformly convergent region of the hypergeometric
series is given by
\begin{eqnarray}
&&\Omega_{_{11}}^a=\left\{(\xi_{_{31}},\;x_{_{21}},\;x_{_{31}}):\left|\begin{array}{l}
\sqrt{|x_{_{21}}|}+\sqrt{|x_{_{31}}|}<1,\;|\xi_{_{31}}|+|x_{_{31}}|<1,\\
|\xi_{_{31}}|+|x_{_{21}}|<1,\;|\xi_{_{31}}|(1+|x_{_{31}}|)<|x_{_{21}}|,\\
|\xi_{_{31}}|(|x_{_{21}}|+|x_{_{31}}|)<|x_{_{21}}|,\\
|\xi_{_{31}}x_{_{31}}|<(|x_{_{21}}|-|\xi_{_{31}}|)(1-|\xi_{_{31}}|)
\end{array}\right.\right\}\;.
\label{1M11-2}
\end{eqnarray}

\item[$\bullet$ 13:] For the isolated singularities defined by the poles of
$\Gamma(-s)$, $\Gamma(3-{D\over2}+s+z_{_1}+z_{_2})$, and $\Gamma(-z_{_2})$,
the induced hypergeometric series is
\begin{eqnarray}
&&C_{_{0,13}}^a(p_{_1}^2,\;p_{_2}^2,\;p_{_3}^2,\;m^2)=C_{_{0,4}}^a(
p_{_2}\rightarrow p_{_1},\;\xi_{_{32}}\rightarrow\xi_{_{31}},\;
x_{_{12}}\rightarrow x_{_{21}},\;x_{_{32}}\rightarrow x_{_{31}})\;.
\label{1M13-1}
\end{eqnarray}
The condition $|p_{_1}^2|>\max(|p_{_2}^2|,\;|p_{_3}^2|,\;m^2)$ should be met
to guarantee the triple series converging. The convergent region
of this series is characterized by Eq.(\ref{1M10-2}) also.

\item[$\bullet$ 14:] For the isolated singularities defined by the poles of
$\Gamma(-s)$, $\Gamma(3-{D\over2}+s+z_{_1}+z_{_2})$, and $\Gamma({D\over2}-2-s-z_{_2})$,
the corresponding hypergeometric series $C_{_{0,14}}^a$ is
\begin{eqnarray}
&&C_{_{0,14}}^a(p_{_1}^2,\;p_{_2}^2,\;p_{_3}^2,\;m^2)=C_{_{0,9}}^a(
p_{_2}\rightarrow p_{_1},\;\xi_{_{32}}\rightarrow\xi_{_{31}},\;
x_{_{12}}\rightarrow x_{_{21}},\;x_{_{32}}\rightarrow x_{_{31}})\;,
\label{1M14-1}
\end{eqnarray}
where the condition $|p_{_1}^2|>\max(|p_{_2}^2|,\;|p_{_3}^2|)$, $|p_{_2}^2|>m^2$
should be satisfied to guarantee the hypergeometric series converging.
The convergent region of the series is also defined by $\Omega_{_{11}}^a$ in Eq.(\ref{1M11-2}).

\item[$\bullet$ 16:] For the isolated singularities defined by the poles of
$\Gamma(-z_{_1})$, $\Gamma(-z_{_2})$, and $\Gamma({D\over2}-2-s-z_{_1})$,
the scalar integral contains some unknown linear combining parameters
originating from the non-isolating singularities.
The essential condition $|p_{_3}^2|>\max(|p_{_2}^2|,\;m^2)$, $m^2>|p_{_1}^2|$
should be satisfied to guarantee the corresponding hypergeometric series converging.

\item[$\bullet$ 17:] For the isolated singularities defined by the poles of
$\Gamma(-z_{_1})$, $\Gamma(-z_{_2})$, and $\Gamma({D\over2}-2-s-z_{_2})$,
the corresponding hypergeometric function contains some unknown linear combining parameters
originating from the non-isolating singularities.
The condition $|p_{_3}^2|>\max(|p_{_1}^2|,\;m^2)$, $m^2>|p_{_2}^2|$
should be satisfied to guarantee the derived hypergeometric series converging.

\item[$\bullet$ 18:] For the isolated singularities defined by the poles of
$\Gamma(-z_{_1})$, $\Gamma(-z_{_2})$, and $\Gamma(1+s+z_{_1}+z_{_2})$,
the hypergeometric series is
\begin{eqnarray}
&&C_{_{0,18}}^a(p_{_1}^2,\;p_{_2}^2,\;p_{_3}^2,\;m^2)=
-{i(m^2)^{D/2-3}\eta_{_{33}}^{D/2-2}\over(4\pi)^{D/2}}\Gamma(2-{D\over2})\Gamma({D\over2}-1)
\nonumber\\
&&\hspace{4.2cm}\times
\sum\limits_{j=0}^\infty\sum\limits_{n_{_1}=0}^\infty\sum\limits_{n_{_2}=0}^\infty
{(-)^{n_{_1}+n_{_2}}\over j!n_{_1}!n_{_2}!}\eta_{_{33}}^j
\eta_{_{13}}^{n_{_1}}\eta_{_{23}}^{n_{_2}}\Gamma({D\over2}-1+j+n_{_1})
\nonumber\\
&&\hspace{4.2cm}\times
{\Gamma(1+j+n_{_1}+n_{_2})\Gamma({D\over2}-1+j+n_{_2})
\over\Gamma(D-2+j+n_{_1}+n_{_2})\Gamma({D\over2}-1+j)}\;,
\label{1M18-1}
\end{eqnarray}
the condition $m^2>\max(|p_{_1}^2|,\;|p_{_2}^2|,\;|p_{_3}^2|)$
should be satisfied to guarantee the power series converging.
The absolutely and uniformly convergent region of the series of Eq.(\ref{1M18-1})
is written as
\begin{eqnarray}
&&\Omega_{_{18}}^a=\left\{(\eta_{_{13}},\;\eta_{_{23}},\;\eta_{_{33}}):\left|\begin{array}{l}
|\eta_{_{13}}|+|\eta_{_{33}}|<1,\;|\eta_{_{23}}|+|\eta_{_{33}}|<1,\\
|\eta_{_{33}}|<(1-|\eta_{_{13}}|)(1-|\eta_{_{23}}|)
\end{array}\right.\right\}\;.
\label{1M18-2}
\end{eqnarray}
Certainly the scalar integral
of Eq.~(\ref{1M18-1}) also satisfies the system of PDEs in Eq.~(\ref{1M-PDEs}).

\item[$\bullet$ 19:] For the isolated singularities defined by the poles of
$\Gamma(-z_{_1})$, $\Gamma(-z_{_2})$, and $\Gamma(3-{D\over2}+s+z_{_1}+z_{_2})$,
correspondingly the hypergeometric series is
\begin{eqnarray}
&&C_{_{0,19}}^a(p_{_1}^2,\;p_{_2}^2,\;p_{_3}^2,\;m^2)=
{i(m^2)^{D/2-3}\over(4\pi)^{D/2}}\Gamma(2-{D\over2})\Gamma({D\over2}-1)
\nonumber\\
&&\hspace{4.2cm}\times
\sum\limits_{j=0}^\infty\sum\limits_{n_{_1}=0}^\infty\sum\limits_{n_{_2}=0}^\infty
{(-)^{n_{_1}+n_{_2}}\over j!n_{_1}!n_{_2}!}
\eta_{_{33}}^{j}\eta_{_{13}}^{n_{_1}}\eta_{_{23}}^{n_{_2}}\Gamma(1+j+n_{_1})
\nonumber\\
&&\hspace{4.2cm}\times
{\Gamma(3-{D\over2}+j+n_{_1}+n_{_2})\Gamma(1+j+n_{_2})
\over\Gamma({D\over2}+j+n_{_1}+n_{_2})\Gamma(3-{D\over2}+j)}\;.
\label{1M19-1}
\end{eqnarray}
The convergent region of Eq.(\ref{1M19-1}) totally coincides with
$\Omega_{_{18}}^a$ in Eq.(\ref{1M18-2}).

\item[$\bullet$ 20:] For the isolated singularities defined by the poles of
$\Gamma(-z_{_1})$, $\Gamma({D\over2}-2-s-z_{_1})$, and $\Gamma({D\over2}-2-s-z_{_2})$,
the derived hypergeometric function contains some unknown linear combination parameters
originating from the non-isolated singularities.
The condition $|p_{_3}^2|>\max(|p_{_1}^2|,\;|p_{_2}^2|,\;m^2)$,
$|p_{_2}^2|>m^2$, $|p_{_3}^2m^2|>|p_{_1}^2p_{_2}^2|$ should be satisfied
to guarantee the corresponding triple series converging.

\item[$\bullet$ 23:] For the isolated singularities defined by the poles of
$\Gamma(-z_{_1})$, $\Gamma({D\over2}-2-s-z_{_1})$, and $\Gamma(1+s+z_{_1}+z_{_2})$,
the corresponding hypergeometric series is
\begin{eqnarray}
&&C_{_{0,23}}^a(p_{_1}^2,\;p_{_2}^2,\;p_{_3}^2,\;m^2)=
{i(-)^{D/2}(p_{_3}^2)^{D/2-3}\over(4\pi)^{D/2}}\Gamma(2-{D\over2})\Gamma({D\over2}-1)
\nonumber\\
&&\hspace{4.2cm}\times
\sum\limits_{j=0}^\infty\sum\limits_{n_{_1}=0}^\infty\sum\limits_{n_{_2}=0}^\infty
\xi_{_{33}}^{D/2-2+j-n_{_1}}x_{_{13}}^{n_{_1}}x_{_{23}}^{1-D/2-j-n_{_2}}
\nonumber\\
&&\hspace{4.2cm}\times
{(-)^{j+n_{_1}}\Gamma({D\over2}-1+n_{_1}+n_{_2})\Gamma({D\over2}-1+j+n_{_2})
\over j!n_{_1}!n_{_2}!\Gamma({D\over2}-1+n_{_2})}
\nonumber\\
&&\hspace{4.2cm}\times
{\Gamma(2-{D\over2}-j+n_{_1})\over\Gamma({D\over2}-1-j+n_{_1})}
\nonumber\\
&&\hspace{3.7cm}=
{i(-)^{D/2}(p_{_2}^2)^{D/2-3}\over(4\pi)^{D/2}}\Gamma(2-{D\over2})\Gamma({D\over2}-1)
\nonumber\\
&&\hspace{4.2cm}\times
\sum\limits_{j=0}^\infty\sum\limits_{n_{_1}=1}^\infty
\sum\limits_{n_{_2}=0}^\infty
\xi_{_{32}}^{D/2-2-n_{_1}}x_{_{12}}^{j+n_{_1}}x_{_{32}}^{D/2-2+n_{_2}}
\nonumber\\
&&\hspace{4.2cm}\times
{(-)^{n_{_1}}\Gamma({D\over2}-1+j+n_{_1}+n_{_2})\Gamma({D\over2}-1+j+n_{_2})
\over j!(j+n_{_1})!n_{_2}!\Gamma({D\over2}-1+n_{_2})}
\nonumber\\
&&\hspace{4.2cm}\times
{\Gamma(2-{D\over2}+n_{_1})\over\Gamma({D\over2}-1+n_{_1})}
\nonumber\\
&&\hspace{4.2cm}
+{i(-)^{D/2}(p_{_2}^2)^{D/2-3}\over(4\pi)^{D/2}}\Gamma(2-{D\over2})\Gamma({D\over2}-1)
\nonumber\\
&&\hspace{4.2cm}\times
\sum\limits_{n_{_1}=0}^\infty\sum\limits_{j=1}^\infty
\sum\limits_{n_{_2}=0}^\infty
\xi_{_{32}}^{D/2-2+j}x_{_{12}}^{n_{_1}}x_{_{32}}^{D/2-2+n_{_2}}
\nonumber\\
&&\hspace{4.2cm}\times
{(-)^{j}\Gamma({D\over2}-1+n_{_1}+n_{_2})\Gamma({D\over2}-1+j+n_{_1}+n_{_2})
\over(j+n_{_1})!n_{_1}!n_{_2}!\Gamma({D\over2}-1+n_{_2})}
\nonumber\\
&&\hspace{4.2cm}\times
{\Gamma(2-{D\over2}+j)\over\Gamma({D\over2}-1+j)}
\nonumber\\
&&\hspace{4.2cm}
+{i(-)^{D/2}(p_{_2}^2)^{D/2-3}\over(4\pi)^{D/2}}\Gamma^2(2-{D\over2})
\nonumber\\
&&\hspace{4.2cm}\times
\sum\limits_{j=0}^\infty\sum\limits_{n_{_2}=0}^\infty
\xi_{_{32}}^{D/2-2}x_{_{12}}^{j}x_{_{32}}^{D/2-2+n_{_2}}
\nonumber\\
&&\hspace{4.2cm}\times
{\Gamma({D\over2}-1+j+n_{_2})\Gamma({D\over2}-1+j+n_{_2})
\over(j!)^2n_{_2}!\Gamma({D\over2}-1+n_{_2})}\;.
\label{1M23-1}
\end{eqnarray}
The condition $|p_{_2}^2|>\max(|p_{_1}^2|,\;|p_{_3}^2|,\;m^2)$,
$m^2>|p_{_1}^2|$ is essential to guarantee the convergence of the hypergeometric series.
By implementing the residue theorem, one derives the first expression above.
The second expression decomposes the summation indices of the hypergeometric series
in the first expression, so that $D=4$ is the pole of the same fixed order
for each term of every hypergeometric series in the second expression above.
In addition, the second expression also satisfies the system of linear PDEs
in Eq.~(\ref{1M-PDEs}).

Correspondingly the convergent region of the series is given by
\begin{eqnarray}
&&\Omega_{_{23}}^a=\left\{(\xi_{_{32}},\;x_{_{12}},\;x_{_{32}}):\left|\begin{array}{l}
|x_{_{12}}|<|\xi_{_{32}}|,\;\sqrt{|x_{_{12}}|}+\sqrt{|x_{_{32}}|}<1,\\
|\xi_{_{32}}|+|x_{_{32}}|<1,\;|\xi_{_{32}}|+|x_{_{12}}|<1,\\
|x_{_{12}}|+|\xi_{_{32}}x_{_{32}}|<|\xi_{_{32}}|,\;|x_{_{12}}|(1+|\xi_{_{32}}|)<|\xi_{_{32}}|
\end{array}\right.\right\}\;.
\label{1M23-2}
\end{eqnarray}

\item[$\bullet$ 24:] For the isolated singularities defined by the poles of
$\Gamma(-z_{_1})$, $\Gamma({D\over2}-2-s-z_{_1})$, and $\Gamma(3-{D\over2}+s+z_{_1}+z_{_2})$,
the corresponding hypergeometric series is
\begin{eqnarray}
&&C_{_{0,24}}^a(p_{_1}^2,\;p_{_2}^2,\;p_{_3}^2,\;m^2)=
-{i(-)^{D/2}(p_{_2}^2)^{D/2-3}\over(4\pi)^{D/2}}\Gamma(2-{D\over2})
\Gamma({D\over2}-1)
\nonumber\\
&&\hspace{4.2cm}\times
\sum\limits_{j=0}^\infty\sum\limits_{n_{_1}=1}^\infty
\sum\limits_{n_{_2}=0}^\infty
\xi_{_{32}}^{D/2-2-n_{_1}}x_{_{12}}^{j+n_{_1}}x_{_{32}}^{n_{_2}}
\nonumber\\
&&\hspace{4.2cm}\times
{(-)^{n_{_1}}\Gamma(1+j+n_{_1}+n_{_2})\Gamma(1+j+n_{_2})
\over j!(j+n_{_1})!n_{_2}!\Gamma(3-{D\over2}+n_{_2})}
\nonumber\\
&&\hspace{4.2cm}\times
{\Gamma(2-{D\over2}+n_{_1})\over\Gamma({D\over2}-1+n_{_1})}
\nonumber\\
&&\hspace{4.2cm}
-{i(-)^{D/2}(p_{_2}^2)^{D/2-3}\over(4\pi)^{D/2}}\Gamma(2-{D\over2})
\Gamma({D\over2}-1)
\nonumber\\
&&\hspace{4.2cm}\times
\sum\limits_{n_{_1}=0}^\infty\sum\limits_{j=1}^\infty
\sum\limits_{n_{_2}=0}^\infty
\xi_{_{32}}^{D/2-2+j}x_{_{12}}^{n_{_1}}x_{_{32}}^{n_{_2}}
\nonumber\\
&&\hspace{4.2cm}\times
{(-)^{j}\Gamma(1+n_{_1}+n_{_2})\Gamma(1+j+n_{_1}+n_{_2})
\over(j+n_{_1})!n_{_1}!n_{_2}!\Gamma(3-{D\over2}+n_{_2})}
\nonumber\\
&&\hspace{4.2cm}\times
{\Gamma(2-{D\over2}+j)\over\Gamma({D\over2}-1+j)}
\nonumber\\
&&\hspace{4.2cm}
-{i(-)^{D/2}(p_{_2}^2)^{D/2-3}\over(4\pi)^{D/2}}\Gamma^2(2-{D\over2})
\nonumber\\
&&\hspace{4.2cm}\times
\sum\limits_{j=0}^\infty\sum\limits_{n_{_2}=0}^\infty
\xi_{_{32}}^{D/2-2}x_{_{12}}^{j}x_{_{32}}^{n_{_2}}
\nonumber\\
&&\hspace{4.2cm}\times
{\Gamma^2(1+j+n_{_2})\over (j!)^2n_{_2}!\Gamma(3-{D\over2}+n_{_2})}\;.
\label{1M24-1}
\end{eqnarray}
The condition $|p_{_2}^2|>\max(|k^2|,\;|p_{_1}^2|,\;m^2)$,
$m^2>|p_{_1}^2|$ is essential to guarantee the multiple power series
converging. In addition, $D=4$ is the first order pole of the first two triple hypergeometric
functions, and the second order pole of the third double hypergeometric
function, respectively.
Certainly the expression satisfies the system of linear PDEs
in Eq.~(\ref{1M-PDEs}).

Applying Horn's study of convergence, one finds that
the absolutely and uniformly convergent region of Eq.(\ref{1M24-1})
is given by the domain $\Omega_{_{23}}^a$ exactly.

\item[$\bullet$ 26:] For the isolated singularities defined by the poles of
$\Gamma(-z_{_2})$, $\Gamma({D\over2}-2-s-z_{_1})$, and $\Gamma({D\over2}-2-s-z_{_2})$,
the derived hypergeometric function contains some unknown linear combining parameters
originating from the non-isolating singularities.
The condition $|p_{_3}^2|>\max(|p_{_1}^2|,\;|p_{_2}^2|,\;m^2)$,
$|p_{_1}^2|>m^2$, $|p_{_3}^2m^2|>|p_{_1}^2p_{_2}^2|$ is necessary
to guarantee the hypergeometric series converging.

\item[$\bullet$ 29:] For the isolated singularities defined by the poles of
$\Gamma(-z_{_2})$, $\Gamma({D\over2}-2-s-z_{_2})$, and $\Gamma(1+s+z_{_1}+z_{_2})$,
the corresponding hypergeometric series is written as
\begin{eqnarray}
&&C_{_{0,29}}^a(p_{_1}^2,\;p_{_2}^2,\;p_{_3}^2,\;m^2)=C_{_{0,23}}^a(
p_{_2}\rightarrow p_{_1},\;\xi_{_{32}}\rightarrow\xi_{_{31}},\;
x_{_{12}}\rightarrow x_{_{21}},\;x_{_{32}}\rightarrow x_{_{31}})\;.
\label{1M29-1}
\end{eqnarray}
The convergent region of the series is similarly given by
\begin{eqnarray}
&&\Omega_{_{29}}^a=\left\{(\xi_{_{31}},\;x_{_{21}},\;x_{_{31}}):\left|\begin{array}{l}
|x_{_{21}}|<|\xi_{_{31}}|,\;\sqrt{|x_{_{21}}|}+\sqrt{|x_{_{31}}|}<1,\\
|\xi_{_{31}}|+|x_{_{21}}|<1,\;|\xi_{_{31}}|+|x_{_{31}}|<1,\\
|\xi_{_{31}}x_{_{31}}|+|x_{_{21}}|<|\xi_{_{31}}|,\;|x_{_{21}}|(1+|\xi_{_{31}}|)<|\xi_{_{31}}|
\end{array}\right.\right\}\;.
\label{1M29-2}
\end{eqnarray}

\item[$\bullet$ 30:] When the isolated singularities are defined by the poles of
$\Gamma(-z_{_2})$, $\Gamma({D\over2}-2-s-z_{_2})$, and
$\Gamma(3-{D\over2}+s+z_{_1}+z_{_2})$, the corresponding hypergeometric series is
\begin{eqnarray}
&&C_{_{0,30}}^a(p_{_1}^2,\;p_{_2}^2,\;p_{_3}^2,\;m^2)=C_{_{0,24}}^a(
p_{_2}\rightarrow p_{_1},\;\xi_{_{32}}\rightarrow\xi_{_{31}},\;
x_{_{12}}\rightarrow x_{_{21}},\;x_{_{32}}\rightarrow x_{_{31}})\;.
\label{1M30-1}
\end{eqnarray}
The necessary condition $|p_{_1}^2|>\max(|p_{_2}^2|,\;|p_{_3}^2|,\;m^2)$,
$m^2>|p_{_2}^2|$ should be satisfied to guarantee the multiple power series
converging. The convergent region of this series is also described by the region
$\Omega_{_{29}}^a$ in Eq.(\ref{1M29-2}).

\item[$\bullet$ 32:] For the isolated singularities defined by the poles of
$\Gamma({D\over2}-2-s-z_{_2})$, $\Gamma({D\over2}-2-s-z_{_2})$, and
$\Gamma(1+s+z_{_1}+z_{_2})$, the corresponding analytic expression
is zero,
\begin{eqnarray}
&&C_{_{0,32}}^a(p_{_1}^2,\;p_{_2}^2,\;p_{_3}^2,\;m^2)\equiv0\;.
\label{1M32-1}
\end{eqnarray}

\item[$\bullet$ 33:] For the isolated singularities defined by the poles of $\Gamma({D\over2}-2-s-z_{_2})$,
$\Gamma({D\over2}-2-s-z_{_2})$, and $\Gamma(3-{D\over2}+s+z_{_1}+z_{_2})$,
correspondingly the derived hypergeometric function is
\begin{eqnarray}
&&C_{_{0,33}}^a(p_{_1}^2,\;p_{_2}^2,\;p_{_3}^2,\;m^2)=
-{i(-)^{D/2}(p_{_3}^2)^{D/2-3}\over(4\pi)^{D/2}}
\Gamma(2-{D\over2})\Gamma({D\over2}-1)\sum\limits_{j=0}^\infty
\sum\limits_{n_{_1}=0}^\infty\sum\limits_{n_{_2}=0}^\infty
\nonumber\\
&&\hspace{4.2cm}\times
{(-)^{n_{_1}+n_{_2}}\over j!n_{_1}!n_{_2}!}
\xi_{_{33}}^{D/2-1+j+n_{_1}+n_{_2}}x_{_{13}}^{-1-j-n_{_1}}x_{_{23}}^{-1-j-n_{_2}}
\nonumber\\
&&\hspace{4.2cm}\times
{\Gamma(3-{D\over2}+j+n_{_1}+n_{_2})\Gamma(1+j+n_{_1})\Gamma(1+j+n_{_2})
\over\Gamma({D\over2}+j+n_{_1}+n_{_2})\Gamma(3-{D\over2}+j)}\;.
\label{1M33-1}
\end{eqnarray}
The necessary condition $m^2<\min(|p_{_1}^2|,\;|p_{_2}^2|)$,
$|p_{_3}^2|m^2<|p_{_1}^2p_{_2}^2|$ should be satisfied
to guarantee the triple series converging.
The convergent region of Eq~(\ref{1M33-1}) is
\begin{eqnarray}
&&\Omega_{_{33}}^a=\left\{(\xi_{_{33}},\;x_{_{13}},\;x_{_{23}}):\left|\begin{array}{l}
|\xi_{_{33}}|<|x_{_{13}}x_{_{23}}|,\;|\xi_{_{33}}|<|x_{_{13}}|,\;|\xi_{_{33}}|<|x_{_{23}}|;\\
|\xi_{_{33}}|(1+|x_{_{13}}|)<|x_{_{13}}x_{_{23}}|,\;|\xi_{_{33}}|(1+|x_{_{23}}|)<|x_{_{13}}x_{_{23}}|,\\
|\xi_{_{33}}|<(|x_{_{13}}|-|\xi_{_{33}}|)(|x_{_{23}}|-|\xi_{_{33}}|)
\end{array}\right.\right\}\;.
\label{1M33-4}
\end{eqnarray}
\end{itemize}

\subsection{The scalar integral in different convergent regions\label{sec3-2}}
\indent\indent
In order to pursue our analysis, we summarize the results presented above.
The absolutely and uniformly convergent domain of
the hypergeometric function $C_{_{0,1}}^a$ is the domain $\Omega_{_{1}}^a$,
that of the hypergeometric function $C_{_{0,2}}^a$ is the domain $\Omega_{_{2}}^a$,
that of the hypergeometric functions $C_{_{0,3}}^a$, $C_{_{0,4}}^a$ is the domain $\Omega_{_{3}}^a$,
that of the hypergeometric function $C_{_{0,6}}^a$ is the domain $\Omega_{_{6}}^a$,
that of the hypergeometric function $C_{_{0,7}}^a$ is the domain $\Omega_{_{7}}^a$,
that of the hypergeometric functions $C_{_{0,8}}^a$ and $C_{_{0,9}}^a$ is the domain $\Omega_{_{8}}^a$,
that of the hypergeometric functions $C_{_{0,10}}^a$ and $C_{_{0,13}}^a$ is the domain $\Omega_{_{10}}^a$,
that of the hypergeometric functions $C_{_{0,11}}^a$ and $C_{_{0,14}}^a$ is the domain $\Omega_{_{11}}^a$,
that of the hypergeometric functions $C_{_{0,18}}^a$ and $C_{_{0,19}}^a$ is the domain $\Omega_{_{18}}^a$,
that of the hypergeometric functions $C_{_{0,23}}^a$ and $C_{_{0,24}}^a$ is the domain $\Omega_{_{23}}^a$,
that of the hypergeometric functions $C_{_{0,29}}^a$ and $C_{_{0,30}}^a$ is the domain $\Omega_{_{29}}^a$,
as well as that of the hypergeometric function $C_{_{0,33}}^a$ is the domain $\Omega_{_{33}}^a$, respectively.
The scalar integral of other parameter space contains some unknown linear combination parameters originating from
the non-isolated singularities. Among those convergent regions, the following domains
compose a set
\begin{eqnarray}
&&\Big\{\Omega_{_{7}}^a,\;\Omega_{_{8}}^a,\;\Omega_{_{11}}^a,\;\Omega_{_{18}}^a,\;
\Omega_{_{23}}^a,\;\Omega_{_{29}}^a\Big\}\;,
\label{1MSum-1}
\end{eqnarray}
whose element satisfies the following constraints simultaneously,
\begin{eqnarray}
&&\Omega_{_{i}}^a\bigcap\Omega_{_{j}}^a=\emptyset,\;i,\;j=7,\;8,\;11,\;18,\;23,\;29,\;i\neq j.
\label{1MSum-2}
\end{eqnarray}
In addition, the relations between the elements of the set and those
convergent regions which do not belong to the set are
\begin{eqnarray}
&&\Omega_{_{7}}^a\subset\Omega_{_{\alpha_{_1}}}^a,\;\alpha_{_1}=1,\;2,\;6,\;33;\;
\Omega_{_{7}}^a\bigcap\Omega_{_{\beta_{_1}}}^a=\emptyset,\;\beta_{_1}=3,\;10;
\nonumber\\
&&\hspace{0.0cm}
\Omega_{_{11}}^a\subset\Omega_{_{\alpha_{_2}}}^a,\;\alpha_{_2}=10,\;33;\;
\Omega_{_{11}}^a\bigcap\Omega_{_{\beta_{_2}}}^a=\emptyset,\;\beta_{_2}=1,\;2,\;3,\;6;
\nonumber\\
&&\hspace{0.0cm}
\Omega_{_{29}}^a\subset\Omega_{_{10}}^a,\;
\Omega_{_{29}}^a\bigcap\Omega_{_{\beta_{_3}}}^a=\emptyset,\;\beta_{_3}=1,\;2,\;3,\;6,\;33;
\nonumber\\
&&\hspace{0.0cm}
\Omega_{_{8}}^a\subset\Omega_{_{\alpha_{_4}}}^a,\;\alpha_{_4}=3,\;33;\;
\Omega_{_{8}}^a\bigcap\Omega_{_{\beta_{_4}}}^a=\emptyset,\;\beta_{_4}=1,\;2,\;6,\;10;
\nonumber\\
&&\hspace{0.0cm}
\Omega_{_{23}}^a\subset\Omega_{_{3}}^a,\;
\Omega_{_{23}}^a\bigcap\Omega_{_{\beta_{_5}}}^a=\emptyset,\;\beta_{_5}=1,\;2,\;6,\;10,\;33;
\nonumber\\
&&\hspace{0.0cm}
\Omega_{_{18}}^a\bigcap\Omega_{_{\beta_{_6}}}^a=\emptyset,\;\beta_{_6}=1,\;2,\;3,\;6,\;10,\;33.
\label{1MSum-3}
\end{eqnarray}
With the preparation above the concrete expression of $C_{_0}^a$ in different parameter space
is respectively presented as following.
\begin{itemize}
\item[$\bullet$ 1:] In the convergent region $\Omega_{_7}^a$,
the scalar integral $C_{_0}^a$ is written as the sum
\begin{eqnarray}
&&C_{_0}^a(p_{_1}^2,\;p_{_2}^2,\;p_{_3}^2,\;m^2)=
\Big\{C_{_{0,1}}^a+C_{_{0,2}}^a+C_{_{0,6}}^a+C_{_{0,7}}^a+C_{_{0,33}}^a\Big\}
(p_{_1}^2,\;p_{_2}^2,\;p_{_3}^2,\;m^2)\;.
\label{1LoopC-1ab1}
\end{eqnarray}
In the limit $m^2\rightarrow0$,
the expression of Eq.(\ref{1LoopC-1ab1}) recovers that presented in Eq.(\ref{Massless-I1}).
Furthermore, we expand the hypergeometric functions $C_{_{0,i}}^a,\;(i=1,\;2,\;6,\;7,\;33)$
around $\varepsilon=0$, and find that $D=4$ is the pole of the second order for
each hypergeometric function individually. Nevertheless the sum presented in
Eq.(\ref{1LoopC-1ab1}) is an analytic function of the dimension in the
neighborhood of $D=4$ because those singularities are canceled clearly,
and the final result is given as
\begin{eqnarray}
&&C_{_0}^a(p_{_1}^2,\;p_{_2}^2,\;p_{_3}^2,\;m^2)=
{i(p_{_3}^2)^{D/2-3}\over(4\pi)^{D/2}}
\Big\{f_{_{\varepsilon}}(x_{_{13}},\;x_{_{23}})
\nonumber\\
&&\hspace{4.0cm}
+2\varepsilon\sum\limits_{j=1}^\infty\sum\limits_{n_{_1}=0}^\infty\sum\limits_{n_{_2}=0}^\infty
\xi_{_{33}}^jx_{_{13}}^{n_{_1}}x_{_{23}}^{n_{_2}}A_{_{j,n_{_1},n_{_2}}}^{\prime(a)}(x_{_{13}},\;x_{_{23}})
\nonumber\\
&&\hspace{4.0cm}\times
{\Gamma(j)\Gamma^2(1+j+n_{_1}+n_{_2})
\over j!n_{_1}!n_{_2}!\Gamma(1+j+n_{_1})\Gamma(1+j+n_{_2})}
\nonumber\\
&&\hspace{4.0cm}
+2\varepsilon\sum\limits_{j=1}^\infty\sum\limits_{n_{_1}=0}^\infty\sum\limits_{n_{_2}=0}^\infty
\xi_{_{32}}^jx_{_{13}}^{n_{_1}}\xi_{_{33}}^{n_{_2}}B_{_{j,n_{_1},n_{_2}}}^{(a)}
\nonumber\\
&&\hspace{4.0cm}\times
{\Gamma(j)\Gamma(j+n_{_2})\Gamma^2(1+n_{_1}+n_{_2})
\over n_{_1}!n_{_2}!\Gamma(1+j+n_{_1}+n_{_2})\Gamma(1+j+n_{_2})}
\nonumber\\
&&\hspace{4.0cm}
+2\varepsilon\sum\limits_{j=1}^\infty\sum\limits_{n_{_1}=0}^\infty\sum\limits_{n_{_2}=0}^\infty
\xi_{_{31}}^j\xi_{_{33}}^{n_{_1}}x_{_{23}}^{n_{_2}}C_{_{j,n_{_1},n_{_2}}}^{(a)}
\nonumber\\
&&\hspace{4.0cm}\times
{\Gamma(j)\Gamma(j+n_{_1})\Gamma^2(1+n_{_1}+n_{_2})
\over n_{_1}!n_{_2}!\Gamma(1+j+n_{_1}+n_{_2})\Gamma(1+j+n_{_1})}
\nonumber\\
&&\hspace{4.0cm}
+\sum\limits_{j=1}^\infty\sum\limits_{n_{_1}=0}^\infty\sum\limits_{n_{_2}=0}^\infty
\Big({\xi_{_{33}}\over x_{_{13}}x_{_{23}}}\Big)^{j}\xi_{_{31}}^{n_{_1}}\xi_{_{32}}^{n_{_2}}
\nonumber\\
&&\hspace{4.0cm}\times
{(-)^{n_{_1}+n_{_2}}\Gamma(j+n_{_1}+n_{_2})\Gamma(j+n_{_1})\Gamma(j+n_{_2})
\over n_{_1}!n_{_2}!\Gamma^2(j)\Gamma(1+j+n_{_1}+n_{_2})}
\nonumber\\
&&\hspace{4.0cm}\times
\Big[\psi(j+n_{_1})+\psi(j+n_{_2})-2\psi(j)
\nonumber\\
&&\hspace{4.0cm}
-{1\over j+n_{_1}+n_{_2}}+\ln{\xi_{_{33}}\over x_{_{13}}x_{_{23}}}
+{\varepsilon\over2} D_{_{j,n_{_1},n_{_2}}}^{(a)}\Big]
\nonumber\\
&&\hspace{4.0cm}
+2\varepsilon\sum\limits_{j=1}^\infty\sum\limits_{n_{_1}=1}^\infty\sum\limits_{n_{_2}=0}^\infty
\xi_{_{31}}^j\xi_{_{32}}^{n_{_1}}\xi_{_{33}}^{n_{_2}}
\nonumber\\
&&\hspace{4.0cm}\times
{(-)^{j+n_{_1}}\Gamma(j)\Gamma(j+n_{_1}+n_{_2})\Gamma(n_{_1})\Gamma^2(1+n_{_2})\over
\Gamma(1+j+n_{_1}+n_{_2})\Gamma(1+n_{_1}+n_{_2})\Gamma(1+j+n_{_2})}
\nonumber\\
&&\hspace{4.0cm}
+{\cal O}(\varepsilon^2)\Big\}\;.
\label{1LoopC-1ab2}
\end{eqnarray}
with
\begin{eqnarray}
&&f_{_{\varepsilon}}(x,y)=
\sum\limits_{n_{_1}=0}^\infty\sum\limits_{n_{_2}=0}^\infty
x^{n_{_1}}y^{n_{_2}}
{\Gamma^2(1+n_{_1}+n_{_2})\over n_{_1}!n_{_2}!\Gamma(1+n_{_1})\Gamma(1+n_{_2})}
\nonumber\\
&&\hspace{2.0cm}\times
\Big[\ln x\ln y-2\ln x\psi(1+n_{_2})-2\ln y\psi(1+n_{_1})
\nonumber\\
&&\hspace{2.0cm}
+2\ln(xy)\psi(1+n_{_1}+n_{_2})
+4\psi(1+n_{_1})\psi(1+n_{_2})
\nonumber\\
&&\hspace{2.0cm}
-4\{\psi(1+n_{_1})+\psi(1+n_{_2})\}\psi(1+n_{_1}+n_{_2})
\nonumber\\
&&\hspace{2.0cm}
+4\psi^2(1+n_{_1}+n_{_2})+2\psi^{(1)}(1+n_{_1}+n_{_2})
-{\varepsilon\over2} A_{_{n_{_1},n_{_2}}}^{(a)}(x,y)\Big]\;.
\label{1LoopC-1ab2a}
\end{eqnarray}
Here the coefficients of $\varepsilon$ above $A_{_{n_{_1},n_{_2}}}^{(a)}(x,y)$,
$A_{_{j,n_{_1},n_{_2}}}^{\prime(a)}(x,y)$, $B_{_{j,n_{_1},n_{_2}}}^{(a)}$,
$C_{_{j,n_{_1},n_{_2}}}^{(a)}$, and $D_{_{j,n_{_1},n_{_2}}}^{(a)}$ can be found
in Eq.(\ref{app2-1}). The absolutely and uniformly converging region $\Omega_{_7}^a$
is a proper subset of the cube
\begin{eqnarray}
&&\Xi_{_{\rm K}}^a=\left\{(\xi_{_{33}},\;x_{_{13}},\;x_{_{23}}):\left|\begin{array}{l}
|\xi_{_{33}}|<1,\;|x_{_{13}}|<1,\;|x_{_{23}}|<1
\end{array}\right.\right\}\;.
\label{1LoopC-1ab3}
\end{eqnarray}
In order to continue the scalar integral $C_{_0}^a$ from the region $\Omega_{_7}^a$
to the whole cube $\Xi_{_{\rm K}}^a$, one can employ the system of linear PDEs in
Eq.(\ref{1M-PDEs}) with the finite element method. We will consider this point
in detail in section \ref{sec6}.

\item[$\bullet$ 2(a):] In the convergent region $\Omega_{_{11}}^a$,
the scalar integral $C_{_0}^a$ is written as the sum
\begin{eqnarray}
&&C_{_0}^a(p_{_1}^2,\;p_{_2}^2,\;p_{_3}^2,\;m^2)=
\Big\{C_{_{0,10}}^a+C_{_{0,11}}^a+C_{_{0,13}}^a+C_{_{0,14}}^a+C_{_{0,33}}^a\Big\}
(p_{_1}^2,\;p_{_2}^2,\;p_{_3}^2,\;m^2)
\nonumber\\
&&\hspace{3.5cm}=
{i(-)^{D/2}(p_{_1}^2)^{D/2-3}\over(4\pi)^{D/2}}
F_{a,p_{_1}}(\xi_{_{31}},\;x_{_{31}},\;x_{_{21}})\;.
\label{1LoopC-2acd1}
\end{eqnarray}
In the limit $m^2\rightarrow0$,
the expression of Eq.(\ref{1LoopC-2acd1}) recovers that presented in Eq.(\ref{Massless-II1}).
Additionally we expand the hypergeometric functions $C_{_{0,i}}^a,\;(i=10,\;11,\;13,\;14,\;33)$
around $\varepsilon=0$, and find that $D=4$ is the pole of the second order for
each hypergeometric function individually. Nevertheless the sum presented in
Eq.(\ref{1LoopC-2acd1}) is an analytic function of the dimension in the
neighborhood of $D=4$, and expansion of $C_{_0}^a$ around $\varepsilon=0$ is
\begin{eqnarray}
&&C_{_0}^a(p_{_1}^2,\;p_{_2}^2,\;p_{_3}^2,\;m^2)=
{i(p_{_1}^2)^{D/2-3}\over(4\pi)^{D/2}}\Big\{
f_{_{\varepsilon}}(x_{_{21}},x_{_{31}})
\nonumber\\
&&\hspace{4.0cm}
+2\varepsilon\sum\limits_{j=1}^\infty\sum\limits_{n_{_1}=0}^\infty
\sum\limits_{n_{_2}=0}^\infty
\xi_{_{31}}^{j}x_{_{31}}^{n_{_1}}x_{_{21}}^{n_{_2}}
E_{_{j,n_{_1},n_{_2}}}^{(a)}
\nonumber\\
&&\hspace{4.0cm}\times
{\Gamma(j)\Gamma(1+j+n_{_1}+n_{_2})\Gamma(1+n_{_1}+n_{_2})
\over\Gamma(1+j)\Gamma(1+j+n_{_2})\Gamma^2(1+n_{_1})\Gamma(1+n_{_2})}
\nonumber\\
&&\hspace{4.0cm}
-2\varepsilon\sum\limits_{j=1}^\infty\sum\limits_{n_{_1}=1}^\infty
\sum\limits_{n_{_2}=0}^\infty
\xi_{_{31}}^{j+n_{_2}}x_{_{31}}^{j+n_{_1}}x_{_{21}}^{-j}
\nonumber\\
&&\hspace{4.0cm}\times
{(-)^{n_{_2}}\Gamma(j)\Gamma(1+j+n_{_1}+n_{_2})\Gamma(j+n_{_2})\Gamma(1+n_{_1})
\over\Gamma^2(1+j+n_{_1})\Gamma(1+j+n_{_2})\Gamma(1+n_{_2})}
\nonumber\\
&&\hspace{4.0cm}\times
\Big(\ln x_{_{31}}+\psi(1+j+n_{_1}+n_{_2})+\psi(1+n_{_1})\Big)
\nonumber\\
&&\hspace{4.0cm}
+\sum\limits_{j=1}^\infty\sum\limits_{n_{_1}=0}^\infty
\sum\limits_{n_{_2}=0}^\infty
\xi_{_{31}}^{j+n_{_1}+n_{_2}}x_{_{31}}^{n_{_1}}x_{_{21}}^{-j-n_{_1}}
\nonumber\\
&&\hspace{4.0cm}\times
{(-)^{j+n_{_2}}\Gamma(j+n_{_1}+n_{_2})\Gamma(j+n_{_1})\Gamma(1+n_{_1}+n_{_2})
\over(j+n_{_1}+n_{_2})!n_{_1}!n_{_2}!\Gamma(j)\Gamma(1+n_{_1})}
\nonumber\\
&&\hspace{4.0cm}\times
\Big[\Big({1\over j+n_{_1}+n_{_2}}-\psi(j+n_{_1})-\psi(1+n_{_1}+n_{_2})
\nonumber\\
&&\hspace{4.0cm}
+2\psi(1+n_{_1})
+\ln{x_{_{21}}\over x_{_{31}}\xi_{_{31}}}\Big)
+{\varepsilon\over2}G_{_{j,n_{_1},n_{_2}}}^{(a)}\Big]+{\cal O}(\varepsilon^2)\Big\}\;,
\label{1LoopC-2acd2}
\end{eqnarray}
where the coefficients of $\varepsilon$ above $E_{_{j,n_{_1},n_{_2}}}^{(a)}$
and $G_{_{j,n_{_1},n_{_2}}}^{(a)}$ are given in Eq.(\ref{app2-2}) also.

\item[$\bullet$ 2(b):] In the convergent region $\Omega_{_{29}}^a$,
the scalar integral $C_{_0}^a$ is formulated as
\begin{eqnarray}
&&C_{_0}^a(p_{_1}^2,\;p_{_2}^2,\;p_{_3}^2,\;m^2)=
\Big\{C_{_{0,10}}^a+C_{_{0,13}}^a+C_{_{0,29}}^a+C_{_{0,30}}^a\Big\}
(p_{_1}^2,\;p_{_2}^2,\;p_{_3}^2,\;m^2)
\nonumber\\
&&\hspace{3.5cm}=
{i(-)^{D/2}(p_{_1}^2)^{D/2-3}\over(4\pi)^{D/2}}
F_{a,p_{_1}}^\prime(\xi_{_{31}},\;x_{_{21}},\;x_{_{31}})\;.
\label{1LoopC-bef1}
\end{eqnarray}
In addition, $D=4$ is the pole of the second order for each hypergeometric function
of $C_{_{0,i}}^a\;(i=10,\;13,\;29,\;30)$ individually.
Nevertheless the sum presented in
Eq.(\ref{1LoopC-bef1}) is an analytic function of the dimension in the
neighborhood of $D=4$, and expansion of $C_{_0}^a$ around $\varepsilon=0$ is
\begin{eqnarray}
&&C_{_0}^a(p_{_1}^2,\;p_{_2}^2,\;p_{_3}^2,\;m^2)=
{i(p_{_1}^2)^{D/2-3}\over(4\pi)^{D/2}}\Big\{
f_{_{\varepsilon}}(x_{_{21}},x_{_{31}})
\nonumber\\
&&\hspace{4.0cm}
+\sum\limits_{j=1}^\infty\sum\limits_{n_{_1}=0}^\infty\sum\limits_{n_{_2}=0}^\infty
\xi_{_{31}}^{j}x_{_{31}}^{n_{_1}}x_{_{21}}^{n_{_2}}
\nonumber\\
&&\hspace{4.0cm}\times
{(-)^{j}\Gamma(1+n_{_1}+n_{_2})\Gamma(1+j+n_{_1}+n_{_2})\Gamma(j)\over
j!n_{_1}!n_{_2}!\Gamma(1+n_{_1})\Gamma(1+j+n_{_2})}
\nonumber\\
&&\hspace{4.0cm}\times
\Big[\psi(1+j+n_{_1}+n_{_2})+\psi(1+n_{_1}+n_{_2})
\nonumber\\
&&\hspace{4.0cm}
-2\psi(1+n_{_1})+\ln y_{_1}-{\varepsilon\over2}H_{_{j,n_{_1},n_{_2}}}^{(a)}\Big]
\nonumber\\
&&\hspace{4.0cm}
+\sum\limits_{j=0}^\infty\sum\limits_{n_{_1}=0}^\infty\sum\limits_{n_{_2}=1}^\infty
\xi_{_{31}}^{-n_{_2}}x_{_{31}}^{j+n_{_1}+n_{_2}}x_{_{21}}^{j+n_{_2}}
\nonumber\\
&&\hspace{4.0cm}\times
{(-)^{n_{_2}}\Gamma(1+j+n_{_1}+n_{_2})\Gamma(1+j+n_{_1})\Gamma(n_{_2})
\over j!n_{_1}!(j+n_{_2})!\Gamma(1+n_{_2})\Gamma(1+n_{_1})}
\nonumber\\
&&\hspace{4.0cm}\times
\Big[2\psi(1+n_{_1})-\psi(1+j+n_{_1}+n_{_2})-\psi(1+j+n_{_1})
\nonumber\\
&&\hspace{4.0cm}
-\ln y_{_1}+{\varepsilon\over2}P_{_{j,n_{_1},n_{_2}}}^{(a)}\Big]
+{\cal O}(\varepsilon^2)\Big\}\;,
\label{1LoopC-bef2}
\end{eqnarray}
where the coefficients of $\varepsilon$ above $H_{_{j,n_{_1},n_{_2}}}^{(a)}$ and
$P_{_{j,n_{_1},n_{_2}}}^{(a)}$ are presented in Eq.(\ref{app2-1}).
In addition, the functions $x_{_{13}}^{D/2-3}F_{a,p_{_1}},\;x_{_{13}}^{D/2-3}F_{a,p_{_1}}^\prime$
both comply with the system of linear PDEs in Eq~(\ref{1M-PDEs}).
Or equivalently, the functions $F_{a,p_{_1}},\;F_{a,p_{_1}}^\prime$ satisfy the following
system of linear PDEs
\begin{eqnarray}
&&\Big\{\Big(3-{D\over2}+\hat{\theta}_{\xi_{_{31}}}+\hat{\theta}_{x_{_{21}}}+\hat{\theta}_{x_{_{31}}}\Big)
(4-D+\hat{\theta}_{\xi_{_{31}}})
\nonumber\\
&&\hspace{0.0cm}
+{1\over\xi_{_{31}}}\hat{\theta}_{\xi_{_{31}}}\Big(2-{D\over2}+\hat{\theta}_{\xi_{_{31}}}
+\hat{\theta}_{x_{_{21}}}\Big)\Big\}F_{a,p_{_1}}(\xi_{_{31}},\;x_{_{21}},\;x_{_{31}})=0
\;,\nonumber\\
&&\Big\{\Big(3-{D\over2}+\hat{\theta}_{\xi_{_{31}}}+\hat{\theta}_{x_{_{21}}}+\hat{\theta}_{x_{_{31}}}\Big)
(1+\hat{\theta}_{x_{_{21}}}+\hat{\theta}_{x_{_{31}}})
\nonumber\\
&&\hspace{0.0cm}
-{1\over x_{_{21}}}\hat{\theta}_{x_{_{21}}}\Big(2-{D\over2}+\hat{\theta}_{\xi_{_{31}}}
+\hat{\theta}_{x_{_{21}}}\Big)\Big\}F_{a,p_{_1}}(\xi_{_{31}},\;x_{_{21}},\;x_{_{31}})=0
\;,\nonumber\\
&&\Big\{\Big(3-{D\over2}+\hat{\theta}_{\xi_{_{31}}}+\hat{\theta}_{x_{_{21}}}+\hat{\theta}_{x_{_{31}}}\Big)
(1+\hat{\theta}_{x_{_{21}}}+\hat{\theta}_{x_{_{31}}})
\nonumber\\
&&\hspace{0.0cm}
-{1\over x_{_{31}}}\hat{\theta}_{x_{_{31}}}\Big(2-{D\over2}+\hat{\theta}_{x_{_{31}}}
\Big)\Big\}F_{a,p_{_1}}(\xi_{_{31}},\;x_{_{21}},\;x_{_{31}})=0\;.
\label{1LoopC-2acd4}
\end{eqnarray}

Obviously the union of the regions $\Omega_{_{11}}^a$ and $\Omega_{_{29}}^a$
is a proper subset of the cube
\begin{eqnarray}
&&\Xi_{_{{\rm p}_{_1}}}^a=\left\{(\xi_{_{31}},\;x_{_{21}},\;x_{_{31}}):\left|\begin{array}{l}
|\xi_{_{31}}|<1,\;|x_{_{21}}|<1,\;|x_{_{31}}|<1
\end{array}\right.\right\}\;.
\label{1LoopC-1acd3}
\end{eqnarray}
In order to continue the scalar integral $C_{_0}^a$ from the region $\Omega_{_{11}}^a\bigcup\Omega_{_{29}}^a$
to the whole cube $\Xi_{_{{\rm p}_{_1}}}^a$, one employs the system of linear PDEs of Eq.~(\ref{1LoopC-2acd4}).

\item[$\bullet$ 3(a):] In the convergent region $\Omega_{_8}^a$,
the scalar integral $C_{_0}^a$ is
\begin{eqnarray}
&&C_{_0}^a(p_{_1}^2,\;p_{_2}^2,\;p_{_3}^2,\;m^2)=
\Big\{C_{_{0,3}}^a+C_{_{0,4}}^a+C_{_{0,8}}^a+C_{_{0,9}}^a+C_{_{0,33}}^a\Big\}
(p_{_1}^2,\;p_{_2}^2,\;p_{_3}^2,\;m^2)
\nonumber\\
&&\hspace{3.5cm}=
{i(p_{_2}^2)^{D/2-3}\over(4\pi)^{D/2}}F_{_{a,p_{_2}}}(\xi_{_{32}},\;x_{_{12}},\;x_{_{32}})\;.
\label{1LoopC-3acd1}
\end{eqnarray}
In the limit $m^2\rightarrow0$,
the expression of Eq.(\ref{1LoopC-3acd1}) recovers that presented in Eq.(\ref{Massless-III1}),
and the expansion of $C_{_0}^a$ around $\varepsilon=0$ is
\begin{eqnarray}
&&C_{_0}^a(p_{_1}^2,\;p_{_2}^2,\;p_{_3}^2,\;m^2)=
{\rm Eq}.(\ref{1LoopC-2acd2})(p_{_1}\rightarrow p_{_2},\;\xi_{_{31}}\rightarrow\xi_{_{32}},\;
x_{_{21}}\rightarrow x_{_{12}},\;x_{_{31}}\rightarrow x_{_{32}})\;.
\label{1LoopC-3acd2}
\end{eqnarray}

\item[$\bullet$ 3(b):] In the convergent region $\Omega_{_{23}}^a$,
the scalar integral $C_{_0}^a$ is written as
\begin{eqnarray}
&&C_{_0}^a(p_{_1}^2,\;p_{_2}^2,\;p_{_3}^2,\;m^2)=
\Big\{C_{_{0,3}}^a+C_{_{0,4}}^a+C_{_{0,23}}^a+C_{_{0,24}}^a\Big\}
(p_{_1}^2,\;p_{_2}^2,\;p_{_3}^2,\;m^2)
\nonumber\\
&&\hspace{3.5cm}=
{i(p_{_2}^2)^{D/2-3}\over(4\pi)^{D/2}}F_{_{a,p_{_2}}}^\prime(\xi_{_{32}},\;x_{_{12}},\;x_{_{32}})\;.
\label{1LoopC-3bef1}
\end{eqnarray}
Correspondingly the expansion of $C_{_0}^a$ around $\varepsilon=0$ is given by
\begin{eqnarray}
&&C_{_0}^a(p_{_1}^2,\;p_{_2}^2,\;p_{_3}^2,\;m^2)=
{\rm Eq}.(\ref{1LoopC-bef2})(p_{_1}\rightarrow p_{_2},\;\xi_{_{31}}\rightarrow\xi_{_{32}},\;
x_{_{21}}\rightarrow x_{_{12}},\;x_{_{31}}\rightarrow x_{_{32}})\;.
\label{1LoopC-3bef2}
\end{eqnarray}
Additionally the functions $x_{_2}^{D/2-3}F_{a,p_{_2}},\;x_{_2}^{D/2-3}F_{a,p_{_2}}^\prime$
both comply with the system of linear PDEs in Eq~(\ref{1M-PDEs}).
Or equivalently, the functions $F_{a,p_{_2}},\;F_{a,p_{_2}}^\prime$ satisfy the
system which is obtained from Eq.(\ref{1LoopC-2acd4}) through
the interchanging $1\leftrightarrow2$.

Obviously the union of the regions $\Omega_{_{8}}^a$ and $\Omega_{_{23}}^a$
is a proper subset of the cube
\begin{eqnarray}
&&\Xi_{_{{\rm p}_{_2}}}^a=\left\{(\xi_{_{32}},\;x_{_{12}},\;x_{_{32}}):\left|\begin{array}{l}
|\xi_{_{32}}|<1,\;|x_{_{12}}|<1,\;|x_{_{32}}|<1
\end{array}\right.\right\}\;.
\label{1LoopC-3acd3}
\end{eqnarray}
In order to continue the scalar integral $C_{_0}^a$ from the region
$\Omega_{_{8}}^a\bigcup\Omega_{_{23}}^a$
to the whole cube $\Xi_{_{{\rm p}_{_1}}}^a$, one employs the correspondingly
holonomic hypergeometric system of linear PDEs.

\item[$\bullet$ 4: ] In the parameter space $\Omega_{_{18}}^a$,
the scalar integral $C_{_0}^a$ is written as
\begin{eqnarray}
&&C_{_0}^a(p_{_1}^2,\;p_{_2}^2,\;p_{_3}^2,\;m^2)=
\Big\{C_{_{0,18}}^a+C_{_{0,19}}^a\Big\}(p_{_1}^2,\;p_{_2}^2,\;p_{_3}^2,\;m^2)
\nonumber\\
&&\hspace{3.5cm}=
{i(m^2)^{D/2-3}\over(4\pi)^{D/2}}F_{_{a,m}}(\eta_{_{13}},\;\eta_{_{23}},\;\eta_{_{33}})\;.
\label{1LoopC4-1}
\end{eqnarray}
The function of Eq~(\ref{1LoopC4-1}) is absolutely and uniformly convergent in the connect region $\Omega_{_{18}}^a$.
Except a convenient factor, the expression coincides with Eq~(27) in the
literature~\cite{Davydychev3}. We expand the hypergeometric functions $C_{_{0,i}}^a,\;(i=18,\;19)$
around $\varepsilon=0$, and find that $D=4$ is the pole of the first order for
each hypergeometric function individually. Nevertheless the sum presented in
Eq.(\ref{1LoopC4-1}) is an analytic function of the dimension in the
neighborhood of $D=4$, and the expansion of $C_{_0}$ around $\varepsilon=0$ is
\begin{eqnarray}
&&C_{_0}^a(p_{_1}^2,\;p_{_2}^2,\;p_{_3}^2,\;m^2)=
{i(m^2)^{D/2-3}\over(4\pi)^{D/2}}\Big\{
\sum\limits_{j=0}^\infty\sum\limits_{n_{_1}=0}^\infty\sum\limits_{n_{_2}=0}^\infty
\eta_{_{33}}^{j}\eta_{_{13}}^{n_{_1}}\eta_{_{23}}^{n_{_2}}
\nonumber\\
&&\hspace{4.0cm}\times
{(-)^{n_{_1}+n_{_2}}\Gamma(1+j+n_{_1}+n_{_2})\Gamma(1+j+n_{_1})\Gamma(1+j+n_{_2})
\over j!n_{_1}!n_{_2}!\Gamma(2+j+n_{_1}+n_{_2})\Gamma(1+j)}
\nonumber\\
&&\hspace{4.0cm}\times
\Big[2\psi(1+j)-\psi(1+j+n_{_1})-\psi(1+j+n_{_2})
\nonumber\\
&&\hspace{4.0cm}
-\ln\eta_{_{33}}
-{1\over1+j+n_{_1}+n_{_2}}+{\varepsilon\over2}Q_{_{j,n_{_1},n_{_2}}}^{(a)}\Big]
+{\cal O}(\varepsilon^2)\Big\}\;,
\label{1LoopC4-2}
\end{eqnarray}
where the concrete expression of $Q_{_{j,n_{_1},n_{_2}}}^{(a)}$ is presented in Eq.(\ref{app2-1}).
Additionally the function $x_{_m}^{D/2-3}F_{_{a,m}}$
complies with the system of linear PDEs in Eq~(\ref{1M-PDEs}).
Or equivalently, the function $F_{_{a,m}}$ satisfies the following
system of linear PDEs
\begin{eqnarray}
&&\Big\{\Big(3-{D\over2}+\sum\limits_{i=1}^3\hat{\theta}_{\eta_{_{i3}}}\Big)
(1+\hat{\theta}_{\eta_{_{33}}}+\hat{\theta}_{\eta_{_{13}}})(1+\hat{\theta}_{\eta_{_{33}}}+\hat{\theta}_{\eta_{_{23}}})
\nonumber\\
&&\hspace{0.0cm}
-{1\over\eta_{_{33}}}\hat{\theta}_{\eta_{_{33}}}\Big({D\over2}-1
+\sum\limits_{i=1}^3\hat{\theta}_{\eta_{_{i3}}}\Big)\Big(2-{D\over2}+\hat{\theta}_{\eta_{_{33}}}\Big)
\Big\}F_{_{a,m}}(\eta_{_{13}},\;\eta_{_{13}},\;\eta_{_{23}})=0
\;,\nonumber\\
&&\Big\{\Big(3-{D\over2}+\sum\limits_{i=1}^3\hat{\theta}_{\eta_{_{i3}}}\Big)
(1+\hat{\theta}_{\eta_{_{33}}}+\hat{\theta}_{\eta_{_{j3}}})
\nonumber\\
&&\hspace{0.0cm}
+{1\over\eta_{_{j3}}}\hat{\theta}_{\eta_{_{j3}}}\Big({D\over2}-1
+\sum\limits_{i=1}^3\hat{\theta}_{\eta_{_{i3}}}\Big)
\Big\}F_{_{a,m}}(\eta_{_{13}},\;\eta_{_{13}},\;\eta_{_{23}})=0,
\;(j=1,\;2).
\label{1LoopC4-4}
\end{eqnarray}

Obviously the absolutely and uniformly convergent region $\Omega_{_{18}}^a$
is a proper subset of the cube
\begin{eqnarray}
&&\Xi_{_{\rm m}}^a=\left\{(\eta_{_{13}},\;\eta_{_{13}},\;\eta_{_{23}}):\left|\begin{array}{l}
|\eta_{_{13}}|<1,\;|\eta_{_{23}}|<1,\;|\eta_{_{33}}|<1
\end{array}\right.\right\}\;.
\label{1LoopC4-3}
\end{eqnarray}
In order to continue the scalar integral $C_{_0}^a$ from the region $\Omega_{_{18}}^a$
to the whole cube $\Xi_{_{\rm m}}^a$, one employs the system of PDEs in Eq.~(\ref{1LoopC4-4}).

\end{itemize}

\section{The $C_{_0}$ function with three equal masses\label{sec5}}
\indent\indent
In this case, the Mellin-Barnes representation of the scalar integral is
\begin{eqnarray}
&&C_{_0}^b(p_{_1}^2,\;p_{_2}^2,\;p_{_3}^2,\;m^2)=
\int{d^Dq\over(2\pi)^D}{1\over[q^2-m^2][(q+p_{_1})^2-m^2]
[(q-p_{_2})^2-m^2]}
\nonumber\\
&&\hspace{3.5cm}=
-{i(-)^{D/2-3}\over(4\pi)^{D/2}(2\pi i)^3}
\int_{-i\infty}^{+i\infty}ds(-m^2)^{s}\Gamma(-s)
\nonumber\\
&&\hspace{4.0cm}\times
\int_{-i\infty}^{+i\infty}dz_{_1}(p_{_1}^2)^{z_{_1}}\Gamma(-z_{_1})
\int_{-i\infty}^{+i\infty}dz_{_2}(p_{_2}^2)^{z_{_2}}\Gamma(-z_{_2})
\nonumber\\
&&\hspace{4.0cm}\times
(p_{_3}^2)^{D/2-3-s-z_{_1}-z_{_2}}\Gamma(3-{D\over2}+s+z_{_1}+z_{_2})
\nonumber\\
&&\hspace{4.0cm}\times
{\Gamma({D\over2}-2-s-z_{_2})\Gamma({D\over2}-2-s-z_{_1})
\Gamma(1+z_{_1}+z_{_2})\over\Gamma(D-3-2s)}\;,
\label{3M}
\end{eqnarray}
which is equivalent to the representation of Eq.~(36) of Ref.~\cite{Davydychev3},
and that of Eq.~(4.5) of Ref.~\cite{Davydychev1991JMP}.
Applying the residue theorem to the  singularities
of $\Gamma$ functions in the numerator of integrand, we formulate the Mellin-Barnes representation
as various hypergeometric functions whose absolutely and uniformly
convergent regions are analyzed through Horn's study of convergence~\cite{Horn1889}.
As the scalar integral is given by
\begin{eqnarray}
&&C_{_{0}}^b(p_{_1}^2,\;p_{_2}^2,\;p_{_3}^2,\;m^2)=
{i(-)^{D/2}(p_{_3}^2)^{D/2-3}\over(4\pi)^{D/2}}
F_{_{b,k}}(\xi_{_{m3}},\;x_{_{13}},\;x_{_{23}})\;,
\label{3Ma}
\end{eqnarray}
with $\xi_{_{mj}}=\xi_{_{ij}},\;(i,\;j=1,\;2,\;3)$. In addition,
the dimensionless function $F_{_{b,k}}$ satisfies the
system of linear PDEs
\begin{eqnarray}
&&\Big\{(3-{D\over2}+\hat{\vartheta}_{\xi_{_{m3}}}+\sum\limits_{i=1}^2\hat{\vartheta}_{x_{_{i3}}})
(5-D+2\hat{\vartheta}_{\xi_{_{m3}}})(4-D+2\hat{\vartheta}_{\xi_{_{m3}}})
\nonumber\\
&&\hspace{0.0cm}
-{1\over\xi_{_{m3}}}\hat{\vartheta}_{\xi_{_{m3}}}\prod\limits_{i=1}^2
(2-{D\over2}+\hat{\vartheta}_{\xi_{_{m3}}}+\hat{\vartheta}_{x_{_{i3}}})\Big\}F_{_{b,p_{_3}}}=0
\;,\nonumber\\
&&\Big\{(3-{D\over2}+\hat{\vartheta}_{\xi_{_{m3}}}+\sum\limits_{i=1}^2\hat{\vartheta}_{x_{_{i3}}})
(1+\sum\limits_{i=1}^2\hat{\vartheta}_{x_{_{i3}}})
\nonumber\\
&&\hspace{0.0cm}
-{1\over x_{_{j3}}}\hat{\vartheta}_{x_{_{j3}}}(2-{D\over2}+\hat{\vartheta}_{\xi_{_{m3}}}
+\hat{\vartheta}_{x_{_{j3}}})\Big\}F_{_{b,p_{_3}}}=0\;,\;(j=1,\;2)\;.
\label{3M-PDEs}
\end{eqnarray}
Now we present our results below in detail.

\subsection{The hypergeometric functions and their convergent regions\label{sec2-1}}
\begin{itemize}
\item[$\bullet$ 1: ] For the isolated singularities defined by the poles of
$\Gamma(-s)$, $\Gamma(-z_{_1})$, and $\Gamma(-z_{_2})$, the
corresponding hypergeometric series is written as:
\begin{eqnarray}
&&C_{_{0,1}}^b(p_{_1}^2,\;p_{_2}^2,\;p_{_3}^2,\;m^2)=
-{i(-)^{D/2-3}(p_{_3}^2)^{D/2-3}\over(4\pi)^{D/2}}
{\Gamma^2(2-{D\over2})\Gamma^2({D\over2}-1)\over\Gamma(D-3)}
\nonumber\\
&&\hspace{4.0cm}\times
\sum\limits_{n_{_1}=0}^\infty\sum\limits_{n_{_2}=0}^\infty
x_{_{13}}^{n_{_1}}x_{_{23}}^{n_{_2}}
{\Gamma(3-{D\over2}+n_{_1}+n_{_2})\Gamma(1+n_{_1}+n_{_2})
\over n_{_1}!n_{_2}!\Gamma(3-{D\over2}+n_{_2})\Gamma(3-{D\over2}+n_{_1})}
\nonumber\\
&&\hspace{4.0cm}
-{i(-)^{D/2-3}(p_{_3}^2)^{D/2-3}\over(4\pi)^{D/2}}
{\Gamma^2(2-{D\over2})\Gamma^2({D\over2}-1)\over\Gamma(D-3)\Gamma(4-D)}
\nonumber\\
&&\hspace{4.0cm}\times
\sum\limits_{j=1}^\infty\sum\limits_{n_{_1}=0}^\infty\sum\limits_{n_{_2}=0}^\infty
\xi_{_{m3}}^jx_{_{13}}^{n_{_1}}x_{_{23}}^{n_{_2}}\Gamma(4-D+2j)
\nonumber\\
&&\hspace{4.0cm}\times
{\Gamma(3-{D\over2}+j+n_{_1}+n_{_2})\Gamma(1+n_{_1}+n_{_2})
\over j!n_{_1}!n_{_2}!\Gamma(3-{D\over2}+j+n_{_2})\Gamma(3-{D\over2}+j+n_{_1})}\;.
\label{3M1-1}
\end{eqnarray}
Here the condition $|p_{_3}^2|>\max(m^2,\;|p_{_1}^2|,\;|p_{_2}^2|)$
should be satisfied to guarantee the convergence of hypergeometric function.
In addition, $D=4$ is the second order pole of the first double hypergeometric
function, and the first order pole of the second triple hypergeometric
function, respectively. Certainly the expression
of Eq.~(\ref{3M1-1}) complies with the system of linear PDEs in Eq.~(\ref{3M-PDEs}).

The absolutely and uniformly convergent region of the hypergeometric function is
\begin{eqnarray}
&&\Omega_{_{1}}^b=\left\{(\xi_{_{m3}},\;x_{_{13}},\;x_{_{23}}):\left|\begin{array}{l}
|\xi_{_{m3}}|<1/4,\\
\sqrt{|x_{_{13}}|}+\sqrt{|x_{_{23}}|}<1
\end{array}\right.\right\}\;.
\label{3M1-5}
\end{eqnarray}

\item[$\bullet$ 2: ] When the isolated singularities are defined by the poles of
$\Gamma(-s)$, $\Gamma(-z_{_1})$, and $\Gamma({D\over2}-2-s-z_{_2})$,
the derived hypergeometric series $C_{_{0,2}}^b$ is given by Eq.~(\ref{3M2-1}), where
the necessary condition $|p_{_3}^2|>\max(m^2,\;|p_{_1}^2|,\;|p_{_2}^2|)$, $|p_{_2}^2|>m^2$
should be met to guarantee the hypergeometric series converging.
By implementing the residue theorem to the Mellin-Barnes representation, one derives the first expression.
The second expression decomposes the summation indices of the hypergeometric series
in the original expression, so $D=4$ is the pole of the same fixed order
for each term of every hypergeometric series of Eq.~(\ref{3M2-1}).
Furthermore, the decomposition of the summation indices is correct and reasonable
because the second expression also satisfies the system of linear PDEs
in Eq.~(\ref{3M-PDEs}).
Horn's study of convergence predicts the absolutely and uniformly convergent region of the
double hypergeometric functions as
\begin{eqnarray}
&&\Omega_{_{2}}^b=\left\{(\xi_{_{m3}},\;x_{_{13}},\;x_{_{23}}):\left|\begin{array}{l}
|\xi_{_{m3}}x_{_{13}}|<|x_{_{23}}|,\;|\xi_{_{m3}}|<|x_{_{23}}|/4,\\
\sqrt{|x_{_{13}}|}+\sqrt{|x_{_{23}}|}<1
\end{array}\right.\right\}\;.
\label{3M2-2}
\end{eqnarray}

\item[$\bullet$ 3: ] For the isolated singularities defined by the poles of
$\Gamma(-s)$, $\Gamma(-z_{_1})$, and $\Gamma(1+z_{_1}+z_{_2})$,
the corresponding hypergeometric series $C_{_{0,3}}^b$ is
\begin{eqnarray}
&&C_{_{0,3}}^b(p_{_1}^2,\;p_{_2}^2,\;p_{_3}^2,\;m^2)=C_{_{0,2}}^b(
p_{_3}\rightarrow p_{_2},\;\xi_{_{m3}}\rightarrow\xi_{_{m2}},\;
x_{_{13}}\rightarrow x_{_{12}},\;x_{_{23}}\rightarrow x_{_{32}})\;,
\label{3M3-1}
\end{eqnarray}
where the condition $|p_{_2}^2|>\max(m^2,\;|p_{_1}^2|,\;|p_{_3}^2|)$, $|p_{_3}^2|>m^2$
is necessary to guarantee the convergence of the hypergeometric functions.
The absolutely and uniformly convergent region of Eq.(\ref{3M3-1}) is similarly given as
\begin{eqnarray}
&&\Omega_{_{3}}^b=\left\{(\xi_{_{m2}},\;x_{_{12}},\;x_{_{32}}):\left|\begin{array}{l}
|\xi_{_{m2}}x_{_{12}}|<|x_{_{32}}|,\;|\xi_{_{m2}}|<|x_{_{32}}|/4,\\
\sqrt{|x_{_{12}}|}+\sqrt{|x_{_{32}}|}<1
\end{array}\right.\right\}\;.
\label{3M3-2}
\end{eqnarray}

\item[$\bullet$ 4: ] For the isolated singularities defined by the poles of
$\Gamma(-s)$, $\Gamma(-z_{_1})$, and $\Gamma(3-{D\over2}+s+z_{_1}+z_{_2})$,
the corresponding hypergeometric series is
\begin{eqnarray}
&&C_{_{0,4}}^b(p_{_1}^2,\;p_{_2}^2,\;p_{_3}^2,\;m^2)=
C_{_{0,1}}^b(p_{_3}\rightarrow p_{_2},\;\xi_{_{m3}}\rightarrow\xi_{_{m2}},\;
x_{_{13}}\rightarrow x_{_{12}},\;x_{_{23}}\rightarrow x_{_{32}})\;.
\label{3M4-1}
\end{eqnarray}
The condition $|p_{_2}^2|>\max(m^2,\;|p_{_1}^2|,\;|p_{_3}^2|)$
is necessary to guarantee the hypergeometric functions converging.
In addition, the absolutely and uniformly convergent region of the term is
\begin{eqnarray}
&&\Omega_{_{4}}^b=\left\{(\xi_{_{m2}},\;x_{_{12}},\;x_{_{32}}):\left|\begin{array}{l}
|\xi_{_{m2}}|<1/4,\\
\sqrt{|x_{_{12}}|}+\sqrt{|x_{_{32}}|}<1
\end{array}\right.\right\}\;.
\label{3M4-2}
\end{eqnarray}

\item[$\bullet$ 6: ] When the isolated singularities are produced by the poles of
$\Gamma(-s)$, $\Gamma(-z_{_2})$, and $\Gamma({D\over2}-2-s-z_{_1})$,
the corresponding hypergeometric series $C_{_{0,6}}^b$ is presented in Eq.~(\ref{3M6-1}), where
the condition $|p_{_3}^2|>\max(m^2,\;|p_{_1}^2|,\;|p_{_2}^2|)$, $|p_{_1}^2|>m^2$
should be satisfied to guarantee the convergence of the hypergeometric functions.
Applying the residue theorem to the Mellin-Barnes representation, one derives the first expression of Eq.~(\ref{3M6-1}).
The second expression decomposes the summation indices of the hypergeometric series
in the original expression, so $D=4$ is the pole of the same fixed order
for each term of every hypergeometric series of the second equation in Eq.~(\ref{3M6-1}).
In addition, the decomposition of the summation indices is correct and reasonable
because the second expression also complies with the system of linear PDEs
in Eq.~(\ref{3M-PDEs}).

The concretely convergent region of the hypergeometric functions is given by
\begin{eqnarray}
&&\Omega_{_{6}}^b=\left\{(\xi_{_{m3}},\;x_{_{13}},\;x_{_{23}}):\left|\begin{array}{l}
|\xi_{_{m3}}x_{_{23}}|<|x_{_{13}}|,\;|\xi_{_{m3}}|<|x_{_{13}}|/4,\\
\sqrt{|x_{_{13}}|}+\sqrt{|x_{_{23}}|}<1
\end{array}\right.\right\}\;.
\label{3M6-2}
\end{eqnarray}

\item[$\bullet$ 7: ] For the isolated singularities defined by the poles of
$\Gamma(-s)$, $\Gamma({D\over2}-2-s-z_{_1})$, and $\Gamma({D\over2}-2-s-z_{_2})$,
the corresponding hypergeometric series $C_{_{0,7}}^b$ is written in Eq.~(\ref{3M7-1}), where
the condition $|p_{_3}^2|>\max(|p_{_1}^2|,\;|p_{_2}^2|),\;
\min(|p_{_1}^2|,\;|p_{_2}^2|)>m^2$,
$|p_{_1}^2p_{_2}^2|>|p_{_3}^2|m^2$
is essential to guarantee the convergence of the hypergeometric functions.
Applying the residue theorem to the Mellin-Barnes representation, one derives the first expression of Eq.~(\ref{3M7-1}).
The second expression decomposes the summation indices of the hypergeometric series
in the original expression, so $D=4$ is the pole of the same fixed order
for each term of every hypergeometric series of the second equation in Eq.~(\ref{3M7-1}).
In addition, the decomposition of the summation indices is correct and reasonable
because the subsequent expression also satisfies the system of linear PDEs
in Eq.~(\ref{3M-PDEs}).

Correspondingly the absolutely and uniformly convergent region of
this hypergeometric function is concretely written as
\begin{eqnarray}
&&\Omega_{_{7}}^b=\left\{(\xi_{_{m3}},\;x_{_{13}},\;x_{_{23}}):\left|\begin{array}{l}
|\xi_{_{m3}}x_{_{13}}|<|x_{_{23}}|,\;|\xi_{_{m3}}x_{_{23}}|<|x_{_{13}}|,\\
|\xi_{_{m3}}|<|x_{_{13}}x_{_{23}}|,\;|\xi_{_{m3}}|<|x_{_{13}}|/4,\\
|\xi_{_{m3}}|<|x_{_{23}}|/4,\;\sqrt{|x_{_{13}}|}+\sqrt{|x_{_{23}}|}<1
\end{array}\right.\right\}\;.
\label{3M7-2}
\end{eqnarray}

\item[$\bullet$ 8: ] For the isolated singularities defined by the poles of
$\Gamma(-s)$, $\Gamma({D\over2}-2-s-z_{_1})$, and $\Gamma(1+z_{_1}+z_{_2})$,
the corresponding hypergeometric series $C_{_{0,8}}^b$ is presented as
\begin{eqnarray}
&&C_{_{0,8}}^b(p_{_1}^2,\;p_{_2}^2,\;p_{_3}^2,\;m^2)=C_{_{0,7}}^b(
p_{_3}\rightarrow p_{_2},\;\xi_{_{m3}}\rightarrow\xi_{_{m2}},\;
x_{_{13}}\rightarrow x_{_{12}},\;x_{_{23}}\rightarrow x_{_{32}})\;.
\label{3M8-1}
\end{eqnarray}
Applying Horn's study of convergence, we write the absolutely and uniformly convergent
region of the hypergeometric functions in Eq.(\ref{3M8-1}) as
\begin{eqnarray}
&&\Omega_{_{8}}^b=\left\{(\xi_{_{m2}},\;x_{_{12}},\;x_{_{32}}):\left|\begin{array}{l}
|\xi_{_{m2}}|<|x_{_{12}}x_{_{32}}|,\;|\xi_{_{m2}}x_{_{32}}|<|x_{_{12}}|,\\
|\xi_{_{m2}}x_{_{12}}|<|x_{_{32}}|,\;,|\xi_{_{m2}}|<|x_{_{12}}|/4,\\
|\xi_{_{m2}}|<|x_{_{32}}|/4,\;\sqrt{|x_{_{12}}|}+\sqrt{|x_{_{32}}|}<1
\end{array}\right.\right\}\;.
\label{3M8-2}
\end{eqnarray}

\item[$\bullet$ 9: ] For the isolated singularities defined by the poles of
$\Gamma(-s)$, $\Gamma({D\over2}-2-s-z_{_1})$, and $\Gamma(3-{D\over2}+s+z_{_1}+z_{_2})$,
the corresponding hypergeometric series $C_{_{0,9}}^b$ is
\begin{eqnarray}
&&C_{_{0,9}}^b(p_{_1}^2,\;p_{_2}^2,\;p_{_3}^2,\;m^2)=
C_{_{0,6}}^b(p_{_3}\rightarrow p_{_2},\;\xi_{_{m3}}\rightarrow\xi_{_{m2}},\;
x_{_{13}}\rightarrow x_{_{12}},\;x_{_{23}}\rightarrow x_{_{32}})\;.
\label{3M9-1}
\end{eqnarray}
The convergent region of those hypergeometric functions is concretely written as
\begin{eqnarray}
&&\Omega_{_{9}}^b=\left\{(\xi_{_{m2}},\;x_{_{12}},\;x_{_{32}}):\left|\begin{array}{l}
|\xi_{_{m2}}x_{_{32}}|<|x_{_{12}}|,\;|\xi_{_{m2}}|<|x_{_{12}}|/4,\\
\sqrt{|x_{_{12}}|}+\sqrt{|x_{_{32}}|}<1
\end{array}\right.\right\}\;.
\label{3M9-2}
\end{eqnarray}

\item[$\bullet$ 10: ] For the isolated singularities defined by the poles of
$\Gamma(-s)$, $\Gamma(1+z_{_1}+z_{_2})$, and $\Gamma(-z_{_2})$,
correspondingly the hypergeometric function $C_{_{0,10}}^b$ is written as
\begin{eqnarray}
&&C_{_{0,10}}^b(p_{_1}^2,\;p_{_2}^2,\;p_{_3}^2,\;m^2)=C_{_{0,6}}^b(
p_{_3}\rightarrow p_{_1},\;\xi_{_{m3}}\rightarrow\xi_{_{m1}},\;
x_{_{13}}\rightarrow x_{_{31}},\;x_{_{23}}\rightarrow x_{_{21}})\;.
\label{3M10-1}
\end{eqnarray}
where the necessary condition $|p_{_1}^2|>\max(m^2,\;|p_{_2}^2|,\;|p_{_3}^2|)$, $|p_{_3}^2|>m^2$
should be satisfied to guarantee the convergence of the hypergeometric functions.
Additionally the expression satisfies the system of linear PDEs
in Eq.~(\ref{3M-PDEs}) also.

Through Horn's study of convergence, the convergent region of the series is similarly given by
\begin{eqnarray}
&&\Omega_{_{10}}^b=\left\{(\xi_{_{m1}},\;x_{_{21}},\;x_{_{31}}):\left|\begin{array}{l}
|\xi_{_{m1}}x_{_{21}}|<|x_{_{31}}|,\;|\xi_{_{m1}}|<|x_{_{31}}|/4,\\
\sqrt{|x_{_{21}}|}+\sqrt{|x_{_{31}}|}<1
\end{array}\right.\right\}\;.
\label{3M10-2}
\end{eqnarray}

\item[$\bullet$ 11: ] For the isolated singularities defined by the poles of
$\Gamma(-s)$, $\Gamma(1+z_{_1}+z_{_2})$, and $\Gamma({D\over2}-2-s-z_{_2})$,
the hypergeometric function $C_{_{0,11}}^b$ is
\begin{eqnarray}
&&C_{_{0,11}}^b(p_{_1}^2,\;p_{_2}^2,\;p_{_3}^2,\;m^2)=C_{_{0,7}}^b(
p_{_3}\rightarrow p_{_1},\;\xi_{_{m3}}\rightarrow\xi_{_{m1}},\;
x_{_{13}}\rightarrow x_{_{31}},\;x_{_{23}}\rightarrow x_{_{21}})\;.
\label{3M11-1}
\end{eqnarray}
Here the necessary condition $|p_{_1}^2|>\max(|p_{_2}^2|,\;|p_{_3}^2|)$,
$|p_{_2}^2p_{_3}^2|>|p_{_1}^2|m^2$
should be met to guarantee the hypergeometric functions converging.
Additionally the expression complies with the system of linear PDEs
in Eq.~(\ref{3M-PDEs}) also.

Using Horn's study of convergence, one formulates the absolutely and uniformly convergent region
of Eq.(\ref{3M11-1}) as
\begin{eqnarray}
&&\Omega_{_{11}}^b=\left\{(\xi_{_{m1}},\;x_{_{21}},\;x_{_{31}}):\left|\begin{array}{l}
|\xi_{_{m1}}|<|x_{_{21}}x_{_{31}}|,\;|\xi_{_{m1}}x_{_{31}}|<|x_{_{21}}|,\\
|\xi_{_{m1}}x_{_{21}}|<|x_{_{31}}|,\;|\xi_{_{m1}}|<|x_{_{21}}|/4,\\
|\xi_{_{m1}}|<|x_{_{31}}|/4,\;\sqrt{|x_{_{21}}|}+\sqrt{|x_{_{31}}|}<1
\end{array}\right.\right\}\;.
\label{3M11-2}
\end{eqnarray}

\item[$\bullet$ 13: ] For the isolated singularities defined by the poles of
$\Gamma(-s)$, $\Gamma(-z_{_2})$, and $\Gamma(3-{D\over2}+s+z_{_1}+z_{_2})$,
the corresponding hypergeometric series is given as
\begin{eqnarray}
&&C_{_{0,13}}^b(p_{_1}^2,\;p_{_2}^2,\;p_{_3}^2,\;m^2)=
C_{_{0,1}}^b(p_{_3}\rightarrow p_{_1},\;\xi_{_{m3}}\rightarrow\xi_{_{m1}},\;
x_{_{13}}\rightarrow x_{_{31}},\;x_{_{23}}\rightarrow x_{_{21}})\;.
\label{3M13-1}
\end{eqnarray}
Similarly the convergent region of the term is
\begin{eqnarray}
&&\Omega_{_{13}}^b=\left\{(\xi_{_{m1}},\;x_{_{21}},\;x_{_{31}}):\left|\begin{array}{l}
|\xi_{_{m1}}|<1/4,\sqrt{|x_{_{21}}|}+\sqrt{|x_{_{31}}|}<1
\end{array}\right.\right\}\;.
\label{3M13-2}
\end{eqnarray}

\item[$\bullet$ 14: ] When the isolated singularities are produced by the poles of
$\Gamma(-s)$, $\Gamma({D\over2}-2-s-z_{_2})$, and $\Gamma(3-{D\over2}+s+z_{_1}+z_{_2})$,
the corresponding hypergeometric series $C_{_{0,14}}^b$ is presented by
\begin{eqnarray}
&&C_{_{0,14}}^b(p_{_1}^2,\;p_{_2}^2,\;p_{_3}^2,\;m^2)=C_{_{0,2}}^b(
p_{_3}\rightarrow p_{_1},\;\xi_{_{m3}}\rightarrow\xi_{_{m1}},\;
x_{_{13}}\rightarrow x_{_{31}},\;x_{_{23}}\rightarrow x_{_{21}})\;,
\label{3M14-1}
\end{eqnarray}
where the necessary condition $|p_{_1}^2|>\max(|p_{_2}^2|,\;|p_{_3}^2|)$, $|p_{_2}^2|>m^2$
should be satisfied to guarantee the convergence of the hypergeometric functions.
Similarly the absolutely and uniformly convergent region of the series is
concretely written as
\begin{eqnarray}
&&\Omega_{_{14}}^b=\left\{(\xi_{_{m1}},\;x_{_{21}},\;x_{_{31}}):\left|\begin{array}{l}
|\xi_{_{m1}}x_{_{31}}|<|x_{_{21}}|,\;|\xi_{_{m1}}|<|x_{_{21}}|/4,\\
\sqrt{|x_{_{21}}|}+\sqrt{|x_{_{31}}|}<1
\end{array}\right.\right\}\;.
\label{3M14-2}
\end{eqnarray}

\item[$\bullet$ 16: ] For the singularities defined by the poles of $\Gamma(-z_{_1})$,
$\Gamma(-z_{_2})$, and $\Gamma({D\over2}-2-s-z_{_1})$, the
scalar integral contains some unknown linear combination parameters
originating from the non-isolated singularities.
The necessary condition $|p_{_3}^2|>\max(|p_{_1}^2|,\;|p_{_2}^2|,\;m^2)$, $m^2>|p_{_1}^2|$
should be satisfied to guarantee the hypergeometric series converging.

\item[$\bullet$ 17: ] For the singularities defined by the poles of $\Gamma(-z_{_1})$, $\Gamma(-z_{_2})$,
and $\Gamma({D\over2}-2-s-z_{_2})$,  the scalar integral contains some unknown linear combination parameters
originating from the non-isolated singularities.
The condition $|p_{_3}^2|>\max(|p_{_1}^2|,\;|p_{_2}^2|,\;m^2)$, $m^2>|p_{_2}^2|$
should be satisfied to guarantee the convergence of the hypergeometric series.

\item[$\bullet$ 19: ] For the isolated singularities defined by the poles of
$\Gamma(-z_{_1})$, $\Gamma(-z_{_2})$, and $\Gamma(3-{D\over2}+s+z_{_1}+z_{_2})$,
the corresponding hypergeometric series is given as
\begin{eqnarray}
&&C_{_{0,19}}^b(p_{_1}^2,\;p_{_2}^2,\;p_{_3}^2,\;m^2)=
-{i(m^2)^{D/2-3}\over(4\pi)^{D/2}}
\sum\limits_{j=0}^\infty\sum\limits_{n_{_1}=0}^\infty\sum\limits_{n_{_2}=0}^\infty
\eta_{_{3m}}^{j}\eta_{_{1m}}^{n_{_1}}\eta_{_{2m}}^{n_{_2}}
\nonumber\\
&&\hspace{4.0cm}\times
{\Gamma(1+j+n_{_1})\Gamma(1+j+n_{_2})
\Gamma(1+n_{_1}+n_{_2})\over j!n_{_1}!n_{_2}!}
\nonumber\\
&&\hspace{4.0cm}\times
{\Gamma(3-{D\over2}+j+n_{_1}+n_{_2})\over\Gamma(3+2j+2n_{_1}+2n_{_2})}\;.
\label{3M19-1}
\end{eqnarray}
The necessary condition $4m^2>\max(|p_{_1}^2|,\;|p_{_2}^2|,\;|p_{_3}^2|)$
should be met to guarantee the convergence of the triple series.
In addition, $D=4$ is the first order pole of the hypergeometric
function. Certainly the expression
of Eq.~(\ref{3M19-1}) complies with the system of linear PDEs in Eq.~(\ref{3M-PDEs}).

Correspondingly the convergent region of the term is
\begin{eqnarray}
&&\Omega_{_{19}}^b=\left\{(\eta_{_{1m}},\;\eta_{_{2m}},\;\eta_{_{3m}}):\left|\begin{array}{l}
|\eta_{_{1m}}|<4,\;|\eta_{_{2m}}|<4,\;|\eta_{_{3m}}|<4
\end{array}\right.\right\}\;.
\label{3M19-2}
\end{eqnarray}

\item[$\bullet$ 20: ]  For the singularities defined by the poles of
 $\Gamma(-z_{_1})$, $\Gamma({D\over2}-2-s-z_{_2})$,
and $\Gamma({D\over2}-2-s-z_{_1})$, the
scalar integral contains some unknown linear combination parameters
originating from the non-isolated singularities.
The condition $|p_{_3}^2|>\max(|p_{_1}^2|,\;|p_{_2}^2|,\;m^2)$,
$|p_{_2}^2|>m^2$, $|p_{_3}^2|m^2>|p_{_1}^2p_{_2}^2|$
should be satisfied to guarantee the convergence of the hypergeometric series.

\item[$\bullet$ 21: ] When the isolated singularities are produced by the poles of
$\Gamma(-z_{_1})$, $\Gamma({D\over2}-2-s-z_{_2})$, and $\Gamma(1+z_{_1}+z_{_2})$,
the corresponding hypergeometric series is given as
\begin{eqnarray}
&&C_{_{0,21}}^b(p_{_1}^2,\;p_{_2}^2,\;p_{_3}^2,\;m^2)=
{i(p_{_3}^2)^{D/2-3}\over(4\pi)^{D/2}}
\Big\{\Gamma(2-{D\over2})\Gamma({D\over2}-1)\Big\}{\xi_{_{m3}}^{D/2-1}\over x_{_{23}}}
\nonumber\\
&&\hspace{4.0cm}\times
\sum\limits_{j=0}^\infty\sum\limits_{n_{_1}=0}^\infty\sum\limits_{n_{_2}=0}^\infty
\xi_{_{m3}}^{j+n_{_1}+n_{_2}}x_{_{13}}^{n_{_1}}x_{_{23}}^{-n_{_1}-n_{_2}}\Gamma(1+j+n_{_1})
\nonumber\\
&&\hspace{4.0cm}\times
{(-)^{n_{_1}}\Gamma(1+n_{_1}+n_{_2})
\Gamma(2+2j+2n_{_1}+2n_{_2})\over j!n_{_1}!n_{_2}!
\Gamma({D\over2}+j+n_{_1}+n_{_2})\Gamma(2+j+2n_{_1}+n_{_2})}
\label{3M21-1}
\end{eqnarray}
The condition $\min(|p_{_2}^2|,\;|p_{_3}^2|)>m^2$,
$|p_{_2}^2p_{_3}^2|>|p_{_1}^2|m^2$
is necessary to guarantee the hypergeometric series converging.
In addition, $D=4$ is the first order pole of the hypergeometric
function. Certainly the expression
of Eq.~(\ref{3M21-1}) satisfies the system of linear PDEs in Eq.~(\ref{3M-PDEs}).

Correspondingly the convergent region of the series is
\begin{eqnarray}
&&\Omega_{_{21}}^b=\left\{(\xi_{_{m3}},\;x_{_{13}},\;x_{_{23}}):\left|\begin{array}{l}
|\xi_{_{m3}}|<1/4,\;|\xi_{_{m3}}x_{_{13}}|<|x_{_{23}}|,\;|\xi_{_{m3}}|<|x_{_{23}}|/4
\end{array}\right.\right\}\;.
\label{3M21-2}
\end{eqnarray}

\item[$\bullet$ 23: ]  For the singularities defined by the poles of
$\Gamma(-z_{_1})$, $\Gamma({D\over2}-2-s-z_{_1})$, and $\Gamma(1+z_{_1}+z_{_2})$, the
scalar integral contains some unknown linear combination parameters
originating from the non-isolated singularities.
The condition $|p_{_3}^2|>\max(|p_{_2}^2|,\;m^2)$,
$|p_{_3}^2|^2>|p_{_1}^2|m^2$, $|p_{_3}^2|m^2>|p_{_1}^2p_{_2}^2|$
should be satisfied to guarantee the convergence of the hypergeometric series.

\item[$\bullet$ 24: ] For the singularities defined by the poles of $\Gamma(-z_{_1})$,
$\Gamma({D\over2}-2-s-z_{_1})$, and $\Gamma(3-{D\over2}+s+z_{_1}+z_{_2})$,
the scalar integral contains some unknown linear combination parameters
originating from the non-isolated singularities.
The condition $|p_{_2}^2|>\max(|p_{_1}^2|,\;|p_{_3}^2|,\;m^2)$,
$m^2>|p_{_1}^2|$ should be satisfied to guarantee the convergence of the hypergeometric series.

\item[$\bullet$ 25: ] For the singularities defined by the poles of $\Gamma(-z_{_1})$,
$\Gamma(1+z_{_1}+z_{_2})$, and $\Gamma(3-{D\over2}+s+z_{_1}+z_{_2})$,
the scalar integral contains some unknown linear combination parameters
originating from the non-isolated singularities.
The necessary condition $|p_{_2}^2|>\max(|p_{_1}^2|,\;|p_{_3}^2|,\;m^2)$,
$m^2>|p_{_3}^2|$ should be satisfied to guarantee the convergence of the hypergeometric series.

\item[$\bullet$ 26: ]  For the singularities defined by the poles of
$\Gamma(-z_{_2})$, $\Gamma({D\over2}-2-s-z_{_1})$, and $\Gamma({D\over2}-2-s-z_{_2})$, the
scalar integral contains some unknown linear combination parameters
originating from the non-isolated singularities.
The condition $|p_{_3}^2|>\max(|p_{_1}^2|,\;|p_{_2}^2|,\;m^2)$,
$|p_{_1}^2|>m^2$, $|p_{_3}^2|m^2>|p_{_1}^2p_{_2}^2|$
should be satisfied to guarantee the convergence of the hypergeometric series.

\item[$\bullet$ 27: ] For the isolated singularities defined by the pole of
$\Gamma(-z_{_2})$, $\Gamma({D\over2}-2-s-z_{_1})$, and $\Gamma(1+z_{_1}+z_{_2})$,
the corresponding hypergeometric series is
\begin{eqnarray}
&&C_{_{0,27}}^b(p_{_1}^2,\;p_{_2}^2,\;p_{_3}^2,\;m^2)=
{i(p_{_3}^2)^{D/2-3}\over(4\pi)^{D/2}}
\Big\{\Gamma(2-{D\over2})\Gamma({D\over2}-1)\Big\}{\xi_{_{m3}}^{D/2-1}\over x_{_{13}}}
\nonumber\\
&&\hspace{4.0cm}\times
\sum\limits_{j=0}^\infty\sum\limits_{n_{_1}=0}^\infty\sum\limits_{n_{_2}=0}^\infty
\xi_{_{m3}}^{j+n_{_1}+n_{_2}}x_{_{23}}^{n_{_2}}x_{_{13}}^{-n_{_1}-n_{_2}}\Gamma(1+j+n_{_2})
\nonumber\\
&&\hspace{4.0cm}\times
{(-)^{n_{_2}}\Gamma(1+n_{_1}+n_{_2})
\Gamma(2+2j+2n_{_1}+2n_{_2})\over j!n_{_1}!n_{_2}!
\Gamma({D\over2}+j+n_{_1}+n_{_2})\Gamma(2+j+n_{_1}+2n_{_2})}\;.
\label{3M27-1}
\end{eqnarray}
The condition $\min(|p_{_1}^2|,\;|p_{_3}^2|)>m^2$, $|p_{_1}^2p_{_3}^2|>|p_{_2}^2|m^2$
should be satisfied to guarantee the convergence of the hypergeometric series.
In addition, $D=4$ is the first order pole of the hypergeometric
function. Certainly the expression
of Eq.~(\ref{3M27-1}) complies with the system of PDEs in Eq.~(\ref{3M-PDEs}).

Correspondingly the convergent region of the series is
\begin{eqnarray}
&&\Omega_{_{27}}^b=\left\{(\xi_{_{m3}},\;x_{_{13}},\;x_{_{23}}):\left|\begin{array}{l}
|\xi_{_{m3}}|<1/4,\;|\xi_{_{m3}}x_{_{23}}|<|x_{_{13}}|,\;|\xi_{_{m3}}|<|x_{_{13}}|/4
\end{array}\right.\right\}\;.
\label{3M27-2}
\end{eqnarray}

\item[$\bullet$ 29: ]  For the singularities defined by the poles of
$\Gamma(-z_{_2})$, $\Gamma({D\over2}-2-s-z_{_2})$, and $\Gamma(1+z_{_1}+z_{_2})$, the
scalar integral contains some unknown linear combination parameters
originating from the non-isolated singularities.
The essential condition $|p_{_3}^2|>\max(|p_{_1}^2|,\;m^2)$,
$|p_{_3}^2|^2>|p_{_2}^2|m^2$, $|p_{_3}^2|m^2>|p_{_1}^2p_{_2}^2|$
should be satisfied to guarantee the convergence of the hypergeometric series.

\item[$\bullet$ 30: ] For the singularities defined by the poles of $\Gamma(-z_{_2})$,
$\Gamma({D\over2}-2-s-z_{_2})$, and $\Gamma(3-{D\over2}+s+z_{_1}+z_{_2})$,
the scalar integral contains some unknown linear combination parameters
originating from the non-isolated singularities.
The condition $|p_{_1}^2|>\max(|p_{_2}^2|,\;|p_{_3}^2|,\;m^2)$,
$m^2>|p_{_2}^2|$ should be satisfied to guarantee the convergence of the hypergeometric series.

\item[$\bullet$ 31: ] For the singularities defined by the poles of $\Gamma(-z_{_2})$,
$\Gamma(1+z_{_1}+z_{_2})$, and $\Gamma(3-{D\over2}+s+z_{_1}+z_{_2})$,
the scalar integral contains some unknown linear combination parameters
originating from the non-isolated singularities.
The condition $|p_{_1}^2|>\max(|p_{_2}^2|,\;|p_{_3}^2|,\;m^2)$,
$m^2>|p_{_3}^2|$ should be satisfied to guarantee the convergence of the hypergeometric series.

\item[$\bullet$ 32: ] For the singularities defined by the poles of $\Gamma({D\over2}-2-s-z_{_1})$,
$\Gamma({D\over2}-2-s-z_{_2})$, and $\Gamma(1+z_{_1}+z_{_2})$,
the scalar integral contains some unknown linear combination parameters
originating from the non-isolated singularities.
The necessary condition $|p_{_3}^2|>\max(|p_{_1}^2|,\;|p_{_2}^2|,\;m^2)$,
$|p_{_1}^2p_{_3}^2|>|p_{_2}^2|m^2$, $|p_{_2}^2p_{_3}^2|>|p_{_1}^2|m^2$
$|p_{_3}^2|m^2>|p_{_1}^2p_{_2}^2|$ should be satisfied to guarantee
 the convergence of the hypergeometric series.

\item[$\bullet$ 33: ] When the isolated singularities are produced by the poles of
$\Gamma(D/2-2-s-z_{_1})$, $\Gamma(D/2-2-s-z_{_2})$, and $\Gamma(3-{D\over2}+s+z_{_1}+z_{_2})$,
the corresponding hypergeometric series is given as
\begin{eqnarray}
&&C_{_{0,33}}^b(p_{_1}^2,\;p_{_2}^2,\;p_{_3}^2,\;m^2)=
{i(p_{_3}^2)^{D/2-3}\over(4\pi)^{D/2}}\Big\{\Gamma(2-{D\over2})\Gamma({D\over2}-1)\Big\}
{\xi_{_{m3}}^{D/2-1}\over x_{_{13}}x_{_{23}}}
\nonumber\\
&&\hspace{4.0cm}\times
\sum\limits_{j=0}^\infty\sum\limits_{n_{_1}=0}^\infty\sum\limits_{n_{_2}=0}^\infty
\xi_{_{m3}}^{j+n_{_1}+n_{_2}}x_{_{13}}^{-j-n_{_2}}x_{_{23}}^{-j-n_{_1}}\Gamma(1+j+n_{_1})
\nonumber\\
&&\hspace{4.0cm}\times
{(-)^{j}\Gamma(1+j+n_{_2})
\Gamma(2+2j+2n_{_1}+2n_{_2})\over j!n_{_1}!n_{_2}!
\Gamma({D\over2}+j+n_{_1}+n_{_2})\Gamma(2+2j+n_{_1}+n_{_2})}\;.
\label{3M33-1}
\end{eqnarray}
The necessary condition $\min(|p_{_1}^2|,\;|p_{_2}^2|)>m^2$,
$|p_{_1}^2p_{_2}^2|>|p_{_3}^2|m^2$ should be met to guarantee
 the convergence of the hypergeometric series.
In addition, $D=4$ is the first order pole of the hypergeometric
function. Certainly the expression of Eq.~(\ref{3M33-1}) complies with
the system of linear PDEs in Eq.~(\ref{3M-PDEs}).

The convergent region is apparently written as
\begin{eqnarray}
&&\Omega_{_{33}}^b=\left\{(\xi_{_{m3}},\;x_{_{13}},\;x_{_{23}}):\left|\begin{array}{l}
|\xi_{_{m3}}|<|x_{_{13}}x_{_{23}}|,\;|\xi_{_{m3}}|<|x_{_{13}}|/4,\;|\xi_{_{m3}}|<|x_{_{23}}|/4
\end{array}\right.\right\}\;.
\label{3M33-2}
\end{eqnarray}

\item[$\bullet$ 34: ]  For the singularities defined by the poles of $\Gamma(D/2-2-s-z_{_1})$,
$\Gamma(1+z_{_1}+z_{_2})$, and $\Gamma(3-{D\over2}+s+z_{_1}+z_{_2})$,
the scalar integral contains some undefined linear combination parameters
originating from the non-isolated singularities.
The necessary condition $|p_{_2}^2|>\max(|p_{_1}^2|,\;|p_{_3}^2|,\;m^2)$,
$|p_{_1}^2p_{_2}^2|>|p_{_3}^2|m^2$, $|p_{_1}^2|>m^2$ should be satisfied to guarantee
 the convergence of the hypergeometric series.

\item[$\bullet$ 35: ]  For the singularities defined by the poles of $\Gamma(D/2-2-s-z_{_2})$,
$\Gamma(1+z_{_1}+z_{_2})$, and $\Gamma(3-{D\over2}+s+z_{_1}+z_{_2})$,
the scalar integral contains some undefined linear combination parameters
originating from the non-isolated singularities.
The necessary condition $|p_{_1}^2|>\max(|p_{_2}^2|,\;|p_{_3}^2|,\;m^2)$,
$|p_{_1}^2p_{_2}^2|>|p_{_3}^2|m^2$, $|p_{_2}^2|>m^2$ should be satisfied to guarantee
 the convergence of the hypergeometric series.
\end{itemize}

\subsection{The scalar integral in different convergent regions\label{sec2-2}}
\indent\indent
The absolutely and uniformly convergent region of
the hypergeometric function $C_{_{0,\alpha}}$ is the domain $\Omega_{_{\alpha}}^b$,
$\alpha=1,\cdots,\;4,\;6,\;\cdots,\;11,\;13,\;14,\;19,\;21,\;27,\;33$.
The scalar integral of other parameter space contains some unknown linear combination parameters originating from
the non-isolated singularities. Among those convergent regions, the following domains
compose a set
\begin{eqnarray}
&&\Big\{\Omega_{_{7}}^b,\;\Omega_{_{8}}^b,\;\Omega_{_{11}}^b,\;\Omega_{_{19}}^b\Big\}\;,
\label{3MSum-1}
\end{eqnarray}
whose element satisfies the following constraints simultaneously,
\begin{eqnarray}
&&\Omega_{_{i}}^b\bigcap\Omega_{_{j}}^b=\emptyset,\;i,\;j=7,\;8,\;11,\;19,\;i\neq j.
\label{3MSum-2}
\end{eqnarray}
Meanwhile, the relations between the elements of the set and those
convergent regions which do not belong to the set are
\begin{eqnarray}
&&\Omega_{_{7}}^b\subset\Omega_{_{\alpha_{_1}}}^b,\;\alpha_{_1}=1,\;2,\;6,\;7,\;21,\;27,\;33;\;
\nonumber\\
&&\hspace{0.0cm}
\Omega_{_{7}}^b\bigcap\Omega_{_{\beta_{_1}}}^b=\emptyset,\;\beta_{_1}=3,\;4,\;8,\;9,\;10,\;11,\;13,\;14,\;19;
\nonumber\\
&&\hspace{0.0cm}
\Omega_{_{8}}^b\subset\Omega_{_{\alpha_{_2}}}^b,\;\alpha_{_2}=3,\;4,\;8,\;9,\;21,\;27,\;33;\;
\nonumber\\
&&\hspace{0.0cm}
\Omega_{_{8}}^b\bigcap\Omega_{_{\beta_{_2}}}^b=\emptyset,\;\beta_{_2}=1,\;2,\;6,\;7,\;10,\;11,\;13,\;14,\;19;
\nonumber\\
&&\hspace{0.0cm}
\Omega_{_{11}}^b\subset\Omega_{_{\alpha_{_3}}}^b,\;\alpha_{_3}=10,\;11,\;13,\;14,\;21,\;27,\;33;\;
\nonumber\\
&&\hspace{0.0cm}
\Omega_{_{11}}^b\bigcap\Omega_{_{\beta_{_3}}}^b=\emptyset,\;\beta_{_3}=1,\;2,\;3,\;4,\;6,\;7,\;8,\;9,\;19;
\nonumber\\
&&\hspace{0.0cm}
\Omega_{_{19}}^b\bigcap\Omega_{_{\beta_{_4}}}^b=\emptyset,\;\beta_{_4}=1,\cdots,\;4,\;6,\cdots,\;11,\;
13,\;14,\;21,\;27,\;33.
\label{3MSum-3}
\end{eqnarray}
With the preparation above the concrete expression of $C_{_0}$ in different parameter space
is respectively presented as following.
\begin{itemize}
\item[$\bullet$ 1:] In the convergent region $\Omega_{_7}^b$,
the scalar integral $C_{_0}^b$ is written as
\begin{eqnarray}
&&C_{_0}^b(p_{_1}^2,\;p_{_2}^2,\;p_{_3}^2,\;m^2)=
\Big\{C_{_{0,1}}^b+C_{_{0,2}}^b+C_{_{0,6}}^b+C_{_{0,7}}^b+C_{_{0,21}}^b
\nonumber\\
&&\hspace{4.0cm}
+C_{_{0,27}}^b+C_{_{0,33}}^b\Big\}(p_{_1}^2,\;p_{_2}^2,\;p_{_3}^2,\;m^2)
\label{1LoopC3-1-1}
\end{eqnarray}
In the limit $m^2\rightarrow0$, the expression of Eq.(\ref{1LoopC3-1-1})
recovers that presented in Eq.(\ref{Massless-I1}).
Furthermore, $D=4$ is the pole of the second order for
each hypergeometric function individually in the above expression. Nevertheless the scalar integral presented in
Eq.(\ref{1LoopC3-1-1}) is an analytic function of the dimension in the
neighborhood of $D=4$ because those singularities are canceled clearly,
and the final result is given as
\begin{eqnarray}
&&C_{_0}^b(p_{_1}^2,\;p_{_2}^2,\;p_{_3}^2,\;m^2)=
{i(p_{_3}^2)^{D/2-3}\over(4\pi)^{D/2}}\Big\{f_{_{\varepsilon}}(x_{_{13}},\;x_{_{23}})
\nonumber\\
&&\hspace{4.0cm}
+2\varepsilon\sum\limits_{j=1}^\infty\sum\limits_{n_{_1}=0}^\infty\sum\limits_{n_{_2}=0}^\infty
\xi_{_{m3}}^jx_{_{13}}^{n_{_1}}x_{_{23}}^{n_{_2}}A_{_{j,n_{_1},n_{_2}}}^{(b)}
\nonumber\\
&&\hspace{4.0cm}\times
{\Gamma(2j)\Gamma(1+j+n_{_1}+n_{_2})\Gamma(1+n_{_1}+n_{_2})\over j!
n_{_1}!n_{_2}!\Gamma(1+j+n_{_1})\Gamma(1+j+n_{_2})}
\nonumber\\
&&\hspace{4.0cm}
+\sum\limits_{j=1}^\infty\sum\limits_{n_{_1}=0}^\infty\sum\limits_{n_{_2}=0}^\infty
\xi_{_{m3}}^{j+n_{_1}+n_{_2}}x_{_{13}}^{n_{_1}}x_{_{23}}^{-j-n_{_1}}
\nonumber\\
&&\hspace{4.0cm}\times
{(-)^{n_{_1}}\Gamma(1+n_{_1}+n_{_2})\Gamma(2j+2n_{_1}+2n_{_2})\Gamma(j+n_{_1})
\over(j+n_{_1}+n_{_2})!n_{_1}!n_{_2}!\Gamma(1+j+2n_{_1}+n_{_2})\Gamma(j)}
\nonumber\\
&&\hspace{4.0cm}\times
\Big[\Big(-2\psi(2j+2n_{_1}+2n_{_2})-\psi(j+n_{_1})
\nonumber\\
&&\hspace{4.0cm}
+\psi(1+j+n_{_1}+n_{_2})+2\psi(1+j+2n_{_1}+n_{_2})+\psi(1+n_{_1})
\nonumber\\
&&\hspace{4.0cm}
-\psi(1+n_{_1}+n_{_2})-\ln{x_{_{13}}\xi_{_{m3}}\over x_{_{23}}}\Big)
+{\varepsilon\over2}B_{_{j,n_{_1},n_{_2}}}^{(b)}\Big]
\nonumber\\
&&\hspace{4.0cm}
+\sum\limits_{j=1}^\infty\sum\limits_{n_{_1}=0}^\infty\sum\limits_{n_{_2}=0}^\infty
\xi_{_{m3}}^{j+n_{_1}+n_{_2}}x_{_{13}}^{-j-n_{_2}}x_{_{23}}^{n_{_2}}
\nonumber\\
&&\hspace{4.0cm}\times
{(-)^{n_{_2}}\Gamma(1+n_{_1}+n_{_2})\Gamma(2j+2n_{_1}+2n_{_2})\Gamma(j+n_{_2})
\over(j+n_{_1}+n_{_2})!n_{_1}!n_{_2}!\Gamma(1+j+n_{_1}+2n_{_2})\Gamma(j)}
\nonumber\\
&&\hspace{4.0cm}\times
\Big[\Big(-2\psi(2j+2n_{_1}+2n_{_2})-\psi(j+n_{_2})
\nonumber\\
&&\hspace{4.0cm}
+\psi(1+j+n_{_1}+n_{_2})+2\psi(1+j+n_{_1}+2n_{_2})+\psi(1+n_{_2})
\nonumber\\
&&\hspace{4.0cm}
-\psi(1+n_{_1}+n_{_2})-\ln{x_{_{23}}\xi_{_{m3}}\over x_{_{13}}}\Big)
+{\varepsilon\over2}C_{_{j,n_{_1},n_{_2}}}^{(b)}\Big]
\nonumber\\
&&\hspace{4.0cm}
+\sum\limits_{j=1}^\infty\sum\limits_{n_{_1}=0}^\infty\sum\limits_{n_{_2}=0}^\infty
\xi_{_{m3}}^{j+n_{_1}+n_{_2}}x_{_{13}}^{-j-n_{_2}}x_{_{23}}^{-j-n_{_1}}
\nonumber\\
&&\hspace{4.0cm}\times
{(-)^{j}\Gamma(j+n_{_2})\Gamma(j+n_{_1})\Gamma(2j+2n_{_1}+2n_{_2})\over\Gamma(j)n_{_1}!n_{_2}!
\Gamma(1+j+n_{_1}+n_{_2})\Gamma(2j+n_{_1}+n_{_2})}
\nonumber\\
&&\hspace{4.0cm}\times
\Big[\Big(2\psi(2j+2n_{_1}+2n_{_2})-2\psi(2j+n_{_1}+n_{_2})-\psi(j)
\nonumber\\
&&\hspace{4.0cm}
+\psi(j+n_{_1})+\psi(j+n_{_2})-\psi(1+j+n_{_1}+n_{_2})
\nonumber\\
&&\hspace{4.0cm}
-\ln{x_{_{13}}x_{_{23}}\over\xi_{_{m3}}}\Big)
+{\varepsilon\over2}D_{_{j,n_{_1},n_{_2}}}^{(b)}\Big]
\nonumber\\
&&\hspace{4.0cm}
+\sum\limits_{j=1}^\infty\sum\limits_{n_{_1}=1}^\infty\sum\limits_{n_{_2}=1}^\infty
\xi_{_m}^{j+n_{_1}+n_{_2}}x_{_{13}}^{-j}x_{_{23}}^{-n_{_1}}
\nonumber\\
&&\hspace{4.0cm}\times
{\Gamma(2j+2n_{_1}+2n_{_2})\Gamma(1+n_{_2})\Gamma(j)
\Gamma(n_{_1})\over(j+n_{_1}+n_{_2})!(n_{_1}+n_{_2})!(j+n_{_2})!
\Gamma(j+n_{_1})}
\nonumber\\
&&\hspace{4.0cm}\times
\Big[1+\varepsilon\Big(2\psi(2j+2n_{_1}+2n_{_2})-2\psi(j+n_{_1})
\nonumber\\
&&\hspace{4.0cm}
-\psi(1+n_{_2})+\psi(j)+\psi(n_{_1})-\ln(x_{_{13}}x_{_{23}})\Big)\Big]
\nonumber\\
&&\hspace{4.0cm}
-2\varepsilon\sum\limits_{j=1}^\infty\sum\limits_{n_{_1}=1}^\infty\sum\limits_{n_{_2}=0}^\infty
\xi_{_m}^{j+n_{_2}}x_{_{13}}^{j+n_{_1}}x_{_{23}}^{-j}
\nonumber\\
&&\hspace{4.0cm}\times
{\Gamma(2j+2n_{_2})\Gamma(j)\Gamma(1+j+n_{_1}+n_{_2})\Gamma(1+n_{_1})
\over(2j+n_{_1}+n_{_2})!(j+n_{_1})!(j+n_{_2})!n_{_2}!}
\nonumber\\
&&\hspace{4.0cm}\times
\Big(\ln x_{_{13}}-\psi(1+2j+n_{_1}+n_{_2})-\psi(1+j+n_{_1})
\nonumber\\
&&\hspace{4.0cm}
+\psi(1+j+n_{_1}+n_{_2})+\psi(1+n_{_2})\Big)
\nonumber\\
&&\hspace{4.0cm}
-2\varepsilon\sum\limits_{j=1}^\infty\sum\limits_{n_{_1}=0}^\infty\sum\limits_{n_{_2}=1}^\infty
\xi_{_m}^{j+n_{_2}}x_{_{13}}^{-j}x_{_{23}}^{j+n_{_2}}
\nonumber\\
&&\hspace{4.0cm}\times
{\Gamma(2j+2n_{_1})\Gamma(j)\Gamma(1+j+n_{_1}+n_{_2})\Gamma(1+n_{_2})
\over(2j+n_{_1}+n_{_2})!(j+n_{_1})!(j+n_{_2})!n_{_1}!}
\nonumber\\
&&\hspace{4.0cm}\times
\Big(\ln x_{_{23}}-\psi(1+2j+n_{_1}+n_{_2})-\psi(1+j+n_{_2})
\nonumber\\
&&\hspace{4.0cm}
+\psi(1+j+n_{_1}+n_{_2})+\psi(1+n_{_1})\Big)
\nonumber\\
&&\hspace{4.0cm}
+2\varepsilon\sum\limits_{j=0}^\infty\sum\limits_{n_{_1}=1}^\infty
\xi_{_{m3}}^{j+n_{_1}}x_{_{13}}^{n_{_1}}x_{_{23}}^{-n_{_1}}
{(-)^{n_{_1}}\Gamma(2j+2n_{_1})\Gamma(n_{_1})
\over n_{_1}!j!\Gamma(1+j+2n_{_1})}E_{_{j,n_{_1}}}^{(b)}(x_{_{13}})
\nonumber\\
&&\hspace{4.0cm}
+2\varepsilon\sum\limits_{j=0}^\infty\sum\limits_{n_{_2}=1}^\infty
\xi_{_{m3}}^{j+n_{_2}}x_{_{13}}^{-n_{_2}}x_{_{23}}^{n_{_2}}
{(-)^{n_{_2}}\Gamma(2j+2n_{_2})\Gamma(n_{_2})
\over n_{_2}!j!\Gamma(1+j+2n_{_2})}E_{_{j,n_{_2}}}^{(b)}(x_{_{23}})
\nonumber\\
&&\hspace{4.0cm}
+\sum\limits_{j=1}^\infty\sum\limits_{n_{_1}=0}^\infty
x_{_m}^{j+n_{_1}}x_{_1}^{-j}x_{_2}^{-n_{_1}}
{\Gamma(2j+2n_{_1})\Gamma(j)\Gamma(n_{_1})
\over j!n_{_1}!(j+n_{_1})!\Gamma(j+n_{_1})}
\nonumber\\
&&\hspace{4.0cm}\times
\Big[1+G_{_{j,n_{_1}}}^{(b)}(x_{_{13}}x_{_{23}})\Big]
\nonumber\\
&&\hspace{4.0cm}
+{\cal O}(\varepsilon^2)\Big\}\;,
\label{1LoopC-3ME-Region1-1}
\end{eqnarray}
where the coefficients of $\varepsilon$ above $A_{_{j,n_{_1},n_{_2}}}^{(b)}$, $B_{_{j,n_{_1},n_{_2}}}^{(b)}$,
$C_{_{j,n_{_1},n_{_2}}}^{(b)}$, $D_{_{j,n_{_1},n_{_2}}}^{(b)}$, $E_{_{j,n}}^{(b)}(x)$,
and $G_{_{j,n}}^{(b)}(x)$ can be found in Eq.(\ref{app2-2}).
The absolutely and uniformly converging region $\Omega_{_7}^b$
is a proper subset of the cube
\begin{eqnarray}
&&\Xi_{_{\rm K}}^b=\left\{(\xi_{_{m3}},\;x_{_{13}},\;x_{_{23}}):\left|\begin{array}{l}
|\xi_{_{m3}}|<1,\;|x_{_{13}}|<1,\;|x_{_{23}}|<1
\end{array}\right.\right\}\;.
\label{1LoopC3-1-2}
\end{eqnarray}
In order to continue the scalar integral $C_{_0}^b$ from the region $\Omega_{_7}^b$
to the cube $\Xi_{_{\rm K}}^b$, one employs the system of PDEs in
Eq.(\ref{3M-PDEs}) with the finite element method. We will consider this
in detail in section \ref{sec6}.

\item[$\bullet$ 2:] In the convergent region $\Omega_{_8}^b$,
the scalar integral $C_{_0}^b$ is
\begin{eqnarray}
&&C_{_0}^b(p_{_1}^2,\;p_{_2}^2,\;p_{_3}^2,\;m^2)=
\Big\{C_{_{0,3}}^b+C_{_{0,4}}^b+C_{_{0,8}}^b+C_{_{0,9}}^b+C_{_{0,21}}^b
\nonumber\\
&&\hspace{4.0cm}
+C_{_{0,27}}^b+C_{_{0,33}}^b\Big\}(p_{_1}^2,\;p_{_2}^2,\;p_{_3}^2,\;m^2)
\nonumber\\
&&\hspace{3.5cm}=
{i(p_{_2}^2)^{D/2-3}\over(4\pi)^{D/2}}F_{_{b,p_{_2}}}(\xi_{_{m2}},\;x_{_{12}},\;x_{_{32}})
\label{1LoopC3-2-1}
\end{eqnarray}
In the limit $m^2\rightarrow0$, the expression of Eq.(\ref{1LoopC3-2-1}) recovers
that presented in Eq.(\ref{Massless-III1}) exactly. The expansion of the scalar integral
around $\varepsilon=0$ is
\begin{eqnarray}
&&C_{_0}^b={\rm Eq}.(\ref{1LoopC-3ME-Region1-1})(p_{_3}\rightarrow p_{_2},\;
\xi_{_{m3}}\rightarrow \xi_{_{m2}},\;
x_{_{13}}\rightarrow x_{_{12}},\;x_{_{23}}\rightarrow x_{_{32}})\;.
\label{1LoopC-3ME-Region2-1}
\end{eqnarray}
Furthermore, the function $x_{_{23}}^{D/2-3}F_{_{b,p_{_2}}}$  complies with
the system of linear PDEs in Eq~(\ref{3M-PDEs}). Or equivalently, the function
$F_{_{b,p_{_2}}}$ satisfies the holonomic hypergeometric system of linear PDEs
obtained through the interchanging $2\leftrightarrow3$ in Eq.(\ref{3M-PDEs}).

Obviously the absolutely and uniformly converging region $\Omega_{_8}^b$
is a proper subset of the cube
\begin{eqnarray}
&&\Xi_{_{{\rm p}_{_2}}}^b=\left\{(\xi_{_{m2}},\;x_{_{12}},\;x_{_{32}}):\left|\begin{array}{l}
|\xi_{_{m2}}|<1,\;|x_{_{12}}|<1,\;|x_{_{32}}|<1
\end{array}\right.\right\}\;.
\label{1LoopC3-2-2}
\end{eqnarray}
In order to continue the scalar integral $C_{_0}^b$ from the region $\Omega_{_8}^b$
to the cube $\Xi_{_{{\rm p}_{_2}}}^b$, one employs the corresponding system of
linear PDEs.

\item[$\bullet$ 3:] In the parameter space $\Omega_{_{11}}^b$,
the scalar integral $C_{_0}^b$ is
\begin{eqnarray}
&&C_{_0}^b(p_{_1}^2,\;p_{_2}^2,\;p_{_3}^2,\;m^2)=
\Big\{C_{_{0,10}}^b+C_{_{0,11}}^b+C_{_{0,13}}^b+C_{_{0,14}}^b+C_{_{0,21}}^b
\nonumber\\
&&\hspace{4.0cm}
+C_{_{0,27}}^b+C_{_{0,33}}^b\Big\}(p_{_1}^2,\;p_{_2}^2,\;p_{_3}^2,\;m^2)
\nonumber\\
&&\hspace{3.5cm}=
{i(p_{_1}^2)^{D/2-3}\over(4\pi)^{D/2}}F_{_{b,p_{_1}}}(\xi_{_{m1}},\;x_{_{21}},\;x_{_{31}})
\label{1LoopC3-3-1}
\end{eqnarray}
In the limit $m^2\rightarrow0$, the expression of Eq.(\ref{1LoopC3-3-1}) recovers
that presented in Eq.(\ref{Massless-II1}). The expansion of the scalar integral
around $\varepsilon=0$ is
\begin{eqnarray}
&&C_{_0}^b={\rm Eq}.(\ref{1LoopC-3ME-Region1-1})(p_{_3}\rightarrow p_{_1},\;
\xi_{_{m3}}\rightarrow x_{_{m1}},\;
x_{_{13}}\rightarrow x_{_{31}},\;x_{_{23}}\rightarrow x_{_{21}})\;.
\label{1LoopC-3ME-Region3-1}
\end{eqnarray}
In addition, the function $x_{_{13}}^{D/2-3}F_{_{b,p_{_1}}}$  complies with
the system of linear PDEs in Eq~(\ref{3M-PDEs}). Or equivalently, the function
$F_{_{b,p_{_1}}}$ satisfies the holonomic hypergeometric system of linear PDEs
obtained through the interchanging $1\leftrightarrow3$ in Eq.(\ref{3M-PDEs}).

Obviously the absolutely and uniformly converging region $\Omega_{_{11}}^b$
is a proper subset of the cube
\begin{eqnarray}
&&\Xi_{_{{\rm p}_{_1}}}^b=\left\{(\xi_{_{m1}},\;x_{_{21}},\;x_{_{31}}):\left|\begin{array}{l}
|\xi_{_{m1}}|<1,\;|x_{_{21}}|<1,\;|x_{_{31}}|<1
\end{array}\right.\right\}\;.
\label{1LoopC3-3-2}
\end{eqnarray}
In order to continue the scalar integral $C_{_0}^b$ from the region $\Omega_{_{11}}^b$
to the cube $\Xi_{_{{\rm p}_{_1}}}^b$, one employs the corresponding system of linear PDEs.

\item[$\bullet$ 4:] In the parameter space $\Omega_{_{19}}^b$,
the scalar integral $C_{_0}^b$ is written as
\begin{eqnarray}
&&C_{_0}^b(p_{_1}^2,\;p_{_2}^2,\;p_{_3}^2,\;m^2)=
C_{_{0,19}}^b(p_{_1}^2,\;p_{_2}^2,\;p_{_3}^2,\;m^2)
\nonumber\\
&&\hspace{3.5cm}=
{i(m^2)^{D/2-3}\over(4\pi)^{D/2}}F_{_{b,m}}(\eta_{_{1m}},\;\eta_{_{2m}},\;\eta_{_{3m}})\;.
\label{1LoopC3-4-1}
\end{eqnarray}
Except a convenient factor, the expression coincides with Eq~(37) of the
literature~\cite{Davydychev3} exactly.
We expand the hypergeometric functions $C_{_{0,19}}^b$
around $\varepsilon=0$ as
\begin{eqnarray}
&&C_{_0}^b(p_{_1}^2,\;p_{_2}^2,\;p_{_3}^2,\;m^2)=
-{i(m^2)^{D/2-3}\over(4\pi)^{D/2}}
\sum\limits_{j=0}^\infty\sum\limits_{n_{_1}=0}^\infty\sum\limits_{n_{_2}=0}^\infty
\eta_{_{3m}}^{j}\eta_{_{1m}}^{n_{_1}}\eta_{_{2m}}^{n_{_2}}
\nonumber\\
&&\hspace{4.0cm}\times
{\Gamma(1+j+n_{_1})\Gamma(1+j+n_{_2})
\Gamma(1+n_{_1}+n_{_2})\over j!n_{_1}!n_{_2}!}
\nonumber\\
&&\hspace{4.0cm}\times
{\Gamma(1+j+n_{_1}+n_{_2})\over\Gamma(3+2j+2n_{_1}+2n_{_2})}
\Big[1+\varepsilon\psi(1+j+n_{_1}+n_{_2})
\nonumber\\
&&\hspace{4.0cm}
+{\cal O}(\varepsilon^2)\Big]\;.
\label{1LoopC-3ME-Region4-1}
\end{eqnarray}
Additionally the function $\xi_{_{m3}}^{D/2-3}F_{_{b,m}}$  complies with
the system of linear PDEs in Eq~(\ref{3M-PDEs}). Or equivalently, the function
$F_{_{b,m}}$ satisfies the following system of linear PDEs
\begin{eqnarray}
&&\Big\{\Big(3-{D\over2}+\sum\limits_{j=1}^3\hat{\theta}_{\eta_{_{jm}}}\Big)
\prod\limits_{j\neq i}^3(1+\hat{\theta}_{y_{_{im}}}+\hat{\theta}_{y_{_{jm}}})
\nonumber\\
&&\hspace{0.0cm}
-{1\over y_{_{im}}}\hat{\theta}_{y_{_{im}}}\Big(1+2\sum\limits_{j=1}^3\hat{\theta}_{\eta_{_{jm}}}\Big)
\Big(2+2\sum\limits_{j=1}^3\hat{\theta}_{\eta_{_{jm}}}\Big)\Big\}F_{_{b,m}}=0,\;(i=1,\;2,\;3)\;.
\label{1LoopC4-6}
\end{eqnarray}

Obviously the cube
\begin{eqnarray}
&&\Xi_{_{\rm m}}^b=\left\{(\eta_{_{1m}},\;\eta_{_{2m}},\;\eta_{_{3m}}):\left|\begin{array}{l}
|\eta_{_{1m}}|<1,\;|\eta_{_{2m}}|<1,\;|\eta_{_{3m}}|<1
\end{array}\right.\right\}\;.
\label{1LoopC4-5}
\end{eqnarray}
is a proper subset of the absolutely and uniformly converging region $\Omega_{_{19}}^b$.
In order to continue the scalar integral $C_{_0}^b$ from the region $\Omega_{_{19}}^b$
to the relevant parameter space, one employs the system of linear PDEs in Eq.~(\ref{1LoopC4-6}).

\end{itemize}

\section{The system of linear PDEs as the stationary
condition of a functional\label{sec6}}
\indent\indent
As stated above, the $C_{_0}$ function of one nonzero mass is formulated
through the hypergeometric functions in some convergent regions of independent variables.
Since there is not the reduction formula for the hypergeometric functions, the scalar integral
cannot be analytically continued outside the convergent regions. Nevertheless
the continuation of the scalar integrals
to the whole parameter space can be done numerically by the systems of linear PDEs
in Eq.(\ref{1M-PDEs}). In order to proceed with our analysis,
we expand $F_{_{a,p_{_3}}}$ around time-space dimensions $D=4$ as
\begin{eqnarray}
&&F_{_{a,p_{_3}}}(\xi_{_{33}},\;x_{_{13}},\;x_{_{23}})=F_{_{a,p_{_3}}}^{(0)}(\xi_{_{33}},\;x_{_{13}},\;x_{_{23}})
+\sum\limits_{i=1}^\infty\varepsilon^i F_{_{a,p_{_3}}}^{(i)}(\xi_{_{33}},\;x_{_{13}},\;x_{_{23}})\;,
\label{C0-expandsion-1}
\end{eqnarray}
where the function $F_{_{a,p_{_3}}}^{(n)}(\xi_{_{33}},\;x_{_{13}},\;x_{_{23}})$
satisfies the system of PDEs
\begin{eqnarray}
&&\Big[\xi_{_{33}}(1+\hat{\vartheta}_{\xi_{_{33}}}+\sum\limits_{i=1}^2\hat{\vartheta}_{x_{_{i3}}})^2
\hat{\vartheta}_{\xi_{_{33}}}
-\prod\limits_{i=1}^2(\hat{\vartheta}_{\xi_{_{33}}}+\hat{\vartheta}_{x_{_{i3}}})
\hat{\vartheta}_{\xi_{_{33}}}\Big]F_{_{a,p_{_3}}}^{(n)}+f_{_{a,0}}^{(n)}=0
\;,\nonumber\\
&&\Big[x_{_{j3}}(1+\hat{\vartheta}_{\xi_{_{33}}}+\sum\limits_{i=1}^2\hat{\vartheta}_{x_{_{i3}}})^2
-\hat{\vartheta}_{x_{_{j3}}}(\hat{\vartheta}_{\xi_{_{33}}}
+\hat{\vartheta}_{x_{_{j3}}})\Big]F_{_{a,p_{_3}}}^{(n)}+f_{_{a,j}}^{(n)}=0\;,\;(j=1,\;2)\;,
\label{1M-PDEs-1}
\end{eqnarray}
with
\begin{eqnarray}
&&f_{_{a,0}}^{(n)}(\xi_{_{33}},\;x_{_{13}},\;x_{_{23}})=\Big[\xi_{_{33}}\Big(2+5\hat{\vartheta}_{\xi_{_{33}}}
+4\sum\limits_{i=1}^2\hat{\vartheta}_{x_{_{i3}}}+(3\hat{\vartheta}_{\xi_{_{33}}}
+\sum\limits_{i=1}^2\hat{\vartheta}_{x_{_{i3}}})(\hat{\vartheta}_{\xi_{_{33}}}
+\sum\limits_{i=1}^2\hat{\vartheta}_{x_{_{i3}}})\Big)
\nonumber\\
&&\hspace{3.5cm}
-2\hat{\vartheta}_{\xi_{_{33}}}^2-\hat{\vartheta}_{\xi_{_{m3}}}\sum\limits_{i=1}^2\hat{\vartheta}_{x_{_{i3}}}
\Big]F_{_{a,p_{_3}}}^{(n-1)}(\xi_{_{33}},\;x_{_{13}},\;x_{_{23}})
\nonumber\\
&&\hspace{3.5cm}
+\Big[2\xi_{_{33}}(1+\hat{\vartheta}_{\xi_{_{33}}}+\sum\limits_{i=1}^2\hat{\vartheta}_{x_{_{i3}}})
-\hat{\vartheta}_{\xi_{_{33}}}\Big]F_{_{a,p_{_3}}}^{(n-2)}(\xi_{_{33}},\;x_{_{13}},\;x_{_{23}})
\;,\nonumber\\
&&f_{_{a,j}}^{(n)}(\xi_{_{33}},\;x_{_{13}},\;x_{_{23}})=\Big[x_{_{j3}}(1+\hat{\vartheta}_{\xi_{_{33}}}
+\sum\limits_{i=1}^2\hat{\vartheta}_{x_{_{i3}}})
-\hat{\vartheta}_{x_{_{j2}}}\Big]F_{_{a,p_{_3}}}^{(n-1)}(\xi_{_{33}},\;x_{_{13}},\;x_{_{23}})
,\;(j=1,\;2)\;.
\label{1M-PDEs-2}
\end{eqnarray}
Through the transformation of variables
\begin{eqnarray}
&&t_{_m}=\ln\xi_{_{33}}
\;,\nonumber\\
&&t_{_1}=\ln x_{_{13}}
\;,\nonumber\\
&&t_{_2}=\ln x_{_{23}}\;,
\label{functional-transformation-1}
\end{eqnarray}
the system of PDEs in Eq.(\ref{1M-PDEs-1}) is recognized as stationary conditions
of the modified functional
\begin{eqnarray}
&&\Pi^*(F_{_{a,p_{_3}}}^{(n)})=\Pi({\partial F_{_{a,p_{_3}}}^{(n)}\over\partial t_{_m}})
\nonumber\\
&&\hspace{2.0cm}
+\sum\limits_{i=1}^2\int\limits_{\Omega}\chi_{_{i}}
\Big\{e^{t_{_i}}\Big(1+{\partial\over\partial t_{_m}}+{\partial\over\partial t_{_1}}
+{\partial\over\partial t_{_2}}\Big)^2F_{_{a,p_{_3}}}^{(n)}
\nonumber\\
&&\hspace{2.0cm}
-{\partial\over\partial t_{_i}}\Big({\partial\over\partial t_{_m}}
+{\partial\over\partial t_{_i}}\Big)F_{_{a,p_{_3}}}^{(n)}+f_{_{a,i}}^{(n)}
\Big\}dt_{_m}dt_{_1}dt_{_2}\;.
\label{functional-1}
\end{eqnarray}
Here $\chi_{_{1}}(t_{_m},t_{_1},t_{_2}),\;\chi_{_{2}}(t_{_m},t_{_1},t_{_2})$
are Lagrange multipliers, $\Omega$ represents the parameter
space where the continuation of the solution is made numerically,
and $\Pi({\partial F_{_{a,k}}^{(n)}\over\partial t_{_m}})$ is the
functional of the first PDE in Eq.(\ref{1M-PDEs-2}):
\begin{eqnarray}
&&\Pi({\partial F_{_{a,p_{_3}}}^{(n)}\over\partial t_{_m}})=\int\limits_{\Omega}\Big\{
(e^{t_{_m}}-1)\Big({\partial^2F_{_{a,p_{_3}}}^{(n)}\over\partial t_{_m}^2}\Big)^2
+e^{t_{_m}}\Big({\partial^2F_{_{a,p_{_3}}}^{(n)}\over\partial t_{_m}\partial t_{_1}}\Big)^2
+e^{t_{_m}}\Big({\partial^2F_{_{a,p_{_3}}}^{(n)}\over\partial t_{_m}\partial t_{_2}}\Big)^2
\nonumber\\
&&\hspace{2.0cm}
+\Big(e^{t_{_m}}-{1\over2}\Big){\partial^2F_{_{a,p_{_3}}}^{(n)}\over\partial t_{_m}^2}
{\partial^2F_{_{a,p_{_3}}}^{(n)}\over\partial t_{_m}\partial t_{_1}}
+\Big(e^{t_{_m}}-{1\over2}\Big){\partial^2F_{_{a,p_{_3}}}^{(n)}\over\partial t_{_m}^2}
{\partial^2F_{_{a,p_{_3}}}^{(n)}\over\partial t_{_m}\partial t_{_2}}
\nonumber\\
&&\hspace{2.0cm}
+\Big(e^{t_{_m}}-{1\over2}\Big){\partial^2F_{_{a,p_{_3}}}^{(n)}\over\partial t_{_m}\partial t_{_1}}
{\partial^2F_{_{a,p_{_3}}}^{(n)}\over\partial t_{_m}\partial t_{_2}}
+e^{t_{_m}}{\partial F_{_{a,p_{_3}}}^{(n)}\over\partial t_{_m}}
{\partial^2F_{_{a,p_{_3}}}^{(n)}\over\partial t_{_m}^2}
\nonumber\\
&&\hspace{2.0cm}
-f_{_{a,0}}^{(n)}{\partial F_{_{a,p_{_3}}}^{(n)}\over\partial t_{_m}}
\Big\}dt_{_m}dt_{_1}dt_{_2}\;.
\label{functional-2}
\end{eqnarray}
Furthermore, the stationary condition
of the second term of Eq.(\ref{functional-1}) is the second PDEs
in Eq.(\ref{1M-PDEs-2}), that of the third term of
Eq.(\ref{functional-1}) is the third PDEs
in Eq.(\ref{1M-PDEs-2}), which are recognized as two restrictions of the system.
Once the solutions $F_{_{a,k}}^{(i)}$, $(i\le n-1,n\ge1)$ are obtained numerically
in the parameter space $\Omega$, the solution $F_{_{a,k}}^{(n)}$ can be numerically continued
from the convergent region $\Omega_{_7}^a$ to the concerned region $\Omega$
with the finite element method~\cite{X.C.Wang03}.

The systems of linear PDEs of Eq.~(\ref{1M-PDEs}), Eq.~(\ref{1LoopC-2acd4}),
and Eq.~(\ref{1LoopC4-4}) are compatible with each other
since they all originate from the Mellin-Barnes representation of
Eq.~(\ref{1M}). Correspondingly a similar functional is constructed with
the system of PDEs in Eq.~(\ref{1LoopC-2acd4}).
Expansion of $F_{_{a,p_{_1}}}$ around space-time dimensions $D=4$ is
\begin{eqnarray}
&&F_{_{a,p_{_1}}}(\xi_{_{31}},\;x_{_{21}},\;x_{_{31}})=F_{_{a,p_{_1}}}^{(0)}(\xi_{_{31}},\;x_{_{21}},\;x_{_{31}})
+\sum\limits_{i=1}^\infty\varepsilon^i F_{_{a,p_{_1}}}^{(i)}(\xi_{_{31}},\;x_{_{21}},\;x_{_{31}})\;,
\label{C0-expandsion-2}
\end{eqnarray}
the solution $F_{_{a,p_{_1}}}^{(n)}$ can be numerically continued
from the convergent region $\Omega_{_{11}}^a\bigcup\Omega_{_{29}}^a$
to the concerned region $\Omega$ with the finite element method.
The compatibility between the systems of linear PDEs of Eq.~(\ref{1M-PDEs})
and Eq.~(\ref{1LoopC-2acd4}) implies
\begin{eqnarray}
&&F_{_{a,p_{_1}}}^{(n)}(\xi_{_{31}},\;x_{_{21}},\;x_{_{31}})
\equiv\sum\limits_{n^\prime=0}^n{(\ln x_{_{13}})^{n^\prime}\over n^\prime !}
F_{_{a,p_{_3}}}^{(n-n^\prime)}(\xi_{_{33}},\;x_{_{13}},\;x_{_{23}})\;,
\label{C0-compatibility-1}
\end{eqnarray}
which can be used to check whether the numerical program runs correctly.

Similarly the $C_{_0}$ function of three equal masses is formulated
through the hypergeometric functions in some convergent regions,
and the continuation of the scalar integrals
to the whole parameter space can be made numerically with the systems of linear PDEs
in Eq.(\ref{3M-PDEs}). In order to proceed with our analysis,
we expand $F_{_{b,p_{_3}}}$ around space-time dimensions $D=4$ as
\begin{eqnarray}
&&F_{_{b,p_{_3}}}(\xi_{_{m3}},\;x_{_{13}},\;x_{_{23}})=F_{_{b,p_{_3}}}^{(0)}(\xi_{_{m3}},\;x_{_{13}},\;x_{_{23}})
+\sum\limits_{i=1}^\infty\varepsilon^i F_{_{b,p_{_3}}}^{(i)}(\xi_{_{m3}},\;x_{_{13}},\;x_{_{23}})\;,
\label{C0-expandsion-3}
\end{eqnarray}
where the function $F_{_{b,p_{_3}}}^{(n)}(\xi_{_{m3}},\;x_{_{13}},\;x_{_{23}})$
satisfies the system of PDEs
\begin{eqnarray}
&&\Big[2\xi_{_{m3}}(1+\hat{\vartheta}_{\xi_{_{m3}}}+\sum\limits_{i=1}^2\hat{\vartheta}_{x_{_{i3}}})
(1+2\hat{\vartheta}_{\xi_{_{m3}}})\hat{\vartheta}_{\xi_{_{m3}}}
\nonumber\\
&&\hspace{0.0cm}
-\prod\limits_{i=1}^2(\hat{\vartheta}_{\xi_{_{m3}}}+\hat{\vartheta}_{x_{_{i3}}})
\hat{\vartheta}_{\xi_{_{m3}}}\Big]F_{_{b,p_{_3}}}^{(n)}+f_{_{b,0}}^{(n)}=0
\;,\nonumber\\
&&\Big[x_{_{j3}}(1+\hat{\vartheta}_{\xi_{_{m3}}}+\sum\limits_{i=1}^2\hat{\vartheta}_{x_{_{i3}}})
(1+\sum\limits_{i=1}^2\hat{\vartheta}_{x_{_{i3}}})
\nonumber\\
&&\hspace{0.0cm}
-\hat{\vartheta}_{x_{_{j3}}}(\hat{\vartheta}_{\xi_{_{m3}}}
+\hat{\vartheta}_{x_{_{j3}}})\Big]F_{_{b,p_{_3}}}^{(n)}+f_{_{b,j}}^{(n)}=0\;,\;(j=1,\;2)\;,
\label{3M-PDEs-1}
\end{eqnarray}
with
\begin{eqnarray}
&&f_{_{b,0}}^{(n)}(\xi_{_{m3}},\;x_{_{13}},\;x_{_{23}})=\Big[2\xi_{_{m3}}(1+6\hat{\vartheta}_{\xi_{_{m3}}}
+4\hat{\vartheta}_{\xi_{_{m3}}}\sum\limits_{i=1}^2\hat{\vartheta}_{x_{_{i3}}}+6\hat{\vartheta}_{\xi_{_{m3}}}^2)
\nonumber\\
&&\hspace{3.5cm}
-2\hat{\vartheta}_{\xi_{_{m3}}}^2-\hat{\vartheta}_{\xi_{_{m3}}}\sum\limits_{i=1}^2\hat{\vartheta}_{x_{_{i3}}}
\Big]F_{_{b,p_{_3}}}^{(n-1)}(\xi_{_{m3}},\;x_{_{13}},\;x_{_{23}})
\nonumber\\
&&\hspace{3.5cm}
+\Big[2\xi_{_{m3}}(3+6\hat{\vartheta}_{\xi_{_{m3}}}+2\sum\limits_{i=1}^2\hat{\vartheta}_{x_{_{i3}}})
-\hat{\vartheta}_{\xi_{_{m3}}}\Big]F_{_{b,p_{_3}}}^{(n-2)}(\xi_{_{m3}},\;x_{_{13}},\;x_{_{23}})
\nonumber\\
&&\hspace{3.5cm}
+\xi_{_{m3}}F_{_{b,p_{_3}}}^{(n-3)}(\xi_{_{m3}},\;x_{_{13}},\;x_{_{23}})
\;,\nonumber\\
&&f_{_{b,j}}^{(n)}(\xi_{_{m3}},\;x_{_{13}},\;x_{_{23}})=\Big[x_{_{j3}}(1+\sum\limits_{i=1}^2\hat{\vartheta}_{x_{_{i3}}})
-\hat{\vartheta}_{x_{_{j3}}}\Big]F_{_{b,p_{_3}}}^{(n-1)}(\xi_{_{m3}},\;x_{_{13}},\;x_{_{23}})\;,\;(j=1,\;2)\;.
\label{3M-PDEs-2}
\end{eqnarray}
Through the transformation of variables in Eq.~(\ref{functional-transformation-1}),
the system of PDEs in Eq.(\ref{3M-PDEs-1}) is recognized as stationary conditions
of the modified functional
\begin{eqnarray}
&&\Pi^*(F_{_{b,p_{_3}}}^{(n)})=\Pi({\partial F_{_{b,p_{_3}}}^{(n)}\over\partial t_{_m}})
\nonumber\\
&&\hspace{2.0cm}
+\sum\limits_{i=1}^2\int\limits_{\Omega}\chi_{_{i}}
\Big\{e^{t_{_i}}\Big(1+{\partial\over\partial t_{_m}}+{\partial\over\partial t_{_1}}
+{\partial\over\partial t_{_2}}\Big)\Big(1+{\partial\over\partial t_{_1}}
+{\partial\over\partial t_{_2}}\Big)F_{_{b,p_{_3}}}^{(n)}
\nonumber\\
&&\hspace{2.0cm}
-{\partial\over\partial t_{_i}}\Big({\partial\over\partial t_{_m}}
+{\partial\over\partial t_{_i}}\Big)F_{_{b,p_{_3}}}^{(n)}+f_{_{b,i}}^{(n)}
\Big\}dt_{_m}dt_{_1}dt_{_2}\;.
\label{functional-3}
\end{eqnarray}
Here $\chi_{_{1}}(t_{_m},t_{_1},t_{_2}),\;\chi_{_{2}}(t_{_m},t_{_1},t_{_2})$
are Lagrange multipliers, $\Omega$ represents the parameter
space where the continuation of the solution is made numerically,
and $\Pi({\partial F_{_{b,k}}^{(n)}\over\partial t_{_m}})$ is the functional of
the first PDE in Eq.(\ref{3M-PDEs-2}):
\begin{eqnarray}
&&\Pi({\partial F_{_{b,p_{_3}}}^{(n)}\over\partial t_{_m}})=\int\limits_{\Omega}\Big\{
(4e^{t_{_m}}-1)\Big({\partial^2F_{_{b,p_{_3}}}^{(n)}\over\partial t_{_m}^2}\Big)^2
+\Big(2e^{t_{_m}}-{1\over2}\Big){\partial^2F_{_{b,p_{_3}}}^{(n)}\over\partial t_{_m}^2}
{\partial^2F_{_{b,p_{_3}}}^{(n)}\over\partial t_{_m}\partial t_{_1}}
\nonumber\\
&&\hspace{2.0cm}
+\Big(2e^{t_{_m}}-{1\over2}\Big){\partial^2F_{_{b,p_{_3}}}^{(n)}\over\partial t_{_m}^2}
{\partial^2F_{_{b,p_{_3}}}^{(n)}\over\partial t_{_m}\partial t_{_2}}
-{1\over2}{\partial^2F_{_{b,p_{_3}}}^{(n)}\over\partial t_{_m}\partial t_{_1}}
{\partial^2F_{_{b,p_{_3}}}^{(n)}\over\partial t_{_m}\partial t_{_2}}
\nonumber\\
&&\hspace{2.0cm}
-e^{t_{_m}}\Big({\partial F_{_{b,p_{_3}}}^{(n)}\over\partial t_{_m}}\Big)^2
-f_{_0}^{(n)}{\partial F_{_{b,p_{_3}}}^{(n)}\over\partial t_{_m}}
\Big\}dt_{_m}dt_{_1}dt_{_2}\;.
\label{functional-4}
\end{eqnarray}
Furthermore, the stationary condition
of the second term of Eq.(\ref{functional-3}) is the second PDEs
in Eq.(\ref{3M-PDEs-2}), that of the third term of
Eq.(\ref{functional-1}) is the third PDEs
in Eq.(\ref{3M-PDEs-2}), which are recognised as two restrictions of the system.
Once the solutions $F_{_k}^{(i)}$ $(i\le n-1,n\ge1)$ are obtained numerically
in the parameter region $\Omega$, the solution $F_{_k}^{(n)}$ can be numerically continued
from the convergent region $\Omega_{_7}^b$ to the concerned region $\Omega$
with the finite element method~\cite{X.C.Wang03}.

The systems of linear PDEs of Eq.~(\ref{3M-PDEs}) and Eq.~(\ref{1LoopC4-6})
are compatible with each other
since every one of them originates from the Mellin-Barnes representation of
Eq.~(\ref{3M}). Correspondingly a similar functional is constructed with
the holonomic hypergeometric system of PDEs obtained through $1\leftrightarrow3$ in Eq.~(\ref{3M-PDEs}).
Expansion of $F_{_{b,p_{_1}}}$ around space-time dimensions $D=4$ is
\begin{eqnarray}
&&F_{_{b,p_{_1}}}(\xi_{_{m1}},\;x_{_{21}},\;x_{_{31}})=F_{_{b,p_{_1}}}^{(0)}(\xi_{_{m1}},\;x_{_{21}},\;x_{_{31}})
+\sum\limits_{i=1}^\infty\varepsilon^i F_{_{b,p_{_1}}}^{(i)}(\xi_{_{m1}},\;x_{_{21}},\;x_{_{31}})\;,
\label{C0-expandsion-4}
\end{eqnarray}
the solution $F_{_{b,p_{_1}}}^{(n)}$ can be numerically continued
from the convergent region $\Omega_{_{11}}^b$
to the concerned region $\Omega$ with the finite element method.
The compatibility between the holonomic hypergeometric systems of linear PDEs implies
\begin{eqnarray}
&&F_{_{b,p_{_1}}}^{(n)}(\xi_{_{m1}},\;x_{_{21}},\;x_{_{31}})
\equiv\sum\limits_{n^\prime=0}^n{(\ln x_{_{13}})^{n^\prime}\over n^\prime !}
F_{_{b,p_{_3}}}^{(n-n^\prime)}(\xi_{_{m3}},\;x_{_{13}},\;x_{_{23}})\;,
\label{C0-compatibility-2}
\end{eqnarray}
which can be used to check whether the numerical program runs correctly.

\section{Summary\label{sec7}}
\indent\indent
Using the $\alpha-$ or Feynman-parameterization, we express any scalar integral
by its Mellin-Barnes representation, and present the system of linear PDEs satisfied
by the scalar integral. Through the residue theorem and homology, those scalar integrals
are written as the multiple hypergeometric functions of the independent variables.
Horn's theory of convergence predicts
the absolutely and uniformly convergent region of each hypergeometric functions.
Several convergent regions of the hypergeometric functions compose a set,
where each definition domain does not intersect with the
others in the set. Additionally each convergent region
of the set either does not intersect with,
or is a proper subset of other convergent region which does not belong to the set.
In each definition domain of the set, the scalar
integral can be written as the sum of those hypergeometric functions
whose convergent regions contain the concerned element entirely.
With the idea above, some well-known results of Ref.~\cite{Davydychev1,Davydychev3} are recovered.
Taking the system of PDEs as stationary conditions of
a functional under some given restrictions, one continues
the scalar integrals to the whole parameter space numerically with the finite element methods.

The approach proposed here can be applied to evaluate any Feynman
integral of one-loop multiple point diagrams directly.
In some connected regions of absolute and uniform convergence,
for example, the $D_{_0}$ function is formulated as
the linear combination of the nonuple hypergeometric functions.
The corresponding system of PDEs is composed by nine linear independent PDEs.

In order to apply the method to the scalar integrals of multi-loop diagrams,
one introduces some auxiliary parameters \cite{Davydychev1993,V.A.Smirnov1999, J.B.Tausk1999}.
For example, the analysis on the scalar integral of two-loop planar vertex introduces
two auxiliary kinematic invariants which are linear combinations of $p_{_1}^2$,
$p_{_2}^2$, and $p_{_3}^2$,
the analysis on the scalar integral of two-loop non-planar vertex introduces
five auxiliary kinematic invariants which are linear combinations of $p_{_1}^2$,
$p_{_2}^2$, and $p_{_3}^2$, respectively.
We will present our analyses elsewhere.

\begin{acknowledgments}
\indent\indent
The work has been supported partly by the National Natural
Science Foundation of China (NNSFC) with Grant No. 11821505,
No. 11447601, No. 11675239, No. 11535002, No. 11705045, and No. 11805140.
Furthermore, the author (C.-H. Chang) is also supported by Key Research
program of Frontier Sciences, CAS, Grant No. QYZDY-SSW-SYS006.
\end{acknowledgments}

\appendix
\section{Some derived hypergeometric functions\label{app1}}
\indent\indent
\begin{eqnarray}
&&C_{_{0,7}}^a(p_{_1}^2,\;p_{_2}^2,\;p_{_3}^2,\;m^2)=
{i(-)^{D/2}(p_{_3}^2)^{D/2-3}\over(4\pi)^{D/2}}
{\Gamma^2(2-{D\over2})\Gamma^2({D\over2}-1)\over\Gamma(D-3)\Gamma(4-D)}
\sum\limits_{j=0}^\infty\sum\limits_{n_{_1}=0}^\infty\sum\limits_{n_{_2}=0}^\infty
\nonumber\\
&&\hspace{4.2cm}\times
\xi_{_{33}}^{j}x_{_{13}}^{D/2-2-j+n_{_1}}x_{_{23}}^{D/2-2-j+n_{_2}}
{\Gamma(4-D+j)\over j!n_{_1}!n_{_2}!}
\nonumber\\
&&\hspace{4.2cm}\times
{\Gamma({D\over2}-1-j+n_{_1}+n_{_2})\Gamma(D-3-j+n_{_1}+n_{_2})
\over\Gamma({D\over2}-1-j+n_{_1})\Gamma({D\over2}-1-j+n_{_2})}
\nonumber\\
&&\hspace{3.7cm}=
{i(-)^{D/2}(p_{_3}^2)^{D/2-3}\over(4\pi)^{D/2}}
{\Gamma^2(2-{D\over2})\Gamma^2({D\over2}-1)\over\Gamma(D-3)}
\nonumber\\
&&\hspace{4.2cm}\times
\sum\limits_{n_{_1}=0}^\infty\sum\limits_{n_{_2}=0}^\infty
x_{_{13}}^{D/2-2+n_{_1}}x_{_{23}}^{D/2-2+n_{_2}}
\nonumber\\
&&\hspace{4.2cm}\times
{\Gamma({D\over2}-1+n_{_1}+n_{_2})\Gamma(D-3+n_{_1}+n_{_2})
\over n_{_1}!n_{_2}!\Gamma({D\over2}-1+n_{_1})\Gamma({D\over2}-1+n_{_2})}
\nonumber\\
&&\hspace{4.2cm}
+{i(-)^{D/2}(p_{_3}^2)^{D/2-3}\over(4\pi)^{D/2}}
{\Gamma(2-{D\over2})\Gamma({D\over2}-1)\over\Gamma(D-3)\Gamma(4-D)}
\nonumber\\
&&\hspace{4.2cm}\times
\sum\limits_{j=1}^\infty\sum\limits_{n_{_1}=1}^\infty
\sum\limits_{n_{_2}=0}^\infty
\xi_{_{33}}^{j+n_{_2}}x_{_{13}}^{D/2-2+n_{_1}}x_{_{23}}^{D/2-2-j}
\nonumber\\
&&\hspace{4.2cm}\times
{(-)^{j}\Gamma(4-D+j+n_{_2})\Gamma(2-{D\over2}+j)
\over(j+n_{_2})!(j+n_{_1}+n_{_2})!n_{_2}!}
\nonumber\\
&&\hspace{4.2cm}\times
{\Gamma({D\over2}-1+n_{_1}+n_{_2})\Gamma(D-3+n_{_1}+n_{_2})
\over\Gamma({D\over2}-1+n_{_1})}
\nonumber\\
&&\hspace{4.2cm}
+{i(-)^{D/2}(p_{_3}^2)^{D/2-3}\over(4\pi)^{D/2}}
{\Gamma^2(2-{D\over2})\Gamma^2({D\over2}-1)\over\Gamma(D-3)\Gamma(4-D)}
\nonumber\\
&&\hspace{4.2cm}\times
\sum\limits_{j=1}^\infty\sum\limits_{n_{_1}=1}^\infty
\sum\limits_{n_{_2}=1}^\infty
\xi_{_{33}}^{j}x_{_{13}}^{D/2-2+n_{_1}}x_{_{23}}^{D/2-2+n_{_2}}
\Gamma(4-D+j)
\nonumber\\
&&\hspace{4.2cm}\times
{\Gamma({D\over2}-1+j+n_{_1}+n_{_2})\Gamma(D-3+j+n_{_1}+n_{_2})
\over j!(j+n_{_1})!(j+n_{_2})!
\Gamma({D\over2}-1+n_{_1})\Gamma({D\over2}-1+n_{_2})}
\nonumber\\
&&\hspace{4.2cm}
+{i(-)^{D/2}(p_{_3}^2)^{D/2-3}\over(4\pi)^{D/2}}
{\Gamma(2-{D\over2})\Gamma({D\over2}-1)\over\Gamma(D-3)\Gamma(4-D)}
\nonumber\\
&&\hspace{4.2cm}\times
\sum\limits_{j=1}^\infty\sum\limits_{n_{_1}=0}^\infty
\sum\limits_{n_{_2}=1}^\infty
\xi_{_{33}}^{j+n_{_1}}x_{_{13}}^{D/2-2-j}x_{_{23}}^{D/2-2+n_{_2}}
\nonumber\\
&&\hspace{4.2cm}\times
{(-)^{j}\Gamma(4-D+j+n_{_1})\Gamma(2-{D\over2}+j)
\over(j+n_{_1})!n_{_1}!(j+n_{_1}+n_{_2})!}
\nonumber\\
&&\hspace{4.2cm}\times
{\Gamma({D\over2}-1+n_{_1}+n_{_2})\Gamma(D-3+n_{_1}+n_{_2})
\over\Gamma({D\over2}-1+n_{_2})}
\nonumber\\
&&\hspace{4.2cm}
+{i(-)^{D/2}(p_{_3}^2)^{D/2-3}\over(4\pi)^{D/2}}
\Gamma(2-{D\over2})\Gamma({D\over2}-1)
\nonumber\\
&&\hspace{4.2cm}\times
\sum\limits_{j=1}^\infty\sum\limits_{n_{_1}=0}^\infty
\sum\limits_{n_{_2}=0}^\infty
\xi_{_{33}}^{j+n_{_1}+n_{_2}}x_{_{13}}^{D/2-2-j-n_{_2}}x_{_{23}}^{D/2-2-j-n_{_1}}
\nonumber\\
&&\hspace{4.2cm}\times
{(-)^{n_{_1}+n_{_2}}\Gamma(4-D+j+n_{_1}+n_{_2})
\Gamma(2-{D\over2}+j+n_{_1})
\over(j+n_{_1}+n_{_2})!n_{_1}!n_{_2}!\Gamma(2-{D\over2}+j)}
\nonumber\\
&&\hspace{4.2cm}\times
{\Gamma(2-{D\over2}+j+n_{_2})\over\Gamma(4-D+j)}
\nonumber\\
&&\hspace{4.2cm}
+{i(-)^{D/2}(p_{_3}^2)^{D/2-3}\over(4\pi)^{D/2}}
{1\over\Gamma(D-3)\Gamma(4-D)}
\nonumber\\
&&\hspace{4.2cm}\times
\sum\limits_{j=1}^\infty\sum\limits_{n_{_1}=1}^\infty
\sum\limits_{n_{_2}=1}^\infty
\xi_{_{33}}^{j+n_{_1}+n_{_2}}x_{_{13}}^{D/2-2-j}x_{_{23}}^{D/2-2-n_{_1}}
\nonumber\\
&&\hspace{4.2cm}\times
{(-)^{j+n_{_1}}
\Gamma({D\over2}-1+n_{_2})\Gamma(D-3+n_{_2})
\over(j+n_{_1}+n_{_2})!(n_{_1}+n_{_2})!(j+n_{_2})!}
\nonumber\\
&&\hspace{4.2cm}\times
\Gamma(4-D+j+n_{_1}+n_{_2})
\Gamma(2-{D\over2}+j)\Gamma(2-{D\over2}+n_{_1})
\nonumber\\
&&\hspace{4.2cm}
+{i(-)^{D/2}(p_{_3}^2)^{D/2-3}\over(4\pi)^{D/2}}
{\Gamma^2(2-{D\over2})\Gamma({D\over2}-1)\over\Gamma(D-3)\Gamma(4-D)}
\nonumber\\
&&\hspace{4.2cm}\times
\sum\limits_{j=1}^\infty\sum\limits_{n_{_1}=1}^\infty
\xi_{_{33}}^{j}x_{_{13}}^{D/2-2+n_{_1}}x_{_{23}}^{D/2-2}
{\Gamma(4-D+j)\over(j!)^2(j+n_{_1})!}
\nonumber\\
&&\hspace{4.2cm}\times
{\Gamma({D\over2}-1+j+n_{_1})\Gamma(D-3+j+n_{_1})
\over\Gamma({D\over2}-1+n_{_1})}
\nonumber\\
&&\hspace{4.2cm}
+{i(-)^{D/2}(p_{_3}^2)^{D/2-3}\over(4\pi)^{D/2}}
{\Gamma(2-{D\over2})\over\Gamma(D-3)\Gamma(4-D)}
\nonumber\\
&&\hspace{4.2cm}\times
\sum\limits_{n_{_1}=1}^\infty\sum\limits_{j=1}^\infty
\xi_{_{33}}^{j+n_{_1}}x_{_{13}}^{D/2-2-j}x_{_{23}}^{D/2-2}
{(-)^{j}\Gamma(4-D+j+n_{_1})\over((j+n_{_1})!)^2n_{_1}!}
\nonumber\\
&&\hspace{4.2cm}\times
\Gamma({D\over2}-1+n_{_1})\Gamma(D-3+n_{_1})\Gamma(2-{D\over2}+j)
\nonumber\\
&&\hspace{4.2cm}
+{i(-)^{D/2}(p_{_3}^2)^{D/2-3}\over(4\pi)^{D/2}}
{\Gamma^2(2-{D\over2})\Gamma({D\over2}-1)\over\Gamma(D-3)\Gamma(4-D)}
\nonumber\\
&&\hspace{4.2cm}\times
\sum\limits_{j=1}^\infty\sum\limits_{n_{_2}=1}^\infty
\xi_{_{33}}^{j}x_{_{13}}^{D/2-2}x_{_{23}}^{D/2-2+n_{_2}}
{\Gamma(4-D+j)\over (j!)^2(j+n_{_2})!}
\nonumber\\
&&\hspace{4.2cm}\times
{\Gamma({D\over2}-1+j+n_{_2})\Gamma(D-3+j+n_{_2})
\over\Gamma({D\over2}-1+n_{_2})}
\nonumber\\
&&\hspace{4.2cm}
+{i(-)^{D/2}(p_{_3}^2)^{D/2-3}\over(4\pi)^{D/2}}
{\Gamma(2-{D\over2})\over\Gamma(D-3)\Gamma(4-D)}
\sum\limits_{n_{_2}=1}^\infty\sum\limits_{j=1}^\infty
\nonumber\\
&&\hspace{4.2cm}\times
\xi_{_{33}}^{j+n_{_2}}x_{_{13}}^{D/2-2}x_{_{23}}^{D/2-2-j}
{(-)^{j}\Gamma(4-D+j+n_{_2})\over ((j+n_{_2})!)^2n_{_2}!}
\nonumber\\
&&\hspace{4.2cm}\times
\Gamma({D\over2}-1+n_{_2})\Gamma(D-3+n_{_2})\Gamma(2-{D\over2}+j)
\nonumber\\
&&\hspace{4.2cm}
+{i(-)^{D/2}(p_{_3}^2)^{D/2-3}\over(4\pi)^{D/2}}
{\Gamma({D\over2}-1)\over\Gamma(4-D)}
\sum\limits_{j=1}^\infty\sum\limits_{n_{_1}=0}^\infty
\nonumber\\
&&\hspace{4.2cm}\times
\xi_{_{33}}^{j+n_{_1}}x_{_{13}}^{D/2-2-j}x_{_{23}}^{D/2-2-n_{_1}}
{(-)^{j+n_{_1}}\Gamma(4-D+j+n_{_1})\over(j+n_{_1})!n_{_1}!j!}
\nonumber\\
&&\hspace{4.2cm}\times
\Gamma(2-{D\over2}+j)\Gamma(2-{D\over2}+n_{_1})
\nonumber\\
&&\hspace{4.2cm}
+{i(-)^{D/2}(p_{_3}^2)^{D/2-3}\over(4\pi)^{D/2}}
{\Gamma(2-{D\over2})\Gamma({D\over2}-1)\over\Gamma(4-D)}
\sum\limits_{j=1}^\infty \xi_{_{33}}^{j}x_{_{13}}^{D/2-2}
\nonumber\\
&&\hspace{4.2cm}\times
x_{_{23}}^{D/2-2-j}
{(-)^j\Gamma(4-D+j)\Gamma(2-{D\over2}+j)\over (j!)^2}\;.
\label{1M7-1}
\end{eqnarray}

\begin{eqnarray}
&&C_{_{0,8}}^a(p_{_1}^2,\;p_{_2}^2,\;p_{_3}^2,\;m^2)=
{i(-)^{D/2}(p_{_3}^2)^{D/2-3}\over(4\pi)^{D/2}}
{\Gamma^2(2-{D\over2})\Gamma^2({D\over2}-1)\over\Gamma(D-3)\Gamma(4-D)}
\sum\limits_{j=0}^\infty\sum\limits_{n_{_1}=0}^\infty\sum\limits_{n_{_2}=0}^\infty
\nonumber\\
&&\hspace{4.2cm}\times
{(-)^{j}\over j!n_{_1}!n_{_2}!}\xi_{_{33}}^{j}x_{_{13}}^{D/2-2-j+n_{_1}}
x_{_{23}}^{1-D/2-n_{_1}-n_{_2}}
\nonumber\\
&&\hspace{4.2cm}\times
{\Gamma(D-3-j+n_{_1}+n_{_2})\Gamma({D\over2}-1+n_{_1}+n_{_2})\Gamma(4-D+j)
\over\Gamma({D\over2}-1-j+n_{_1})\Gamma({D\over2}-1+n_{_2})}
\nonumber\\
&&\hspace{3.7cm}=
{i(-)^{D/2}(p_{_2}^2)^{D/2-3}\over(4\pi)^{D/2}}
{\Gamma^2(2-{D\over2})\Gamma^2({D\over2}-1)\over\Gamma(D-3)}
\nonumber\\
&&\hspace{4.2cm}\times
\sum\limits_{n_{_1}=0}^\infty\sum\limits_{n_{_2}=0}^\infty
x_{_{12}}^{D/2-2+n_{_1}}x_{_{32}}^{D/2-2+n_{_2}}
\nonumber\\
&&\hspace{4.2cm}\times
{\Gamma(D-3+n_{_1}+n_{_2})\Gamma({D\over2}-1+n_{_1}+n_{_2})
\over n_{_1}!n_{_2}!\Gamma({D\over2}-1+n_{_1})\Gamma({D\over2}-1+n_{_2})}
\nonumber\\
&&\hspace{4.2cm}
+{i(-)^{D/2}(p_{_2}^2)^{D/2-3}\over(4\pi)^{D/2}}
{\Gamma^2(2-{D\over2})\Gamma^2({D\over2}-1)\over\Gamma(D-3)\Gamma(4-D)}
\nonumber\\
&&\hspace{4.2cm}\times
\sum\limits_{j=1}^\infty\sum\limits_{n_{_1}=1}^\infty\sum\limits_{n_{_2}=0}^\infty
\xi_{_{32}}^{j}x_{_{12}}^{D/2-2+n_{_1}}x_{_{32}}^{D/2-2+n_{_2}}
\nonumber\\
&&\hspace{4.2cm}\times
{(-)^{j}\Gamma({D\over2}-1+j+n_{_1}+n_{_2})\Gamma(4-D+j)\over j!(j+n_{_1})!n_{_2}!
\Gamma({D\over2}-1+n_{_2})}
\nonumber\\
&&\hspace{4.2cm}\times
{\Gamma(D-3+n_{_1}+n_{_2})\over\Gamma({D\over2}-1+n_{_1})}
\nonumber\\
&&\hspace{4.2cm}
+{i(-)^{D/2}(p_{_2}^2)^{D/2-3}\over(4\pi)^{D/2}}
{\Gamma(2-{D\over2})\Gamma({D\over2}-1)\over\Gamma(D-3)\Gamma(4-D)}
\nonumber\\
&&\hspace{4.2cm}\times
\sum\limits_{j=1}^\infty\sum\limits_{n_{_1}=0}^\infty\sum\limits_{n_{_2}=1}^\infty
\xi_{_{32}}^{j+n_{_1}}x_{_{12}}^{D/2-2-j}x_{_{32}}^{D/2-1+j+n_{_2}}
\nonumber\\
&&\hspace{4.2cm}\times
{(-)^{n_{_1}}\Gamma({D\over2}-1+j+n_{_1}+n_{_2})\Gamma(4-D+j+n_{_1})
\over(j+n_{_1})!n_{_1}!(j+n_{_2})!\Gamma({D\over2}-1+j+n_{_2})}
\nonumber\\
&&\hspace{4.2cm}\times
\Gamma(D-3+n_{_2})\Gamma(2-{D\over2}+j)
\nonumber\\
&&\hspace{4.2cm}
+{i(-)^{D/2}(p_{_2}^2)^{D/2-3}\over(4\pi)^{D/2}}
\Gamma(2-{D\over2})\Gamma({D\over2}-1)
\nonumber\\
&&\hspace{4.2cm}\times
\sum\limits_{j=1}^\infty\sum\limits_{n_{_1}=0}^\infty\sum\limits_{n_{_2}=0}^\infty
\xi_{_{32}}^{j+n_{_1}+n_{_2}}x_{_{12}}^{D/2-2-j-n_{_2}}x_{_{32}}^{D/2-2+n_{_2}}
\nonumber\\
&&\hspace{4.2cm}\times
{(-)^{j+n_{_1}}\Gamma({D\over2}-1+n_{_1}+n_{_2})\Gamma(4-D+j+n_{_1}+n_{_2})
\over(j+n_{_1}+n_{_2})!n_{_1}!n_{_2}!\Gamma({D\over2}-1+n_{_2})}
\nonumber\\
&&\hspace{4.2cm}\times
{\Gamma(2-{D\over2}+j+n_{_2})\over\Gamma(4-D+j)}
\nonumber\\
&&\hspace{4.2cm}
+{i(-)^{D/2}(p_{_2}^2)^{D/2-3}\over(4\pi)^{D/2}}
{\Gamma(2-{D\over2})\Gamma({D\over2}-1)\over\Gamma(4-D)}
\nonumber\\
&&\hspace{4.2cm}\times
\sum\limits_{j=1}^\infty\sum\limits_{n_{_1}=0}^\infty
\xi_{_{32}}^{j+n_{_1}}x_{_{12}}^{D/2-2-j}x_{_{32}}^{D/2-2+j}
\nonumber\\
&&\hspace{4.2cm}\times
{(-)^{n_{_1}}\Gamma({D\over2}-1+j+n_{_1})\Gamma(4-D+j+n_{_1})
\over(j+n_{_1})!n_{_1}!j!\Gamma({D\over2}-1+j)}
\nonumber\\
&&\hspace{4.2cm}\times
\Gamma(2-{D\over2}+j)
\nonumber\\
&&\hspace{4.2cm}
+{i(-)^{D/2}(p_{_2}^2)^{D/2-3}\over(4\pi)^{D/2}}
{\Gamma^2(2-{D\over2})\Gamma({D\over2}-1)\over\Gamma(D-3)\Gamma(4-D)}
\nonumber\\
&&\hspace{4.2cm}\times
\sum\limits_{j=1}^\infty\sum\limits_{n_{_2}=0}^\infty
\xi_{_{32}}^{j}x_{_{12}}^{D/2-2}x_{_{32}}^{D/2-2+n_{_2}}
\nonumber\\
&&\hspace{4.2cm}\times
{(-)^{j}\Gamma({D\over2}-1+j+n_{_2})\Gamma(4-D+j)\Gamma(D-3+n_{_2})
\over (j!)^2n_{_2}!\Gamma({D\over2}-1+n_{_2})}\;.
\label{1M8-1}
\end{eqnarray}

\begin{eqnarray}
&&C_{_{0,9}}^a(p_{_1}^2,\;p_{_2}^2,\;p_{_3}^2,\;m^2)=
-{i(-)^{D/2}(p_{_3}^2)^{D/2-3}\over(4\pi)^{D/2}}
{\Gamma^2(2-{D\over2})\Gamma^2({D\over2}-1)\over\Gamma(D-3)\Gamma(4-D)}
\nonumber\\
&&\hspace{4.2cm}\times
\sum\limits_{n_{_2}=0}^\infty\sum\limits_{j=0}^\infty\sum\limits_{n_{_1}=0}^\infty
{(-)^{j}\over j!n_{_1}!n_{_2}!}\xi_{_{33}}^{j}x_{_{13}}^{D/2-2-j+n_{_1}}
x_{_{23}}^{-1-n_{_1}-n_{_2}}
\nonumber\\
&&\hspace{4.2cm}\times
{\Gamma(1+n_{_1}+n_{_2})\Gamma(4-D+j)\Gamma({D\over2}-1-j+n_{_1}+n_{_2})
\over\Gamma(3-{D\over2}+n_{_2})\Gamma({D\over2}-1-j+n_{_1})}
\nonumber\\
&&\hspace{3.7cm}=
-{i(-)^{D/2}(p_{_2}^2)^{D/2-3}\over(4\pi)^{D/2}}
{\Gamma^2(2-{D\over2})\Gamma^2({D\over2}-1)\over\Gamma(D-3)\Gamma(3-{D\over2})}
\sum\limits_{n_{_1}=0}^\infty x_{_{12}}^{D/2-2+n_{_1}}
\nonumber\\
&&\hspace{4.2cm}
-{i(-)^{D/2}(p_{_2}^2)^{D/2-3}\over(4\pi)^{D/2}}
{\Gamma^2(2-{D\over2})\Gamma^2({D\over2}-1)\over\Gamma(D-3)\Gamma(4-D)\Gamma(3-{D\over2})}
\nonumber\\
&&\hspace{4.2cm}\times
\sum\limits_{j=1}^\infty\sum\limits_{n_{_1}=0}^\infty
\xi_{_{32}}^{j}x_{_{12}}^{D/2-2-j+n_{_1}}
{(-)^{j}\Gamma(4-D+j)\over j!}
\nonumber\\
&&\hspace{4.2cm}
-{i(-)^{D/2}(p_{_2}^2)^{D/2-3}\over(4\pi)^{D/2}}
{\Gamma^2(2-{D\over2})\Gamma^2({D\over2}-1)\over\Gamma(D-3)\Gamma(4-D)}
\nonumber\\
&&\hspace{4.2cm}\times
\sum\limits_{n_{_2}=1}^\infty\sum\limits_{n_{_1}=0}^\infty
\sum\limits_{j=1}^\infty
\xi_{_{32}}^{j+n_{_1}+n_{_2}}x_{_{12}}^{D/2-2-j-n_{_2}}x_{_{32}}^{n_{_2}}
\nonumber\\
&&\hspace{4.2cm}\times
{(-)^{j+n_{_1}}\Gamma(1+n_{_1}+n_{_2})\Gamma(4-D+j+n_{_1}+n_{_2})
\over(j+n_{_1}+n_{_2})!n_{_1}!n_{_2}!\Gamma(3-{D\over2}+n_{_2})}
\nonumber\\
&&\hspace{4.2cm}\times
{\Gamma(2-{D\over2}+j+n_{_2})\over\Gamma(2-{D\over2}+j)}
\nonumber\\
&&\hspace{4.2cm}
-{i(-)^{D/2}(p_{_2}^2)^{D/2-3}\over(4\pi)^{D/2}}
{\Gamma(2-{D\over2})\Gamma({D\over2}-1)\over\Gamma(D-3)\Gamma(4-D)}
\nonumber\\
&&\hspace{4.2cm}\times
\sum\limits_{n_{_2}=1}^\infty\sum\limits_{n_{_1}=0}^\infty
\sum\limits_{j=1}^\infty
\xi_{_{32}}^{j+n_{_1}}x_{_{12}}^{D/2-2-j}x_{_{32}}^{j+n_{_2}}
\nonumber\\
&&\hspace{4.2cm}\times
{(-)^{n_{_1}}\Gamma(1+j+n_{_1}+n_{_2})\Gamma(4-D+j+n_{_1})
\over(j+n_{_1})!n_{_1}!(j+n_{_2})!\Gamma(3-{D\over2}+j+n_{_2})}
\nonumber\\
&&\hspace{4.2cm}\times
\Gamma({D\over2}-1+n_{_2})\Gamma(2-{D\over2}+j)
\nonumber\\
&&\hspace{4.2cm}
-{i(-)^{D/2}(p_{_2}^2)^{D/2-3}\over(4\pi)^{D/2}}
{\Gamma^2(2-{D\over2})\Gamma^2({D\over2}-1)\over\Gamma(D-3)\Gamma(4-D)}
\nonumber\\
&&\hspace{4.2cm}\times
\sum\limits_{n_{_2}=1}^\infty\sum\limits_{j=1}^\infty
\sum\limits_{n_{_1}=1}^\infty \xi_{_{32}}^{j}x_{_{12}}^{D/2-2+n_{_1}}
x_{_{32}}^{n_{_2}}
\nonumber\\
&&\hspace{4.2cm}\times
{(-)^{j}\Gamma(1+j+n_{_1}+n_{_2})\Gamma(4-D+j)\Gamma({D\over2}-1+n_{_1}+n_{_2})
\over j!(j+n_{_1})!n_{_2}!\Gamma(3-{D\over2}+n_{_2})\Gamma({D\over2}-1+n_{_1})}
\nonumber\\
&&\hspace{4.2cm}
-{i(-)^{D/2}(p_{_2}^2)^{D/2-3}\over(4\pi)^{D/2}}
{\Gamma^2(2-{D\over2})\Gamma^2({D\over2}-1)\over\Gamma(D-3)}
\nonumber\\
&&\hspace{4.2cm}\times
\sum\limits_{n_{_2}=1}^\infty\sum\limits_{n_{_1}=1}^\infty
x_{_{12}}^{D/2-2+n_{_1}}x_{_{32}}^{n_{_2}}
\nonumber\\
&&\hspace{4.2cm}\times
{\Gamma(1+n_{_1}+n_{_2})\Gamma({D\over2}-1+n_{_1}+n_{_2})
\over n_{_1}!n_{_2}!\Gamma(3-{D\over2}+n_{_2})\Gamma({D\over2}-1+n_{_1})}
\nonumber\\
&&\hspace{4.2cm}
-{i(-)^{D/2}(p_{_2}^2)^{D/2-3}\over(4\pi)^{D/2}}
{\Gamma^2(2-{D\over2})\Gamma({D\over2}-1)\over\Gamma(D-3)\Gamma(4-D)}
\nonumber\\
&&\hspace{4.2cm}\times
\sum\limits_{n_{_2}=1}^\infty\sum\limits_{n_{_1}=1}^\infty
\xi_{_{32}}^{n_{_1}}x_{_{12}}^{D/2-2}x_{_{32}}^{n_{_2}}
\nonumber\\
&&\hspace{4.2cm}\times
{(-)^{n_{_1}}\Gamma(1+n_{_1}+n_{_2})\Gamma(4-D+n_{_1})\Gamma({D\over2}-1+n_{_2})
\over(n_{_1}!)^2n_{_2}!\Gamma(3-{D\over2}+n_{_2})}
\nonumber\\
&&\hspace{4.2cm}
-{i(-)^{D/2}(p_{_2}^2)^{D/2-3}\over(4\pi)^{D/2}}
{\Gamma^2(2-{D\over2})\Gamma({D\over2}-1)\over\Gamma(D-3)}
\nonumber\\
&&\hspace{4.2cm}\times
\sum\limits_{n_{_2}=1}^\infty x_{_{12}}^{D/2-2}x_{_{32}}^{n_{_2}}
{\Gamma({D\over2}-1+n_{_2})\over\Gamma(3-{D\over2}+n_{_2})}
\nonumber\\
&&\hspace{4.2cm}
-{i(-)^{D/2}(p_{_2}^2)^{D/2-3}\over(4\pi)^{D/2}}
{\Gamma(2-{D\over2})\Gamma^2({D\over2}-1)\over\Gamma(D-3)\Gamma(4-D)}
\nonumber\\
&&\hspace{4.2cm}\times
\sum\limits_{n_{_2}=1}^\infty\sum\limits_{n_{_1}=0}^\infty
\xi_{_{32}}^{n_{_1}+n_{_2}}x_{_{12}}^{D/2-2-n_{_2}}x_{_{32}}^{n_{_2}}
\nonumber\\
&&\hspace{4.2cm}\times
{(-)^{n_{_1}}\Gamma(4-D+n_{_1}+n_{_2})\Gamma(2-{D\over2}+n_{_2})
\over n_{_1}!n_{_2}!\Gamma(3-{D\over2}+n_{_2})}\;.
\label{1M9-1}
\end{eqnarray}

\begin{eqnarray}
&&C_{_{0,2}}^b(p_{_1}^2,\;p_{_2}^2,\;p_{_3}^2,\;m^2)=
{i(-)^{D/2-3}(p_{_3}^2)^{D/2-3}\over(4\pi)^{D/2}}
{\Gamma^2(2-{D\over2})\Gamma^2({D\over2}-1)\over\Gamma(D-3)\Gamma(4-D)}x_{_{23}}^{D/2-2}
\nonumber\\
&&\hspace{4.0cm}\times
\sum\limits_{j=0}^\infty\sum\limits_{n_{_1}=0}^\infty\sum\limits_{n_{_2}=0}^\infty
\xi_{_{m3}}^jx_{_{13}}^{n_{_1}}x_{_{23}}^{-j+n_{_2}}\Gamma(4-D+2j)
\nonumber\\
&&\hspace{4.0cm}\times
{\Gamma(1+n_{_1}+n_{_2})\Gamma({D\over2}-1-j+n_{_1}+n_{_2})
\over j!n_{_1}!n_{_2}!\Gamma(3-{D\over2}+j+n_{_1})\Gamma({D\over2}-1-j+n_{_2})}
\nonumber\\
&&\hspace{3.5cm}=
{i(-)^{D/2-3}(p_{_3}^2)^{D/2-3}\over(4\pi)^{D/2}}
{\Gamma^2(2-{D\over2})\Gamma^2({D\over2}-1)\over\Gamma(D-3)}x_{_{23}}^{D/2-2}
\nonumber\\
&&\hspace{4.0cm}\times
\sum\limits_{n_{_1}=0}^\infty\sum\limits_{n_{_2}=0}^\infty
x_{_{13}}^{n_{_1}}x_{_{23}}^{n_{_2}}
{\Gamma(1+n_{_1}+n_{_2})\Gamma({D\over2}-1+n_{_1}+n_{_2})
\over n_{_1}!n_{_2}!\Gamma(3-{D\over2}+n_{_1})\Gamma({D\over2}-1+n_{_2})}
\nonumber\\
&&\hspace{4.0cm}
+{i(-)^{D/2-3}(p_{_3}^2)^{D/2-3}\over(4\pi)^{D/2}}
{\Gamma^2(2-{D\over2})\Gamma^2({D\over2}-1)\over\Gamma(D-3)\Gamma(4-D)}x_{_{23}}^{D/2-2}
\nonumber\\
&&\hspace{4.0cm}\times
\sum\limits_{j=1}^\infty\sum\limits_{n_{_1}=0}^\infty\sum\limits_{n_{_2}=1}^\infty
\xi_{_{m3}}^jx_{_{13}}^{n_{_1}}x_{_{23}}^{n_{_2}}
\nonumber\\
&&\hspace{4.0cm}\times
{\Gamma(1+j+n_{_1}+n_{_2})\Gamma(4-D+2j)\Gamma({D\over2}-1+n_{_1}+n_{_2})
\over j!n_{_1}!(j+n_{_2})!\Gamma(3-{D\over2}+j+n_{_1})\Gamma({D\over2}-1+n_{_2})}
\nonumber\\
&&\hspace{4.0cm}
+{i(-)^{D/2-3}(p_{_3}^2)^{D/2-3}\over(4\pi)^{D/2}}
{\Gamma(2-{D\over2})\Gamma({D\over2}-1)\over\Gamma(D-3)\Gamma(4-D)}x_{_{23}}^{D/2-2}
\nonumber\\
&&\hspace{4.0cm}\times
\sum\limits_{j=1}^\infty\sum\limits_{n_{_2}=0}^\infty\sum\limits_{n_{_1}=1}^\infty
\xi_{_{m3}}^{j+n_{_2}}x_{_{13}}^{j+n_{_1}}x_{_{23}}^{-j}\Gamma(2-{D\over2}+j)
\Gamma({D\over2}-1+n_{_1})
\nonumber\\
&&\hspace{4.0cm}\times
{(-)^{j}\Gamma(1+j+n_{_1}+n_{_2})\Gamma(4-D+2j+2n_{_2})
\over (j+n_{_2})!(j+n_{_1})!n_{_2}!\Gamma(3-{D\over2}+2j+n_{_1}+n_{_2})}
\nonumber\\
&&\hspace{4.0cm}
+{i(-)^{D/2-3}(p_{_3}^2)^{D/2-3}\over(4\pi)^{D/2}}
{\Gamma^2(2-{D\over2})\Gamma^2({D\over2}-1)\over\Gamma(D-3)\Gamma(4-D)}x_{_{23}}^{D/2-2}
\nonumber\\
&&\hspace{4.0cm}\times
\sum\limits_{j=1}^\infty\sum\limits_{n_{_1}=0}^\infty
\sum\limits_{n_{_2}=0}^\infty
\xi_{_{m3}}^{j+n_{_1}+n_{_2}}x_{_{13}}^{n_{_1}}x_{_{23}}^{-j-n_{_1}}
{\Gamma(2-{D\over2}+j+n_{_1})\over\Gamma(2-{D\over2}+j)}
\nonumber\\
&&\hspace{4.0cm}\times
{(-)^{n_{_1}}\Gamma(1+n_{_1}+n_{_2})\Gamma(4-D+2j+2n_{_1}+2n_{_2})
\over(j+n_{_1}+n_{_2})!n_{_1}!n_{_2}!\Gamma(3-{D\over2}+j+2n_{_1}+n_{_2})}
\nonumber\\
&&\hspace{4.0cm}
+{i(-)^{D/2-3}(p_{_3}^2)^{D/2-3}\over(4\pi)^{D/2}}
{\Gamma(2-{D\over2})\Gamma^2({D\over2}-1)\over\Gamma(D-3)\Gamma(4-D)}x_{_{23}}^{D/2-2}
\nonumber\\
&&\hspace{4.0cm}\times
\sum\limits_{j=0}^\infty\sum\limits_{n_{_1}=1}^\infty
\xi_{_{m3}}^{j+n_{_1}}x_{_{13}}^{n_{_1}}x_{_{23}}^{-n_{_1}}
\nonumber\\
&&\hspace{4.0cm}\times
{(-)^{n_{_1}}\Gamma(4-D+2j+2n_{_1})\Gamma(2-{D\over2}+n_{_1})
\over n_{_1}!j!\Gamma(3-{D\over2}+j+2n_{_1})}
\nonumber\\
&&\hspace{4.0cm}
+{i(-)^{D/2-3}(p_{_3}^2)^{D/2-3}\over(4\pi)^{D/2}}
{\Gamma^2(2-{D\over2})\Gamma({D\over2}-1)\over\Gamma(D-3)\Gamma(4-D)}x_{_{23}}^{D/2-2}
\nonumber\\
&&\hspace{4.0cm}\times
\sum\limits_{j=1}^\infty\sum\limits_{n_{_1}=1}^\infty
\xi_{_{m3}}^jx_{_{13}}^{n_{_1}}
{\Gamma(1+j+n_{_1})\Gamma(4-D+2j)
\over (j!)^2n_{_1}!\Gamma(3-{D\over2}+j+n_{_1})}
\nonumber\\
&&\hspace{4.0cm}\times
\Gamma({D\over2}-1+n_{_1})
\nonumber\\
&&\hspace{4.0cm}
+{i(-)^{D/2-3}(p_{_3}^2)^{D/2-3}\over(4\pi)^{D/2}}
{\Gamma^2(2-{D\over2})\Gamma^2({D\over2}-1)\over\Gamma(D-3)\Gamma(4-D)}x_{_{23}}^{D/2-2}
\nonumber\\
&&\hspace{4.0cm}\times
\sum\limits_{j=1}^\infty\xi_{_{m3}}^j
{\Gamma(4-D+2j)\over j!\Gamma(3-{D\over2}+j)}\;.
\label{3M2-1}
\end{eqnarray}

\begin{eqnarray}
&&C_{_{0,6}}^b(p_{_1}^2,\;p_{_2}^2,\;p_{_3}^2,\;m^2)=
{i(-)^{D/2-3}(p_{_3}^2)^{D/2-3}\over(4\pi)^{D/2}}
{\Gamma^2(2-{D\over2})\Gamma^2({D\over2}-1)\over\Gamma(D-3)\Gamma(4-D)}x_{_{13}}^{D/2-2}
\nonumber\\
&&\hspace{4.0cm}\times
\sum\limits_{j=0}^\infty\sum\limits_{n_{_1}=0}^\infty\sum\limits_{n_{_2}=0}^\infty
\xi_{_{m3}}^jx_{_{13}}^{-j+n_{_1}}x_{_{23}}^{n_{_2}}
\nonumber\\
&&\hspace{4.0cm}\times
{\Gamma(1+n_{_1}+n_{_2})\Gamma(4-D+2j)\Gamma({D\over2}-1-j+n_{_1}+n_{_2})
\over j!n_{_1}!n_{_2}!\Gamma(3-{D\over2}+j+n_{_2})\Gamma({D\over2}-1-j+n_{_1})}
\nonumber\\
&&\hspace{3.5cm}=
{i(-)^{D/2-3}(p_{_3}^2)^{D/2-3}\over(4\pi)^{D/2}}
{\Gamma^2(2-{D\over2})\Gamma^2({D\over2}-1)\over\Gamma(D-3)}x_{_{13}}^{D/2-2}
\nonumber\\
&&\hspace{4.0cm}\times
\sum\limits_{n_{_1}=0}^\infty\sum\limits_{n_{_2}=0}^\infty
x_{_{13}}^{n_{_1}}x_{_{23}}^{n_{_2}}
{\Gamma(1+n_{_1}+n_{_2})\Gamma({D\over2}-1+n_{_1}+n_{_2})
\over n_{_1}!n_{_2}!\Gamma({D\over2}-1+n_{_1})\Gamma(3-{D\over2}+n_{_2})}
\nonumber\\
&&\hspace{4.0cm}
+{i(-)^{D/2-3}(p_{_3}^2)^{D/2-3}\over(4\pi)^{D/2}}
{\Gamma^2(2-{D\over2})\Gamma^2({D\over2}-1)\over\Gamma(D-3)\Gamma(4-D)}x_{_{13}}^{D/2-2}
\nonumber\\
&&\hspace{4.0cm}\times
\sum\limits_{j=1}^\infty\sum\limits_{n_{_2}=0}^\infty\sum\limits_{n_{_1}=1}^\infty
\xi_{_{m3}}^jx_{_{13}}^{n_{_1}}x_{_{23}}^{n_{_2}}
\nonumber\\
&&\hspace{4.0cm}\times
{\Gamma(1+j+n_{_1}+n_{_2})\Gamma(4-D+2j)\Gamma({D\over2}-1+n_{_1}+n_{_2})
\over j!n_{_2}!(j+n_{_1})!\Gamma(3-{D\over2}+j+n_{_2})\Gamma({D\over2}-1+n_{_1})}
\nonumber\\
&&\hspace{4.0cm}
+{i(-)^{D/2-3}(p_{_3}^2)^{D/2-3}\over(4\pi)^{D/2}}
{\Gamma(2-{D\over2})\Gamma({D\over2}-1)\over\Gamma(D-3)\Gamma(4-D)}x_{_{13}}^{D/2-2}
\nonumber\\
&&\hspace{4.0cm}\times
\sum\limits_{j=1}^\infty\sum\limits_{n_{_1}=0}^\infty\sum\limits_{n_{_2}=1}^\infty
\xi_{_{m3}}^{j+n_{_1}}x_{_{13}}^{-j}x_{_{23}}^{j+n_{_2}}\Gamma(2-{D\over2}+j)
\Gamma({D\over2}-1+n_{_2})
\nonumber\\
&&\hspace{4.0cm}\times
{(-)^{j}\Gamma(1+j+n_{_1}+n_{_2})\Gamma(4-D+2j+2n_{_1})
\over (j+n_{_1})!(j+n_{_2})!n_{_1}!\Gamma(3-{D\over2}+2j+n_{_1}+n_{_2})}
\nonumber\\
&&\hspace{4.0cm}
+{i(-)^{D/2-3}(p_{_3}^2)^{D/2-3}\over(4\pi)^{D/2}}
{\Gamma^2(2-{D\over2})\Gamma^2({D\over2}-1)\over\Gamma(D-3)\Gamma(4-D)}x_{_{13}}^{D/2-2}
\nonumber\\
&&\hspace{4.0cm}\times
\sum\limits_{j=1}^\infty\sum\limits_{n_{_1}=0}^\infty\sum\limits_{n_{_2}=0}^\infty
\xi_{_{m3}}^{j+n_{_1}+n_{_2}}x_{_{13}}^{-j-n_{_2}}x_{_{23}}^{n_{_2}}
{\Gamma(2-{D\over2}+j+n_{_2})\over\Gamma(2-{D\over2}+j)}
\nonumber\\
&&\hspace{4.0cm}\times
{(-)^{n_{_2}}\Gamma(1+n_{_1}+n_{_2})\Gamma(4-D+2j+2n_{_1}+2n_{_2})
\over(j+n_{_1}+n_{_2})!n_{_1}!n_{_2}!\Gamma(3-{D\over2}+j+n_{_1}+2n_{_2})}
\nonumber\\
&&\hspace{4.0cm}
+{i(-)^{D/2-3}(p_{_3}^2)^{D/2-3}\over(4\pi)^{D/2}}
{\Gamma(2-{D\over2})\Gamma^2({D\over2}-1)\over\Gamma(D-3)\Gamma(4-D)}x_{_{13}}^{D/2-2}
\nonumber\\
&&\hspace{4.0cm}\times
\sum\limits_{j=0}^\infty\sum\limits_{n_{_2}=1}^\infty
\xi_{_{m3}}^{j+n_{_2}}x_{_{13}}^{-n_{_2}}x_{_{23}}^{n_{_2}}
\nonumber\\
&&\hspace{4.0cm}\times
{(-)^{n_{_2}}\Gamma(4-D+2j+2n_{_2})\Gamma(2-{D\over2}+n_{_2})
\over n_{_2}!j!\Gamma(3-{D\over2}+j+2n_{_2})}
\nonumber\\
&&\hspace{4.0cm}
+{i(-)^{D/2-3}(p_{_3}^2)^{D/2-3}\over(4\pi)^{D/2}}
{\Gamma^2(2-{D\over2})\Gamma({D\over2}-1)\over\Gamma(D-3)\Gamma(4-D)}x_{_{13}}^{D/2-2}
\nonumber\\
&&\hspace{4.0cm}\times
\sum\limits_{j=1}^\infty\sum\limits_{n_{_2}=1}^\infty
\xi_{_{m3}}^jx_{_{23}}^{n_{_2}}
\nonumber\\
&&\hspace{4.0cm}\times
{\Gamma(1+j+n_{_2})\Gamma(4-D+2j)\Gamma({D\over2}-1+n_{_2})
\over (j!)^2n_{_2}!\Gamma(3-{D\over2}+j+n_{_2})}
\nonumber\\
&&\hspace{4.0cm}
+{i(-)^{D/2-3}(p_{_3}^2)^{D/2-3}\over(4\pi)^{D/2}}
{\Gamma^2(2-{D\over2})\Gamma^2({D\over2}-1)\over\Gamma(D-3)\Gamma(4-D)}x_{_{13}}^{D/2-2}
\nonumber\\
&&\hspace{4.0cm}\times
\sum\limits_{j=1}^\infty \xi_{_{m3}}^j
{\Gamma(4-D+2j)\over j!\Gamma(3-{D\over2}+j)}\;.
\label{3M6-1}
\end{eqnarray}

\begin{eqnarray}
&&C_{_{0,7}}^b(p_{_1}^2,\;p_{_2}^2,\;p_{_3}^2,\;m^2)=
-{i(-)^{D/2-3}(p_{_3}^2)^{D/2-3}\over(4\pi)^{D/2}}
{\Gamma^2(2-{D\over2})\Gamma^2({D\over2}-1)\over\Gamma(D-3)\Gamma(4-D)}
(x_{_{13}}x_{_{23}})^{D/2-2}
\nonumber\\
&&\hspace{4.0cm}\times
\sum\limits_{j=0}^\infty\sum\limits_{n_{_1}=0}^\infty\sum\limits_{n_{_2}=0}^\infty
\xi_{_{m3}}^jx_{_{13}}^{-j+n_{_1}}x_{_{23}}^{-j+n_{_2}}
{\Gamma(4-D+2j)\over j!n_{_1}!n_{_2}!}
\nonumber\\
&&\hspace{4.0cm}\times
{\Gamma({D\over2}-1-j+n_{_1}+n_{_2})\Gamma(D-3-2j+n_{_1}+n_{_2})\over
\Gamma({D\over2}-1-j+n_{_1})\Gamma({D\over2}-1-j+n_{_2})}
\nonumber\\
&&\hspace{3.5cm}=
-{i(-)^{D/2-3}(p_{_3}^2)^{D/2-3}\over(4\pi)^{D/2}}
{\Gamma^2(2-{D\over2})\Gamma^2({D\over2}-1)\over\Gamma(D-3)}(x_{_{13}}x_{_{23}})^{D/2-2}
\nonumber\\
&&\hspace{4.0cm}\times
\sum\limits_{n_{_1}=0}^\infty\sum\limits_{n_{_2}=0}^\infty
x_{_{13}}^{n_{_1}}x_{_{23}}^{n_{_2}}
{\Gamma({D\over2}-1+n_{_1}+n_{_2})\Gamma(D-3+n_{_1}+n_{_2})\over n_{_1}!n_{_2}!
\Gamma({D\over2}-1+n_{_1})\Gamma({D\over2}-1+n_{_2})}
\nonumber\\
&&\hspace{4.0cm}
-{i(-)^{D/2-3}(p_{_3}^2)^{D/2-3}\over(4\pi)^{D/2}}
{\Gamma(2-{D\over2})\Gamma({D\over2}-1)\over\Gamma(D-3)\Gamma(4-D)}(x_{_{13}}x_{_{23}})^{D/2-2}
\nonumber\\
&&\hspace{4.0cm}\times
\sum\limits_{j=1}^\infty\sum\limits_{n_{_2}=0}^\infty\sum\limits_{n_{_1}=1}^\infty
\xi_{_{m3}}^{j+n_{_2}}x_{_{13}}^{j+n_{_1}}x_{_{23}}^{-j}
{(-)^{j}\Gamma(4-D+2j+2n_{_2})\over(j+n_{_2})!(2j+n_{_1}+n_{_2})!n_{_2}!}
\nonumber\\
&&\hspace{4.0cm}\times
{\Gamma({D\over2}-1+j+n_{_1}+n_{_2})\Gamma(D-3+n_{_1})
\Gamma(2-{D\over2}+j)\over\Gamma({D\over2}-1+j+n_{_1})}
\nonumber\\
&&\hspace{4.0cm}
-{i(-)^{D/2-3}(p_{_3}^2)^{D/2-3}\over(4\pi)^{D/2}}
{\Gamma^2(2-{D\over2})\Gamma^2({D\over2}-1)\over\Gamma(D-3)\Gamma(4-D)}(x_{_{13}}x_{_{23}})^{D/2-2}
\nonumber\\
&&\hspace{4.0cm}\times
\sum\limits_{j=1}^\infty\sum\limits_{n_{_1}=1}^\infty\sum\limits_{n_{_2}=1}^\infty
\xi_{_{m3}}^jx_{_{13}}^{n_{_1}}x_{_{23}}^{n_{_2}}\Gamma(4-D+2j)
\nonumber\\
&&\hspace{4.0cm}\times
{\Gamma({D\over2}-1+j+n_{_1}+n_{_2})\Gamma(D-3+n_{_1}+n_{_2})
\over j!(j+n_{_1})!(j+n_{_2})!\Gamma({D\over2}-1+n_{_1})\Gamma({D\over2}-1+n_{_2})}
\nonumber\\
&&\hspace{4.0cm}
-{i(-)^{D/2-3}(p_{_3}^2)^{D/2-3}\over(4\pi)^{D/2}}
{\Gamma(2-{D\over2})\Gamma({D\over2}-1)\over\Gamma(D-3)\Gamma(4-D)}(x_{_{13}}x_{_{23}})^{D/2-2}
\nonumber\\
&&\hspace{4.0cm}\times
\sum\limits_{j=1}^\infty\sum\limits_{n_{_1}=0}^\infty\sum\limits_{n_{_2}=1}^\infty
\xi_{_{m3}}^{j+n_{_1}}x_{_{13}}^{-j}x_{_{23}}^{j+n_{_2}}
{(-)^{j}\Gamma(4-D+2j+2n_{_1})\over(j+n_{_1})!n_{_1}!(2j+n_{_1}+n_{_2})!}
\nonumber\\
&&\hspace{4.0cm}\times
{\Gamma({D\over2}-1+j+n_{_1}+n_{_2})\Gamma(D-3+n_{_2})
\Gamma(2-{D\over2}+j)\over\Gamma({D\over2}-1+j+n_{_2})}
\nonumber\\
&&\hspace{4.0cm}
-{i(-)^{D/2-3}(p_{_3}^2)^{D/2-3}\over(4\pi)^{D/2}}
\Gamma(2-{D\over2})\Gamma({D\over2}-1)(x_{_{13}}x_{_{23}})^{D/2-2}
\nonumber\\
&&\hspace{4.0cm}\times
\sum\limits_{j=1}^\infty\sum\limits_{n_{_1}=0}^\infty\sum\limits_{n_{_2}=1}^\infty
\xi_{_{m3}}^{j+n_{_1}+n_{_2}}x_{_{13}}^{-j-n_{_2}}x_{_{23}}^{n_{_2}}
{\Gamma({D\over2}-1+n_{_1}+n_{_2})\over\Gamma(4-D+j)}
\nonumber\\
&&\hspace{4.0cm}\times
{(-)^{n_{_2}}\Gamma(4-D+2j+2n_{_1}+2n_{_2})\Gamma(2-{D\over2}+j+n_{_2})\over
(j+n_{_1}+n_{_2})!n_{_1}!(j+n_{_1}+2n_{_2})!\Gamma({D\over2}-1+n_{_2})}
\nonumber\\
&&\hspace{4.0cm}
-{i(-)^{D/2-3}(p_{_3}^2)^{D/2-3}\over(4\pi)^{D/2}}
\Gamma(2-{D\over2})\Gamma({D\over2}-1)(x_{_{13}}x_{_{23}})^{D/2-2}
\nonumber\\
&&\hspace{4.0cm}\times
\sum\limits_{j=1}^\infty\sum\limits_{n_{_1}=0}^\infty\sum\limits_{n_{_2}=0}^\infty
\xi_{_{m3}}^{j+n_{_1}+n_{_2}}x_{_{13}}^{-j-n_{_2}}x_{_{23}}^{-j-n_{_1}}
{\Gamma(2-{D\over2}+j+n_{_1})\over\Gamma(2-{D\over2}+j)}
\nonumber\\
&&\hspace{4.0cm}\times
{(-)^{j}\Gamma(4-D+2j+2n_{_1}+2n_{_2})\Gamma(2-{D\over2}+j+n_{_2})\over
(j+n_{_1}+n_{_2})!n_{_1}!n_{_2}!\Gamma(4-D+2j+n_{_1}+n_{_2})}
\nonumber\\
&&\hspace{4.0cm}
-{i(-)^{D/2-3}(p_{_3}^2)^{D/2-3}\over(4\pi)^{D/2}}(x_{_{13}}x_{_{23}})^{D/2-2}
\nonumber\\
&&\hspace{4.0cm}\times
\sum\limits_{j=1}^\infty\sum\limits_{n_{_1}=1}^\infty\sum\limits_{n_{_2}=1}^\infty
\xi_{_{m3}}^{j+n_{_1}+n_{_2}}x_{_{13}}^{-j}x_{_{23}}^{-n_{_1}}
{\Gamma(2-{D\over2}+j)\over\Gamma(4-D+j+n_{_1})}
\nonumber\\
&&\hspace{4.0cm}\times
{\Gamma(4-D+2j+2n_{_1}+2n_{_2})\Gamma({D\over2}-1+n_{_2})
\Gamma(2-{D\over2}+n_{_1})\over(j+n_{_1}+n_{_2})!(n_{_1}+n_{_2})!(j+n_{_2})!}
\nonumber\\
&&\hspace{4.0cm}
-{i(-)^{D/2-3}(p_{_3}^2)^{D/2-3}\over(4\pi)^{D/2}}
\Gamma(2-{D\over2})\Gamma({D\over2}-1)(x_{_{13}}x_{_{23}})^{D/2-2}
\nonumber\\
&&\hspace{4.0cm}\times
\sum\limits_{j=1}^\infty\sum\limits_{n_{_2}=0}^\infty\sum\limits_{n_{_1}=1}^\infty
\xi_{_{m3}}^{j+n_{_1}+n_{_2}}x_{_{13}}^{n_{_1}}x_{_{23}}^{-j-n_{_1}}
{\Gamma(2-{D\over2}+j+n_{_1})\over\Gamma(4-D+j)}
\nonumber\\
&&\hspace{4.0cm}\times
{(-)^{n_{_1}}\Gamma(4-D+2j+2n_{_1}+2n_{_2})\Gamma({D\over2}-1+n_{_1}+n_{_2})\over
(j+n_{_1}+n_{_2})!(j+2n_{_1}+n_{_2})!n_{_2}!\Gamma({D\over2}-1+n_{_1})}
\nonumber\\
&&\hspace{4.0cm}
-{i(-)^{D/2-3}(p_{_3}^2)^{D/2-3}\over(4\pi)^{D/2}}
\Gamma({D\over2}-1)(x_{_{13}}x_{_{23}})^{D/2-2}
\nonumber\\
&&\hspace{4.0cm}\times
\sum\limits_{j=1}^\infty\sum\limits_{n_{_1}=0}^\infty
\xi_{_{m3}}^{j+n_{_1}}x_{_{13}}^{-j}x_{_{23}}^{-n_{_1}}
{\Gamma(2-{D\over2}+j)\Gamma(2-{D\over2}+n_{_1})
\over j!n_{_1}!(j+n_{_1})!}
\nonumber\\
&&\hspace{4.0cm}\times
{\Gamma(4-D+2j+2n_{_1})\over\Gamma(4-D+j+n_{_1})}
\nonumber\\
&&\hspace{4.0cm}
-{i(-)^{D/2-3}(p_{_3}^2)^{D/2-3}\over(4\pi)^{D/2}}
{\Gamma(2-{D\over2})\Gamma({D\over2}-1)\over\Gamma(4-D)}(x_{_{13}}x_{_{23}})^{D/2-2}
\nonumber\\
&&\hspace{4.0cm}\times
\sum\limits_{j=1}^\infty\sum\limits_{n_{_1}=0}^\infty
\xi_{_{m3}}^{j+n_{_1}}x_{_{13}}^{-j}x_{_{23}}^{j}\Gamma(4-D+2j+2n_{_1})
\nonumber\\
&&\hspace{4.0cm}\times
{(-)^{j}\Gamma({D\over2}-1+j+n_{_1})\Gamma(2-{D\over2}+j)
\over(j+n_{_1})!n_{_1}!\Gamma(1+2j+n_{_1})\Gamma({D\over2}-1+j)}
\nonumber\\
&&\hspace{4.0cm}
-{i(-)^{D/2-3}(p_{_3}^2)^{D/2-3}\over(4\pi)^{D/2}}
{\Gamma(2-{D\over2})\Gamma({D\over2}-1)\over\Gamma(4-D)}(x_{_{13}}x_{_{23}})^{D/2-2}
\nonumber\\
&&\hspace{4.0cm}\times
\sum\limits_{j=0}^\infty\sum\limits_{n_{_1}=1}^\infty
\xi_{_{m3}}^{j+n_{_1}}x_{_{13}}^{n_{_1}}x_{_{23}}^{-n_{_1}}\Gamma(4-D+2j+2n_{_1})
\nonumber\\
&&\hspace{4.0cm}\times
{(-)^{n_{_1}}\Gamma({D\over2}-1+j+n_{_1})\Gamma(2-{D\over2}+n_{_1})
\over(j+n_{_1})!(j+2n_{_1})!\Gamma(1+j)\Gamma({D\over2}-1+n_{_1})}
\nonumber\\
&&\hspace{4.0cm}
-{i(-)^{D/2-3}(p_{_3}^2)^{D/2-3}\over(4\pi)^{D/2}}
\Gamma(2-{D\over2})(x_{_{13}}x_{_{23}})^{D/2-2}
\nonumber\\
&&\hspace{4.0cm}\times
\sum\limits_{j=1}^\infty\sum\limits_{n_{_1}=1}^\infty
\xi_{_{m3}}^{j+n_{_1}}x_{_{13}}^{-j}\Gamma(4-D+2j+2n_{_1})
\nonumber\\
&&\hspace{4.0cm}\times
{\Gamma({D\over2}-1+n_{_1})\Gamma(2-{D\over2}+j)\over((j+n_{_1})!)^2n_{_1}!
\Gamma(4-D+j)}
\nonumber\\
&&\hspace{4.0cm}
-{i(-)^{D/2-3}(p_{_3}^2)^{D/2-3}\over(4\pi)^{D/2}}
{\Gamma^2(2-{D\over2})\Gamma({D\over2}-1)\over\Gamma(D-3)\Gamma(4-D)}(x_{_{13}}x_{_{23}})^{D/2-2}
\nonumber\\
&&\hspace{4.0cm}\times
\sum\limits_{j=1}^\infty\sum\limits_{n_{_1}=0}^\infty
\xi_{_{m3}}^jx_{_{13}}^{n_{_1}}\Gamma(4-D+2j)
\nonumber\\
&&\hspace{4.0cm}\times
{\Gamma({D\over2}-1+j+n_{_1})\Gamma(D-3+n_{_1})
\over(j!)^2(j+n_{_1})!\Gamma({D\over2}-1+n_{_1})}
\nonumber\\
&&\hspace{4.0cm}
-{i(-)^{D/2-3}(p_{_3}^2)^{D/2-3}\over(4\pi)^{D/2}}
\Gamma(2-{D\over2})(x_{_{13}}x_{_{23}})^{D/2-2}
\nonumber\\
&&\hspace{4.0cm}\times
\sum\limits_{j=1}^\infty\sum\limits_{n_{_2}=1}^\infty
\xi_{_{m3}}^{j+n_{_2}}x_{_{23}}^{-j}\Gamma(4-D+2j+2n_{_2})
\nonumber\\
&&\hspace{4.0cm}\times
{\Gamma({D\over2}-1+n_{_2})\Gamma(2-{D\over2}+j)
\over((j+n_{_2})!)^2n_{_2}!\Gamma(4-D+j)}
\nonumber\\
&&\hspace{4.0cm}
-{i(-)^{D/2-3}(p_{_3}^2)^{D/2-3}\over(4\pi)^{D/2}}
{\Gamma^2(2-{D\over2})\Gamma({D\over2}-1)\over\Gamma(D-3)\Gamma(4-D)}(x_{_{13}}x_{_{23}})^{D/2-2}
\nonumber\\
&&\hspace{4.0cm}\times
\sum\limits_{j=1}^\infty\sum\limits_{n_{_2}=0}^\infty
\xi_{_{m3}}^jx_{_{23}}^{n_{_2}}\Gamma(4-D+2j)
\nonumber\\
&&\hspace{4.0cm}\times
{\Gamma({D\over2}-1+j+n_{_2})\Gamma(D-3+n_{_2})\over(j!)^2(j+n_{_2})!
\Gamma({D\over2}-1+n_{_2})}
\nonumber\\
&&\hspace{4.0cm}
+{i(-)^{D/2-3}(p_{_3}^2)^{D/2-3}\over(4\pi)^{D/2}}
{\Gamma^2(2-{D\over2})\over\Gamma(4-D)}(x_{_{13}}x_{_{23}})^{D/2-2}
\nonumber\\
&&\hspace{4.0cm}\times
\sum\limits_{j=1}^\infty \xi_{_{m3}}^j
{\Gamma(4-D+2j)\Gamma({D\over2}-1+j)\over (j!)^3}
\nonumber\\
&&\hspace{4.0cm}
-{i(-)^{D/2-3}(p_{_3}^2)^{D/2-3}\over(4\pi)^{D/2}}
\Gamma(2-{D\over2})\Gamma({D\over2}-1)(x_{_{13}}x_{_{23}})^{D/2-2}
\nonumber\\
&&\hspace{4.0cm}\times
\sum\limits_{j=1}^\infty \xi_{_{m3}}^jx_{_{23}}^{-j}
{\Gamma(4-D+2j)\Gamma(2-{D\over2}+j)\over(j!)^2\Gamma(4-D+j)}
\nonumber\\
&&\hspace{4.0cm}
-{i(-)^{D/2-3}(p_{_3}^2)^{D/2-3}\over(4\pi)^{D/2}}
\Gamma(2-{D\over2})\Gamma({D\over2}-1)(x_{_{13}}x_{_{23}})^{D/2-2}
\nonumber\\
&&\hspace{4.0cm}\times
\sum\limits_{j=1}^\infty \xi_{_{m3}}^jx_{_{13}}^{-j}
{\Gamma(4-D+2j)\Gamma(2-{D\over2}+j)\over(j!)^2\Gamma(4-D+j)}\;.
\label{3M7-1}
\end{eqnarray}

\section{Some coefficients in $\varepsilon$ expansions\label{app2}}
\indent\indent
\begin{eqnarray}
&&A_{_{n_{_1},n_{_2}}}^{(a)}(x,y)=
4\gamma_{_{\rm E}}\ln x\ln y+\ln^2x\ln y
+\ln x\ln^2y+8\psi^3(1+n_{_1}+n_{_2})
\nonumber\\
&&\hspace{2.6cm}
+8\psi^2(1+n_{_1}+n_{_2})\Big[2\gamma_{_{\rm E}}+\ln x+\ln y
-\psi(1+n_{_2})\Big]
\nonumber\\
&&\hspace{2.6cm}
-8\gamma_{_{\rm E}}\ln x\psi( 1 + n_{_2})
-2\ln^2x\psi(1+n_{_2})-2\ln x\ln y\psi(1+n_{_2})
\nonumber\\
&&\hspace{2.6cm}
+8\gamma_{_{\rm E}} \psi^\prime( 1 + n_{_1}+n_{_2})+4\ln x\psi^\prime(1+n_{_1}+n_{_2})
+4\ln y\psi^\prime(1+n_{_1}+n_{_2})
\nonumber\\
&&\hspace{2.6cm}
-4\psi(1+n_{_2})\psi^\prime(1+n_{_1}+n_{_2})
-2\psi(1+n_{_1})\Big[4\gamma_{_{\rm E}}\ln y+\ln x\ln y
\nonumber\\
&&\hspace{2.6cm}
+\ln^2y+4\psi(1+n_{_1}+n_{_2})^2+2\psi(1+n_{_1}+n_{_2})\Big(4\gamma_{_{\rm E}}+\ln x
+2\ln y
\nonumber\\
&&\hspace{2.6cm}
-2\psi(1+n_{_2})\Big)-2\Big(4\gamma_{_{\rm E}}+\ln x+\ln y\Big)\psi(1+n_{_2})
+2\psi^\prime(1+n_{_1}+n_{_2})\Big]
\nonumber\\
&&\hspace{2.6cm}
+2\psi(1+n_{_1}+n_{_2})\Big[4\gamma_{_{\rm E}}\ln x+\ln^2x+4\gamma_{_{\rm E}}\ln y
+3\ln x\ln y+\ln^2y
\nonumber\\
&&\hspace{2.6cm}
-2\Big(4\gamma_{_{\rm E}}+2\ln x+\ln y\Big)\psi(1+n_{_2})
+6\psi^\prime(1+n_{_1}+n_{_2})\Big]
\nonumber\\
&&\hspace{2.6cm}
+2\psi^{\prime\prime}(1 + n_{_1}+n_{_2})\;,
\nonumber\\
&&A_{_{j,n_{_1},n_{_2}}}^{\prime(a)}(x,y)=
\ln x\ln y+4\psi^2(1+j+n_{_1}+n_{_2})-\ln x\psi(1+j+n_{_2})
-\ln y\psi(1+n_{_1})
\nonumber\\
&&\hspace{2.7cm}
+\psi(1+j+n_{_2})\psi(1+n_{_1})+2\psi(1+j+n_{_1}+n_{_2})\Big[\ln(xy)
-\psi(1 +j+n_{_2})
\nonumber\\
&&\hspace{2.7cm}
-\psi(1+n_{_1})-\psi(1 +n_{_2})\Big]-\ln x\psi(1+n_{_2})
+\psi(1+n_{_1})\psi(1+n_{_2})
\nonumber\\
&&\hspace{2.7cm}
+\psi(1+j+n_{_1})\Big[-\ln y-2\psi(1+j+n_{_1}+n_{_2})+\psi(1+j+n_{_2})
\nonumber\\
&&\hspace{2.7cm}
+\psi(1+n_{_2})\Big]+2\psi^\prime(1+j+n_{_1}+n_{_2})\;,
\nonumber\\
&&B_{_{j,n_{_1},n_{_2}}}^{(a)}=
-\ln x_{_{13}}+\psi(1+j+n_{_1}+n_{_2})+\psi(1+n_{_1})-2\psi(1+n_{_1}+n_{_2})\;,
\nonumber\\
&&C_{_{j,n_{_1},n_{_2}}}^{(a)}=
-\ln x_{_{23}}+\psi(1+j+n_{_1}+n_{_2})+\psi(1+n_{_2})-2\psi(1+n_{_1}+n_{_2})\;,
\nonumber\\
&&D_{_{j,n_{_1},n_{_2}}}^{(a)}=
\ln^2(x_{_{13}}x_{_{23}})-\ln^2\xi_{_{33}}
+8\psi^2(j)+\psi^2(j+n_{_1})
\nonumber\\
&&\hspace{2.0cm}
-\psi(j+n_{_1}+n_{_2})\Big[4\ln(x_{_{13}}x_{_{23}})
-2\ln\xi_{_{33}}-3\psi(j+n_{_1}+n_{_2})\Big]
\nonumber\\
&&\hspace{2.0cm}
+\psi(1+j+n_{_1}+n_{_2})\Big[2\ln\xi_{_{33}}
-2\psi(j+n_{_1}+n_{_2})-\psi(1+j+n_{_1}+n_{_2})\Big]
\nonumber\\
&&\hspace{2.0cm}
+2\psi(j)\Big[3\ln(x_{_{13}}x_{_{23}})-\ln\xi_{_{33}}-3\psi(j+n_{_1})
-5\psi(j+n_{_1}+n_{_2})
\nonumber\\
&&\hspace{2.0cm}
+\psi(1+j+n_{_1}+n_{_2})-3\psi(j+n_{_2})\Big]-4\psi^\prime(j)+3\psi^\prime(j+n_{_1}+n_{_2})
\nonumber\\
&&\hspace{2.0cm}
-2\psi(j+n_{_1})\Big[\ln(x_{_{13}}x_{_{23}})-2\psi(j+n_{_1}+n_{_2})
-\psi(j+n_{_2})\Big]
\nonumber\\
&&\hspace{2.0cm}
-\psi(j+n_{_2})\Big[2\ln(x_{_{13}}x_{_{23}})-4\psi(j+n_{_1}+n_{_2})-\psi(j+n_{_2})\Big]
\nonumber\\
&&\hspace{2.0cm}
+\psi^\prime(j+n_{_1})+\psi^\prime(1+j+n_{_1}+n_{_2})+\psi^\prime(j+n_{_2})
\;,\nonumber\\
&&E_{_{j,n_{_1},n_{_2}}}^{(a)}=
\ln x_{_{31}}\ln x_{_{21}}+\psi^2(1+j+n_{_1}+n_{_2})-2\ln x_{_{21}}\psi(1+n_{_1})
\nonumber\\
&&\hspace{2.0cm}
+\ln x_{_{31}}\psi(1+n_{_1}+n_{_2})+\ln x_{_{21}}\psi(1+n_{_1}+n_{_2})
\nonumber\\
&&\hspace{2.0cm}
-2\psi(1+n_{_1})\psi(1+n_{_1}+n_{_2})+\psi^2(1+n_{_1}+n_{_2})
\nonumber\\
&&\hspace{2.0cm}
-\psi(1+j+n_{_2})\Big[\ln x_{_{31}}-2\psi(1+n_{_1})+\psi(1+n_{_1}+n_{_2})\Big]
\nonumber\\
&&\hspace{2.0cm}
+\psi(1+j+n_{_1}+n_{_2})\Big[\ln(x_{_{31}}x_{_{21}})-\psi(1+j+n_{_2})-2\psi(1+n_{_1})
\nonumber\\
&&\hspace{2.0cm}
+2\psi(1+n_{_1}+n_{_2})-\psi(1+n_{_2})\Big]-\ln x_{_{31}}\psi(1+n_{_2})
\nonumber\\
&&\hspace{2.0cm}
+2\psi(1+n_{_1})\psi(1+n_{_2})-\psi(1+n_{_1}+n_{_2})\psi(1+n_{_2})
\nonumber\\
&&\hspace{2.0cm}
+\psi^\prime(1+j+n_{_1}+n_{_2})+\psi^\prime(1+n_{_1}+n_{_2})
\;,\nonumber\\
&&G_{_{j,n_{_1},n_{_2}}}^{(a)}=
2\pi^2+\ln^2x_{_{31}}+2\ln x_{_{31}}\ln x_{_{21}}-\ln^2x_{_{21}} + \ln^2\xi_{_{31}}
+2\psi^2(j)-\psi^2(j+n_{_1})
\nonumber\\
&&\hspace{2.0cm}
-\psi(j+n_{_1}+n_{_2})\Big[4\ln x_{_{31}}-4\ln x_{_{21}}
+2\ln\xi_{_{31}}+3\psi(j+n_{_1}+n_{_2})\Big]
\nonumber\\
&&\hspace{2.0cm}
-\psi(1+j+n_{_1}+n_{_2})\Big[2\ln\xi_{_{31}}-2\psi(j+n_{_1}+n_{_2})
-\psi(1+j+n_{_1}+n_{_2})\Big]
\nonumber\\
&&\hspace{2.0cm}
-2\psi(1+n_{_1})\Big[\ln x_{_{31}}+3\ln x_{_{21}}-\ln\xi_{_{31}}
-5\psi(j+n_{_1}+n_{_2})
\nonumber\\
&&\hspace{2.0cm}
+\psi(1+j+n_{_1}+n_{_2})\Big]
+\psi(1+n_{_1}+n_{_2})\Big[2\ln(x_{_{31}}x_{_{21}})
-4\psi(j+n_{_1}+n_{_2})
\nonumber\\
&&\hspace{2.0cm}
-2\psi(1+n_{_1})+\psi(1+n_{_1}+n_{_2})\Big]
-2\psi(j+n_{_1})\Big[\ln(x_{_{31}}/x_{_{21}})
\nonumber\\
&&\hspace{2.0cm}
+2\psi(j+n_{_1}+n_{_2})-3\psi(1+n_{_1})+\psi(1+n_{_1}+n_{_2})\Big]
\nonumber\\
&&\hspace{2.0cm}
+4\psi(j)\Big[\ln x_{_{31}}-2\psi(1+n_{_1})+\psi(1+n_{_1}+n_{_2})\Big]
-2\psi^\prime(j)-\psi^\prime(j+n_{_1})
\nonumber\\
&&\hspace{2.0cm}
-3\psi^\prime(j+n_{_1}+n_{_2})
-\psi^\prime(1+j+n_{_1}+n_{_2})+\psi^\prime(1+n_{_1}+n_{_2})
\;,\nonumber\\
&&H_{_{j,n_{_1},n_{_2}}}^{(a)}=
\ln x_{_{31}}^2-2\ln x_{_{31}}\ln\xi_{_{31}}
-\psi(1+j+n_{_1}+n_{_2})\Big[2\ln x_{_{31}}+2\ln \xi_{_{31}}-6\psi(1+n_{_1})
\nonumber\\
&&\hspace{2.0cm}
+3\psi(1 +j+n_{_1}+n_{_2})-4\psi(1 +j+n_{_2})
+2\psi(1+n_{_1}+n_{_2})\Big]
\nonumber\\
&&\hspace{2.0cm}
+\psi(1+n_{_1})\Big[4\ln\xi_{_{31}}-2\ln x_{_{31}}-8\psi(1+j+n_{_2})\Big]
+4\ln x_{_{31}}\psi(1+j+n_{_2})
\nonumber\\
&&\hspace{2.0cm}
+\psi(1+n_{_1}+n_{_2})\Big[2\ln x_{_{31}}-2\ln\xi_{_{31}}+4\psi(1+j+n_{_2})
-2\psi(1 +n_{_1})
\nonumber\\
&&\hspace{2.0cm}
+\psi(1+n_{_1}+n_{_2})\Big]
-6\psi(j)\Big[\ln x_{_{31}}+\psi(1+j+n_{_1}+n_{_2})-2\psi(1+n_{_1})
\nonumber\\
&&\hspace{2.0cm}
+\psi(1+n_{_1}+n_{_2})\Big]
+2\psi(1+j)\Big[\ln x_{_{31}}+\psi(1+j+n_{_1}+n_{_2})-2\psi(1+n_{_1})
\nonumber\\
&&\hspace{2.0cm}
+\psi(1+n_{_1}+n_{_2})\Big]
-3\psi^\prime(1+j+n_{_1}+n_{_2})+\psi^\prime(1+n_{_1}+n_{_2})
\;,\nonumber\\
&&P_{_{j,n_{_1},n_{_2}}}^{(a)}=
\ln x_{_{31}}^2+2\ln x_{_{31}}\ln\xi_{_{31}}+\psi^2(1+j+n_{_1})+\psi^2(1+j+n_{_1}+n_{_2})
\nonumber\\
&&\hspace{2.0cm}
-2\ln x_{_{31}}\psi(1+n_{_1})-4\ln\xi_{_{31}}\psi(1+n_{_1})-2\ln x_{_{31}}\psi(n_{_2})
+4\psi(1+n_{_1})\psi(n_{_2})
\nonumber\\
&&\hspace{2.0cm}
+2\psi(1+j+n_{_1}+n_{_2})\Big[\ln x_{_{31}}+\ln\xi_{_{31}}-\psi(1+n_{_1})
-\psi(n_{_2})-\psi( 1 +n_{_2})\Big]
\nonumber\\
&&\hspace{2.0cm}
+2\psi(1+j+n_{_1})\Big[\ln(x_{_{31}}\xi_{_{31}})+\psi(1+j+n_{_1}+n_{_2})
-\psi(1+n_{_1})-\psi(n_{_2})
\nonumber\\
&&\hspace{2.0cm}
-\psi(1 +n_{_2})\Big]-2\ln x_{_{31}}\psi(1+n_{_2})+4\psi(1+n_{_1})\psi(1+n_{_2})
\nonumber\\
&&\hspace{2.0cm}
+\psi^\prime(1+j+n_{_1})+\psi^\prime(1+j+n_{_1}+n_{_2})
\;,\nonumber\\
&&Q_{_{j,n_{_1},n_{_2}}}^{(a)}=
\ln^2\eta_{_{33}}+\psi^2(1+j+n_{_1})-\psi^2(1+j+n_{_1}+n_{_2})
-4\ln\eta_{_{33}}\psi(2+j+n_{_1}+n_{_2})
\nonumber\\
&&\hspace{2.0cm}
-2\psi(1+j+n_{_1}+n_{_2})\psi(2+j+n_{_1}+n_{_2})
+3\psi^2(2+j+n_{_1}+n_{_2})
\nonumber\\
&&\hspace{2.0cm}
+2\ln\eta_{_{33}}\psi(1+j+n_{_2})
-4\psi(2+j+n_{_1}+n_{_2})\psi(1+j+n_{_2})+\psi^2(1+j+n_{_2})
\nonumber\\
&&\hspace{2.0cm}
-2\psi(1+j)\Big[\ln\eta_{_{33}}+\psi(1+j+n_{_1})-\psi(1+j+n_{_1}+n_{_2})
\nonumber\\
&&\hspace{2.0cm}
-3\psi(2+j+n_{_1}+n_{_2})+\psi(1+j+n_{_2})\Big]
\nonumber\\
&&\hspace{2.0cm}
+2\psi(1+j+n_{_1})\Big[\ln\eta_{_{33}}-2\psi(2+j+n_{_1}+n_{_2})
+\psi(1+j+n_{_2})\Big]
\nonumber\\
&&\hspace{2.0cm}
+\psi^\prime(1+j+n_{_1})-\psi^\prime(1+j+n_{_1}+n_{_2})
-3\psi^\prime(2+j+n_{_1}+n_{_2})
\nonumber\\
&&\hspace{2.0cm}
+\psi^\prime(1+j+n_{_2})
\label{app2-1}
\end{eqnarray}

\begin{eqnarray}
&&A_{_{j,n_{_1},n_{_2}}}^{(b)}=
\ln x_{_{13}}\ln x_{_{23}}+\psi^2(1+j+n_{_1}+n_{_2})-\ln x_{_{13}}\psi(1+j+n_{_2})
 \nonumber\\
&&\hspace{2.0cm}
-\ln x_{_{23}}\psi(1+n_{_1})+\psi(1+j+n_{_2})\psi(1+n_{_1})
+\ln x_{_{13}}\psi(1+n_{_1}+n_{_2})
\nonumber\\
&&\hspace{2.0cm}
+\ln x_{_{23}}\psi(1+n_{_1}+n_{_2})-\psi(1+j+n_{_2})\psi(1+n_{_1}+n_{_2})
 \nonumber\\
&&\hspace{2.0cm}
-\psi(1+n_{_1})\psi(1+n_{_1}+n_{_2})+\psi^2(1+n_{_1}+n_{_2})
-\psi(1+j+n_{_1})\Big[\ln x_{_{23}}
\nonumber\\
&&\hspace{2.0cm}
+\psi(1+j+n_{_1}+n_{_2})-\psi(1+j+n_{_2})+\psi(1+n_{_1}+n_{_2})-\psi(1+n_{_2})\Big]
\nonumber\\
&&\hspace{2.0cm}
+\psi(1+j+n_{_1}+n_{_2})\Big[\ln x_{_{13}}+\ln x_{_{23}}-\psi(1+j+n_{_2})-\psi(1+n_{_1})
\nonumber\\
&&\hspace{2.0cm}
+2\psi(1+n_{_1}+n_{_2})-\psi(1+n_{_2})\Big]-\ln x_{_{13}}\psi(1+n_{_2})
+\psi(1+n_{_1})\psi(1+n_{_2})
\nonumber\\
&&\hspace{2.0cm}
-\psi(1+n_{_1}+n_{_2})\psi(1+n_{_2})+\psi^\prime(1+j+n_{_1}+n_{_2})
+\psi^\prime(1+n_{_1}+n_{_2})
\;,\nonumber\\
&&B_{_{j,n_{_1},n_{_2}}}^{(b)}=
2\pi^2 + \ln^2x_{_{13}}+2\ln x_{_{13}}\ln x_{_{23}}-\ln^2x_{_{23}}+\ln^2\xi_{_{m3}}
+2\psi^2(j)-\psi^2(j+n_{_1})
\nonumber\\
&&\hspace{2.0cm}
-4\ln x_{_{13}}\psi(2j+2n_{_1}+2n_{_2})+4\ln x_{_{23}}\psi(2j+2n_{_1}+2n_{_2})
\nonumber\\
&&\hspace{2.0cm}
-4\psi^2(2j+2n_{_1}+2n_{_2})-2\ln\xi_{_{m3}}\psi(1+j+n_{_1}+n_{_2})
+\psi^2(1+j+n_{_1}+n_{_2})
\nonumber\\
&&\hspace{2.0cm}
-4\ln x_{_{23}}\psi(1+j+2n_{_1}+n_{_2})+8\psi(2j+2n_{_1}+2n_{_2})\psi(1+j+2n_{_1}+n_{_2})
\nonumber\\
&&\hspace{2.0cm}
-2\psi^2(1+j+2n_{_1}+n_{_2})-2\ln x_{_{13}}\psi(1+n_{_1})-2\ln x_{_{23}}\psi(1+ n_{_1})
\nonumber\\
&&\hspace{2.0cm}
+4\psi(2j+2n_{_1}+2n_{_2})\psi(1+n_{_1})+\psi^2(1+n_{_1})+2\ln x_{_{13}}\psi(1+n_{_1}+n_{_2})
\nonumber\\
&&\hspace{2.0cm}
+2\ln x_{_{23}}\psi(1+n_{_1}+n_{_2})-4\psi(2j+2n_{_1}+2n_{_2})\psi(1+n_{_1}+n_{_2})
\nonumber\\
&&\hspace{2.0cm}
-2\psi(1+n_{_1})\psi(1+n_{_1}+n_{_2})+\psi^2(1+n_{_1}+n_{_2})
-2\psi(j+n_{_1})\Big[\ln x_{_{13}}
\nonumber\\
&&\hspace{2.0cm}
-\ln x_{_{23}}+2\psi(2j+2n_{_1}+2n_{_2})-2\psi(1+j+2n_{_1}+n_{_2})-\psi(1+n_{_1})
\nonumber\\
&&\hspace{2.0cm}
+\psi(1+n_{_1}+n_{_2})\Big]+4\psi(j)\Big[\ln x_{_{13}}-\psi(1+j+2n_{_1}+n_{_2})
-\psi(1+n_{_1})
\nonumber\\
&&\hspace{2.0cm}
+\psi(1+n_{_1}+n_{_2})\Big]-2\psi^\prime(j)-\psi^\prime(j+n_{_1})
-4\psi^\prime(2j+2n_{_1}+2n_{_2})
\nonumber\\
&&\hspace{2.0cm}
-\psi^\prime(1+j+n_{_1}+n_{_2})
+2\psi^\prime(1+j+2n_{_1}+n_{_2})-\psi^\prime(1+n_{_1})
\nonumber\\
&&\hspace{2.0cm}
+\psi^\prime(1+n_{_1}+n_{_2})
\;,\nonumber\\
&&C_{_{j,n_{_1},n_{_2}}}^{(b)}=B_{_{j,n_{_2},n_{_1}}}^{(b)}(x_{_{13}}\leftrightarrow
x_{_{23}})
\;,\nonumber\\
&&D_{_{j,n_{_1},n_{_2}}}^{(b)}=
\ln^2(x_{_{13}}x_{_{23}})-\ln^2x_{_{m3}}
+4\psi^2(2j+n_{_1}+n_{_2})+\psi^2(j)
\nonumber\\
&&\hspace{2.0cm}
-2\ln x_{_{13}}\psi(j+n_{_1})-2\ln x_{_{23}}\psi(j+n_{_1})
+\psi^2(j+n_{_1})
\nonumber\\
&&\hspace{2.0cm}
-4\ln x_{_{13}}\psi(2j+2n_{_1}+2n_{_2})-4\ln x_{_{23}}\psi(2j+2n_{_1}+2n_{_2})
\nonumber\\
&&\hspace{2.0cm}
+4\psi(j+n_{_1})\psi(2j+2n_{_1}+2n_{_2})+4\psi^2(2j+2n_{_1}+2n_{_2})
\nonumber\\
&&\hspace{2.0cm}
+2\ln x_{_{m3}}\psi(1+j+n_{_1}+n_{_2})-\psi^2(1+j+n_{_1}+n_{_2})
\nonumber\\
&&\hspace{2.0cm}
+2\psi(j)\Big[\ln(x_{_{13}}x_{_{23}})-\psi(j+n_{_1})
-2\psi(2j+2n_{_1}+2n_{_2})-\psi(j+n_{_2})\Big]
\nonumber\\
&&\hspace{2.0cm}
+4\psi(2j+n_{_1}+n_{_2})\Big[\ln(x_{_{13}}x_{_{23}})+\psi(j)-\psi(j+n_{_1})
\nonumber\\
&&\hspace{2.0cm}
-2\psi(2j+2n_{_1}+2n_{_2})
-\psi(j+n_{_2})\Big]-2\ln(x_{_{13}}x_{_{23}})\psi(j+n_{_2})
\nonumber\\
&&\hspace{2.0cm}
+2\psi(j+n_{_1})\psi( j+n_{_2})+4\psi(2j+2n_{_1}+2n_{_2})\psi(j+n_{_2})+\psi^2(j+n_{_2})
\nonumber\\
&&\hspace{2.0cm}
-4\psi^\prime(2j+n_{_1}+n_{_2})-\psi^\prime(j)+\psi^\prime(j+n_{_1})
+4\psi^\prime(2j+2n_{_1}+2n_{_2})
\nonumber\\
&&\hspace{2.0cm}
+\psi^\prime(1+j+n_{_1}+n_{_2})+\psi^\prime(j+n_{_2})\;,
\;,\nonumber\\
&&E_{_{j,n}}^{(b)}(x)=
\gamma_{_{\rm E}}-\ln x-\psi(1+j+n)+\psi(1+j+2n)+\psi(1+n)
\;,\nonumber\\
&&G_{_{j,n}}^{(b)}(x)=
\gamma_{_{\rm E}}-\ln x+\psi(j)-2\psi(j+n)+2\psi(2j+2n)+\psi(n)\;.
\label{app2-2}
\end{eqnarray}

\end{document}